\documentclass[11pt,a4paper]{article}

\usepackage[utf8]{inputenc}
\usepackage[T1]{fontenc}
\usepackage{lmodern}
\usepackage{geometry}
\usepackage{times}
\usepackage{graphicx}
\usepackage{xcolor}
\usepackage{float}
\usepackage{wrapfig}
\usepackage{caption}
\usepackage{subcaption}
\usepackage{amsmath, amssymb}
\usepackage{booktabs}
\usepackage{adjustbox}
\usepackage{multirow}
\usepackage{tabularx}
\usepackage{threeparttable}
\usepackage{placeins}
\usepackage{gensymb}
\usepackage{textcomp}
\usepackage{siunitx}
\usepackage{marginnote}
\usepackage{soulutf8}  % For highlighting and utf8 text
\usepackage{txfonts}   % Astronomy/astrophysics fonts (safe for arXiv)
\usepackage[authoryear]{natbib}
\usepackage{url}
\usepackage{rotating}
\usepackage{pdflscape}
\UseRawInputEncoding

\usepackage[colorlinks = true,
linkcolor = blue,
urlcolor  = blue,
citecolor = blue,
anchorcolor = blue,
breaklinks = true]{hyperref}

\defcitealias{kodric2018_CepheidsM31PAndromeda}{K18}

\title{\textbf{Using Type-II Cepheids as Extragalactic Standard Candles:}\\
Distances to M31}

\author{
  V.~D.~Pipwala$^{1,2,3,4}$\thanks{Corresponding author: 
  \href{mailto:vasudipakkumarpipwala@gmail.com}{vasudipakkumarpipwala@gmail.com}; 
  \href{mailto:vasu.pipwala@stud.uni-heidelberg.de}{vasu.pipwala@stud.uni-heidelberg.de}; 
  \href{mailto:vasudipakkumar.pipwala@students.uniroma2.eu}{vasudipakkumar.pipwala@students.uniroma2.eu}} \and
  H.~N.~Lala$^{5}$ \and
  B.~Lemasle$^{6}$ \and
  E.~K.~Grebel$^{1}$ \and
  G.~Bono$^{2,4}$ \and
  G.~Fiorentino$^{2}$
}

\date{
  \small
  $^{1}$Astronomisches Rechen-Institut, Zentrum f\"ur Astronomie der Universit\"at Heidelberg,\\
M\"onchhofstr.~12--14, D--69120 Heidelberg, Germany
\\
$^{2}$Dipartimento di Fisica, Universit\`a di Roma ``Tor Vergata'',\\
Via della Ricerca Scientifica~1, 00133~Roma, Italy
\\
$^{3}$
Dipartimento di Fisica, Sapienza Universit\`a di Roma,\\
Piazzale A.~Moro~5, 00185~Roma, Italy
\\
$^{4}$
INAF--Osservatorio Astronomico di Roma,\\
Via Frascati~33, I--00078~Monte Porzio Catone, Italy
\\
$^{5}$
CricViz, Mumbai, India
\\
$^{6}$
cosnova GmbH, Am Limespark~2, 65843~Sulzbach, Germany
\\[0.5em]
  \today
}

\begin{document}
\maketitle

\begin{abstract}
Several standard candles have been tested and used to measure accurate extragalactic distances over the past decades. There have been discussions regarding the possibility of using Type-II Cepheids (T2Cs) as an alternative tool, but rarely was this ever implemented. The aim of this project is to assert the use of T2Cs as a new avenue for calibrating the extragalactic distance scale, by using M31 as a benchmark galaxy. Since Ordinary Least Squares regression methods are not immune to outliers and offer an incomplete treatment of the uncertainties, we favor a Bayesian robust regression model to compute new Period--Luminosity (PL) and Period--Wesenheit (PW) relations calibrated using $\sim$100 T2Cs, $\sim$1000 fundamental-mode and $\sim$750 first-overtone classical Cepheids (CCs) in the LMC. Using these relations, we employ a classification routine based on Bhattacharyya distances to filter out any contaminants from the M31 sample. We validate our method by verifying that we retrieve an accurate distance for the LMC. We find a distance to M31 of $24.487\pm0.001$ (statistical) $\pm0.052$ (systematic) mag using CCs and of $24.409\pm0.025$ (statistical) $\pm0.156$ (systematic) mag using T2Cs. Both values are in excellent agreement with literature values derived from meta-analyses, from Hubble Space Telescope (HST) observations of CCs, from the Tip of the Red Giant Branch method, and from HST observations of RR Lyrae. In almost all cases, we reach a relative accuracy better than 98\%, although the archival ground-based data we use cannot compare with HST photometry. We demonstrate that T2Cs can also be used as accurate tracers for determining extragalactic distances, thereby making them excellent candidates for JWST, LSST, and ELT observations. These stars allow us to probe galaxies deprived of young populations and are beyond the reach of the fainter RR Lyrae.
\end{abstract}


% ------------------------------------------------
% MAIN TEXT
% ------------------------------------------------
\newpage
\section{Introduction}
\label{intoduction}

\par Cosmic distance measurements serve as a fundamental pillar for local group archaeology and observational cosmology. Under the local group archaeological studies, these distance measurements are crucial for studying the dynamics, interactions, and evolution of galaxies within our cosmic neighborhood. Observational cosmological studies require accurate and precise extragalactic distances to understand the scale, expansion, composition (dark energy, dark and baryonic matter), and evolution of the universe by constraining the standard $\Lambda$CDM model. In particular, the Hubble constant ($H_{0}$) is the parameter that quantifies the expansion rate of the universe and is an integral part of the $\Lambda$CDM cosmology.
\DeclareSIUnit \parsec {pc}
\par A model-dependent method to estimate the $H_{0}$ value based on cosmic microwave background (CMB) measurements has become available since the year 2000. Under the $\Lambda$CDM paradigm, the latest evaluated $H_{0}$ value is 67.4~$\pm$~0.5~\SI{}{\kilo\meter\per\second\per\mega\parsec} \citep{planckcollaboration2020_Planck2018Results}. On the other hand, a model-independent baseline result of 73.04~$\pm$~1.04~\SI{}{\kilo\meter\per\second\per\mega\parsec} has been determined by the SH0ES team \citep{riess2022_ComprehensiveMeasurementLocal} by using Hubble Space Telescope (HST) observations. Since the CMB evaluation takes into account the change in expansion rate since the ``Early universe", their $H_{0}$ should be comparable to the ``Late universe" assessment. However, the 5$\sigma$ discrepancy in these "direct" and "indirect" determinations, triggers the so-called Hubble tension.

\par Over the past decades, several standard candles have been tested to measure extragalactic distances. The most famous one is the distance estimation using Cepheid variables. These stars are broadly divided into three types on the basis of their distinctly different masses, ages, and evolutionary histories: Classical or Type-I Cepheids (hereafter CCs), Type-II Cepheids (hereafter T2Cs), and Anomalous Cepheids (hereafter ACEPs). CCs are young, evolved intermediate-mass and massive stars during central helium burning phases, that make them excellent tracers of young populations in the Milky Way \citep[e.g.,][]{lemasle2013_GalacticAbundanceGradients, dasilva2022_NewHomogeneousMetallicity, trentin2023_CepheidMetallicityLeavitt}, the Magellanic Clouds \citep[e.g.,][]{lemasle2017_DetailedChemicalComposition, romaniello2022_IronOxygenContent}, and in nearby dwarf irregular galaxies \citep[e.g.,][]{neeley2021_VariableStarsLocal}. They are usually found in the thin disk of spiral galaxies, which allowed \cite{lemasle2022_TracingMilkyWay} to investigate the location of spiral arms in the Milky Way. On the contrary, T2Cs are old, low-mass Population-II stars which evolved from the horizontal branch (HB) in the asymptotic giant branch (AGB) and the post-AGB star \citep[e.g.,][]{gingold1985_EvolutionaryStatusType,bono1997_EvolutionaryScenarioMetalpoora,bono2020_EvolutionaryPulsationProperties}. They are therefore located in old Galactic components (halo, bulge, globular clusters) and they have been found both in early and late-type galaxies. Finally, ACEPs are considered be the aftermath either of old (binaries) or 
of single intermediate-age stars that are believed to be in their core-Helium burning phase \citep{demers1971_PhotometryVariablesGlobular, hirshfeld1980_StellarContentDwarf, fiorentino2006_SyntheticPropertiesBright, fiorentino2012_CentralHeliumburningVariable}. Some properties of Cepheid variables are summarized in Table~\ref{Table: CCs vs ACs vs T2Cs}.
 
\par They all follow distinct Period-Luminosity (PL) relations or Leavitt's laws \citep{leavitt1912_Periods25Variable}, with the PL relations of CCs being better-defined thanks to a larger number of stars. As an alternative to these mono-band relations, \citet{madore1982_PeriodluminosityRelationIV} proposed to use two-band Period-Wesenheit (PW) relations, where the Wesenheit index $W$ is a pseudo-magnitude which is reddening-free by construction. \cite{riess2011_SolutionDeterminationHubble} extended this idea on an empirical basis and constructed potentially more robust three-band PW relations, which they used in \cite{riess2022_ComprehensiveMeasurementLocal}.
%. Using this relation and HST observations of CCs located in 37 galaxies hosting 42 Type-Ia Supernovae (SNe Ia), \cite{riess2022_ComprehensiveMeasurementLocal} calibrated their distances with an accuracy up to \bl{TBD} and a precision up to \bl{TBD} under the SH0ES project \vp{they don't provide any value}.
It is strongly suspected that CC's PL/PW relations notably depend on metallicity \citep[][and references therein]{breuval2022_ImprovedCalibrationWavelength}, which acts as an additional variable necessary to accurately quantify the local value of $H_{0}$.\\

\par Several studies \citep{beaton2018_OldAgedPrimaryDistance,bhardwaj2020_HighprecisionDistanceMeasurements, bhardwaj2022_RRLyraeType} discussed the potential of T2Cs as extragalactic distance indicators. Compared to CCs, T2Cs are $\sim$1.5\,mag intrinsically fainter, and their PL relations are less tightly constrained. Several endeavors based on near-infrared (NIR, JHK bands) observations of T2Cs observed in the globular clusters \citep[GCs,][]{matsunaga2006_PeriodluminosityRelationType}, the galactic bulge \citep{groenewegen2008_DistanceGalacticCentre, bhardwaj2017_GalacticBulgePopulation,braga2018_StructureKinematicsType}, and the Magellanic Clouds \citep{matsunaga2009_PeriodluminosityRelationsType,ciechanowska2010_AraucariaProjectDistancea,matsunaga2011_PeriodluminosityRelationsType,ripepi2015_VMCSurveyXIII,bhardwaj2017_LargeMagellanicCloud,soszynski2018_OGLECollectionVariable,wielgorski2022_AbsoluteCalibrationNearinfrared}, indicate that T2Cs PL/PW relations become tighter in the NIR regime. Moreover, they suggest that the PL/PW relations of T2Cs show marginal to no metallicity-dependence according to the spectral window used \citep[see among others,][and references therein]{dicriscienzo2007_SyntheticPropertiesBright,    groenewegen2017_PeriodluminosityPeriodradiusRelations,    bhardwaj2020_HighprecisionDistanceMeasurements,bhardwaj2022_RRLyraeType,das2021_TheoreticalFrameworkBL, ngeow2022_ZwickyTransientFacility}. In contrast, \citet{wielgorski2022_AbsoluteCalibrationNearinfrared} report a significant metallicity dependence of $\sim$-0.2\,mag/dex in each of the $JHK$ bands They also determined a LMC distance modulus of $\mu$=18.540$\pm$0.026 (statistical) $\pm$0.034 (systematic) mag.

\par PL/PW relations for T2Cs in optical bands were also obtained by \cite{demers1971_PhotometryVariablesGlobular,breger1975_PeriodluminositycolorRelationsPulsation,nemec1994_PeriodLuminosityMetallicityRelationsPulsation,alcock1998_MACHOProjectLMC,pritzl2003_HubbleSpaceTelescope,kubiak2003_OpticalGravitationalLensing,matsunaga2011_PeriodluminosityRelationsType,groenewegen2017_PeriodluminosityPeriodradiusRelations,ripepi2019_ReclassificationCepheidsGaia,ripepi2022_GaiaDR3Specific}. The first ever derivation of PL/PW relations for T2Cs in the $gri$ bands was delivered by \citet{ngeow2022_ZwickyTransientFacility} using archival data of T2Cs in 18 GCs compiled in \cite{bhardwaj2022_RRLyraeType} and time-series observations from the Zwicky Transient Facility project \citep[ZTF, ][]{graham2019_ZwickyTransientFacility}. By using [Fe/H] value for the host GCs from the GlObular clusTer Homogeneous Abundances Measurements survey \citep[GOTHAM,][and references therein]{vasquez2018_HomogeneousMetallicitiesRadial}, they also report a negligible metallicity dependence. Through their PW relations, they further estimated the M31 distance modulus to $\mu$=24.423 $\pm$0.026 mags using a sample of $\sim$270 T2Cs from the PAndromeda project \citep[][hereafter \citetalias{kodric2018_CepheidsM31PAndromeda}]{kodric2018_CepheidsM31PAndromeda}.\\
\newcommand{\about}[1]{$\sim$\,#1}
\par Another well-known old (> 10 Gyr), Population-II standard candles are RR~Lyrae stars. Although they are on the horizontal branch, they are very similar to Cepheid variables. In particular, they follow similar PL/PW relations as T2Cs in the near-infrared (NIR) regime \citep{matsunaga2006_PeriodluminosityRelationType, braga2020_SeparationRRLyrae}. However, they are much more numerous for instance, the Magellanic Clouds contain \about{49\,000} RR~Lyrae and only \about{550} T2Cs (Lala et al., private communication). Thus, RR~Lyrae stars are better standard candles from a statistical point of view compared to T2Cs. However, they are a few magnitudes intrinsically fainter than T2Cs, the exact value depends on the pulsation periods and photometric bands considered. This limits their usage as old population tracers to systems that are located at large distances.

\par An alternative, precise, and accurate, primary distance indicator used to derive the local value of $H_{0}$ is the Tip of the Red Giant Branch (TRGB) \citep{freedman2019_CarnegieChicagoHubbleProgram, freedman2020_CalibrationTipRed}. The TRGB method solely relies on detecting (using star counts in a color-magnitude diagram) the upper limit of the tip of the Red Giant branch sequence in a given system \citep{sandage1971_DistanceLocalGroupGalaxy, lee1993_TipRedGiant, madore1995_TipRedGiant}.
TRGB stars are also Population-II stars, but they are on average more luminous than T2Cs and are non-variable in nature.\newline
\indent Lower-mass stellar systems may have sparsely populated RGBs, which may in turn limit the robustness of the method in such systems \citep{madore1995_TipRedGiant}. Otherwise, the TRGB method is extremely robust regarding color and metallicity variations \citep{beaton2018_OldAgedPrimaryDistance}. \cite{freedman2020_CalibrationTipRed} estimated a value of the local Hubble constant at $H_{0}$=69.6 $\pm$ 0.8 (statistical) $\pm$ 1.7 (systematic) \SI{}{\kilo\meter\per\second\per\mega\parsec}. This value holds $\sim$ 1.7$\sigma$ difference with the $H_{0}$ estimated by the SH0ES team, raising another tension between these two model-independent estimates of $H_{0}$. 
 
\par In this study, we intend to test the accuracy of the extragalactic distance scale based on T2Cs, taking advantage of state-of-the-art machine learning methods. Since the Vera Rubin Observatory Legacy Survey of Space and Time \citep[VRO, LSST, ][]{ivezic2019_LSSTScienceDrivers} will observe the Southern sky using the $ugrizy$ filters, we will focus on photometry obtained in these bands from which unfortunately only $gri$ photometric data is available for our M31 sample of CCs, T2Cs, and ACEPs.
Since \cite{pietrzynski2019_DistanceLargeMagellanic} determined the 1\% precise geometrical distance to the LMC from Eclipsing binary systems, we leverage this Cepheid-rich neighbor as an anchor galaxy for the calibration of their PL/PW relations. The inference of robust $gri$ bands PL and $W_{ri}$ PW relations for these LMC Cepheid variables (thanks to a Bayesian Probabilistic approach) is described in Sect.~\ref{LMC}. To investigate the accuracy and precision of our derived PL/PW relations for CCs and T2Cs, we also considered M31 as a benchmark galaxy using PAndromeda data, but we adopted a new outlier rejection method in order to include in our test sample only those stars with a robust classification. This stage is detailed in Sect. \ref{M31}. In Sect. \ref{discussion}, we discuss our results and compare them with previous studies. In Sect. \ref{conclusion}, we summarize our findings.

\begin{table}[!htbp]
\centering
\caption{Age, mass range, and evolutionary stages of different classes of Cepheid variables.}
\label{Table: CCs vs ACs vs T2Cs}
\smallskip
\begin{adjustbox}{max width=\columnwidth}
\begin{tabular}{lccc}
\toprule\toprule
\textbf{Type} & \textbf{Age (Gyr)} & \textbf{Mass\textsuperscript{1} ($\mathrm{M}_\odot$)} & \textbf{Evolutionary Stage} \\
\midrule
Classical Cepheids (CCs) & $<0.5$ & $3$--$15$ & Central helium burning (blue loop phase) \\
\midrule
Type-II Cepheids (T2Cs) & $>10$ & $0.5$--$0.6$ & Asymptotic Giant Branch (AGB) and post-AGB \\
\midrule
Anomalous Cepheids (ACEPs)\textsuperscript{2} & $1$--$6$ or $>10$ & $1.3$--$2.3$ & Central helium burning \\
\bottomrule
\end{tabular}
\end{adjustbox}

\smallskip
\begin{minipage}{0.97\columnwidth}
\footnotesize
\textbf{Notes.} 
\textsuperscript{1} The mass range depends on metallicity. 
\textsuperscript{2} A subset of Type-II Cepheids (notably the peculiar W~Virginis stars) and some Anomalous Cepheids originate from binary mergers rather than single-star evolution. In such cases, ACEPs can be $>10$~Gyr old, while peculiar W~Virginis stars are typically younger than the bulk of the W~Virginis population.
\end{minipage}
\end{table}

%-------------------------------------------------------------------------

\section{Calibration of PL/PW relations}
\label{LMC}

\subsection{Data: the LMC as an anchor galaxy}
\label{LMC_data}

\par To test the reliability of T2Cs as extragalactic distance indicators, we plan to estimate the distance to M31 using both CCs and T2Cs from the \citetalias{kodric2018_CepheidsM31PAndromeda} sample, which is the only large, homogeneous, non-HST sample of M31 Cepheid variables available in the public domain. The \citetalias{kodric2018_CepheidsM31PAndromeda} sample is based on the complete Pan-STARRS1 survey of Andromeda (PAndromeda) in the \textit{gri} bands. To achieve this goal, we derive robust PL/PW relations (see Sect.~\ref{LMC_method}) using Cepheid variables in the LMC. Calibrating PL relations on LMC Cepheids is a common practice in the field, and indeed this choice of our proximate dwarf galaxy is advantageous due to two central reasons. First, all Cepheids in the LMC can be assumed to be at the same distance in the sky, as its inclination is only $\sim$25$^{\circ}$ \citep{inno2016_PanchromaticView, pietrzynski2019_DistanceLargeMagellanic, ripepi2022_VMCSurveyXLVIII}, where a face-on galaxy would have an inclination of 0$^{\circ}$. Moreover, this distance is known with great accuracy \citep[1\%,][thanks to double eclipsing binaries]{pietrzynski2019_DistanceLargeMagellanic}. Second, the reddening in the LMC is low \citep[$<E(V-I)>\,=0.100 \pm 0.043$\,mag,][measured using red clump stars as tracers]{skowron2021_OGLEingMagellanicSystem} and LMC-based PL relations are less impacted by reddening than those drawn from Galactic Cepheid variables, for which the reddening is rapidly changing $\it{wrt}$ the line of sight, especially in the case of CCs that are located in the thin disk.

\par Since we consider the LMC an anchor galaxy in this paper, we need photometric data in the \textit{gri} bands for Cepheids in the LMC. Lala et al. provides an extensive catalog of 5083 LMC Cepheid variables. This catalog is an assembly of data that were reported either in the Optical Gravitational Lensing Experiment \citep[OGLE,][]{soszynski2018_OGLECollectionVariable}, or in the Gaia photometric survey \citep{ripepi2023_GaiaDataRelease, clementini2023_GaiaDataRelease},  which are both designed to access variability properties thanks to their time-domain capabilities.
% https://iopscience.iop.org/article/10.1086/323099

\par
\begin{figure}[!htbp]
\centering
\includegraphics[width=0.7\columnwidth, keepaspectratio]{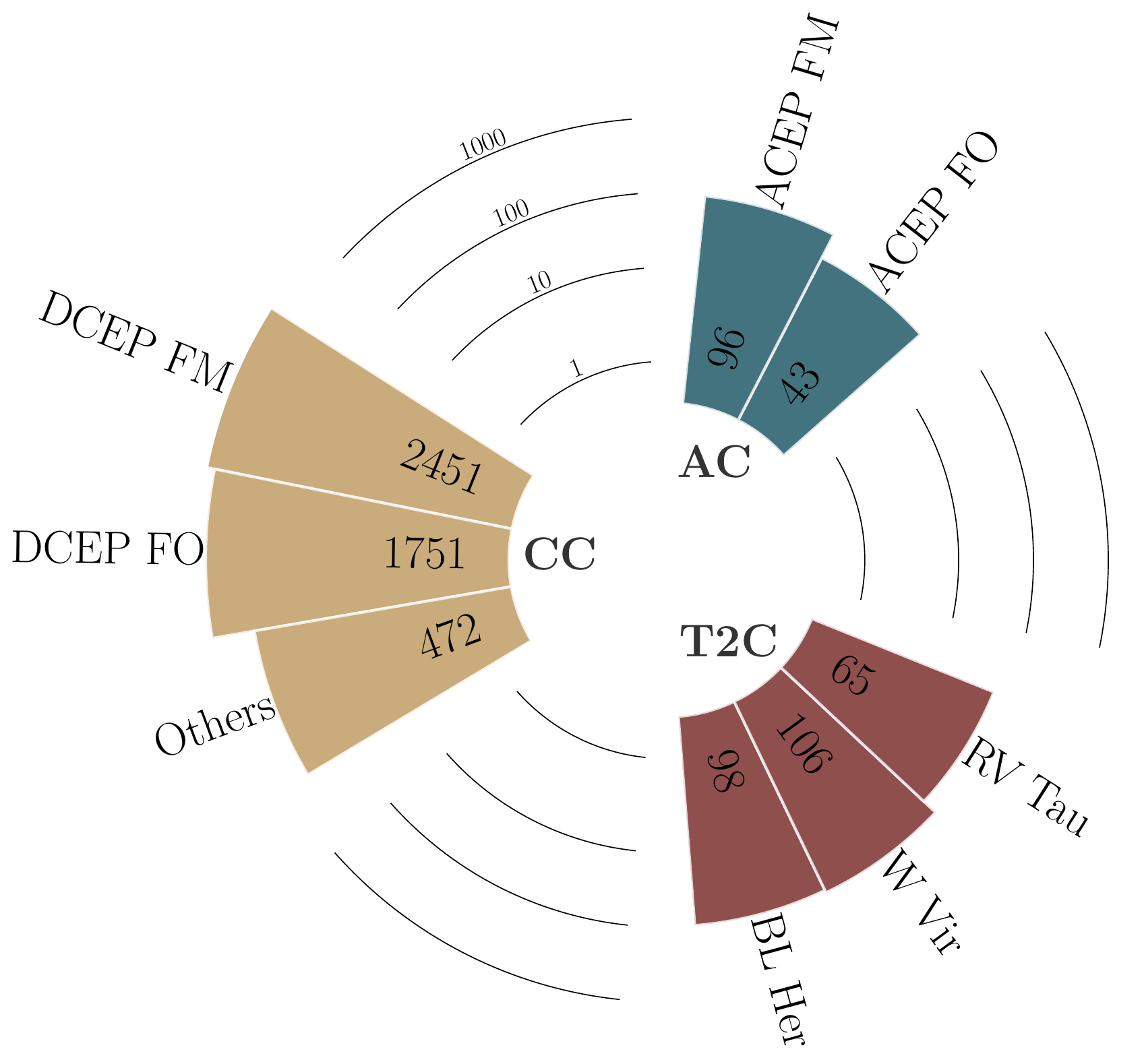}
\caption{Number of LMC stars in different classes of Cepheid variables, as compiled by Lala et al. from numerous photometric surveys.} 
\label{Fig: all_lmc_stars_groupby_type_class}
\end{figure}

 Lala et al. validated the classification of LMC Cepheids based on internal and external comparisons (see their Sect. 4 and 5). The subtype distribution of their validated sample is presented in Fig.~\ref{Fig: all_lmc_stars_groupby_type_class}. The primary classification performed by OGLE/Gaia underlines their various radial pulsation modes: CCs that pulsate in the fundamental radial mode have a period ranging from $\approx$1 to $\approx$100 days and are commonly coined as CCs or $\delta$ Cepheids in the literature. Throughout this paper, we refer to these stars as \textit{DCEP~FM}. Short-period, small-amplitude Cepheids that pulsate in the first-overtone mode (hereafter, \textit{DCEP~FO}) usually have periods ranging from 0.25 to 6 days (the sample from Lala et al. contains 1751 of them). 

\par Moreover, the data from Lala et al. also contain small samples of several multi-mode CCs present in the LMC. We cumulatively categorize these stars as "Others" as shown in Fig.~\ref{Fig: all_lmc_stars_groupby_type_class}. Although it is always possible to "fundamentalize" their periods according to relations found in the literature, we prefer to keep our samples as clean as possible and therefore do not consider these stars further.

\par In contrast, the classification of T2Cs is based on the empirical criteria of their range of variability periods \citep{gingold1985_EvolutionaryStatusType}. The three classes of T2Cs are BL Herculis (\textit{BL Her}), W Virginis (\textit{W Vir}), and RV Tauri (\textit{RV Tau}). Of these classes, \textit{BL Her} stars are the faintest with a period ranging from $\approx$1 to $\approx$5 days. While \textit{W Vir} stars pulsate with periods between $\approx$5 and $\approx$20 days, the brightest T2Cs, the \textit{RV Tau} stars, have a period greater than $\approx$20 days. Through OGLE and Gaia survey data, Lala et al. collectively provides 269 of these T2Cs in the LMC. 

\par Another class of Cepheids, the Anomalous Cepheids (ACEPs) populate the same region on the color-magnitude diagram (CMD) as the short-period CCs. They obey to different PL/PW relations than CCs and T2Cs, and their luminosity, at a fixed period, falls between CCs and T2Cs. These properties of ACEPs are evident in Fig.~1 and 2 from \cite{monelli2022_RRLyraeStars} where they show the location of variable stars on the V-I, V CMD, and on the $W_{VI}$ PW plot respectively. Having masses of $\sim$ 1 to $\sim$ 2 $M_{\odot}$, their main oscillation modes are fundamental and first overtone, similarly to CCs. The Lala et al. catalog provides 96 fundamental-mode ACEPs (hereafter, \textit{ACEP FM}) and 43 first-overtone ACEPs (hereafter, \textit{ACEP FO}) in the LMC, their pulsation periods fall between $\sim$ 0.5 and $\sim$ 2.5 days. 

\begin{table*}[!htbp]
%\begin{adjustbox}{width=0.992\textwidth, keepaspectratio}
\centering
\caption{Generation of our caLMC calibrating data starting from the Lala et al. catalog. The different lines indicate the number of variable stars in different classes remaining after various cross-matching and processing stages.}
\label{Table: Generation of caLMC data}
\begin{tabular}{cccccccc}
\toprule\toprule
\multicolumn{1}{c}{\multirow{2}{*}{\textbf{XMatch}}}                                                 & \multicolumn{7}{c}{\textbf{$N_{matched}$}}      \\ 
\\
%\midrule 
\multicolumn{1}{c}{}                                                                        & \multicolumn{1}{l}{DCEP FM} & \multicolumn{1}{l}{DCEP FO} & \multicolumn{1}{l}{BL Her} & \multicolumn{1}{l}{W Vir} & \multicolumn{1}{l}{RV Tau} & \multicolumn{1}{l}{ACEP FM} & \multicolumn{1}{l}{ACEP FO} \\
\midrule
\multicolumn{1}{c}{\begin{tabular}[c]{@{}c@{}}Gaia DR3\\ neighborhood\\ table\end{tabular}} & 2165                        & 1515                        & 74                         & 90                        & 59                         & 84                          & 40                          \\
\midrule
SMSS DR2                                                                                    & 2137                        & 1492                        & 71                         & 85                        & 57                         & 74                          & 36                          \\
\midrule
\begin{tabular}[c]{@{}l@{}}OGLE-IV RC\\ reddening map\end{tabular}                          & 2134                        & 1491                        & 69                         & 84                        & 57                         & 66                          & 33                          \\
\bottomrule
\\
Final caLMC data                                                                            & 2134                        & 1491                        & 69                         & 84                        & 57                         & 66                          & 33       
\\
\bottomrule              
\end{tabular}
%\end{adjustbox}
\end{table*}

\par From the coordinates ($\alpha$,$\delta$) of these pulsating stars in Lala et al., we retrieve the Cepheids' photometry in the $gri$ bands from the SkyMapper Southern Survey \citep[SMSS, ][]{onken2019_SkyMapperSouthernSurvey}. SMSS covers the entire southern hemisphere in six optical filters (\textit{u, v, g, r, i, z}). The Data Release 2 (DR2) of SMSS provides 21,000 deg$^2$ of data which delivers complete coverage of the LMC. This was achieved in two steps: first, we cross-matched the Lala et al. catalog with the Gaia DR3 neighborhood tables created by \cite{marrese2019_GaiaDataRelease, marrese2022_GaiaDR3Documentation} using the Gaia source ID.  In order to build this neighborhood table, they propagated the coordinates of Gaia stars to the epoch of SMSS DR2, ensuring that even high proper-motion stars can be matched to their best counterparts. Moreover, they ranked every potential neighbor in the on-sky neighborhood of a star based on the local stellar density. This figure of merit helped us determine the best matching star for every star in the Lala et al. catalog. In this way, we minimize the effects of crowding in the resultant matches of stars and also partially account for the difference in the PSF FWHM between Gaia and SMSS DR2. We then retrieved the photometric data in the \textit{gri} bands from SMSS DR2 with a search radius of $\SI{4}{\arcsecond}$. For these retrieved data, we further combine the $E(V-I)$ reddening map, which is derived from the color properties of LMC Red Clump (RC) stars by \cite{skowron2021_OGLEingMagellanicSystem}. Such an approach to build our calibration data set (hereafter caLMC data) ensures minimal spurious matches between SMSS DR2 and Gaia DR3. The number of stars available after the various cross-matching stages is indicated in Table~\ref{Table: Generation of caLMC data}.

\begin{figure}[!htbp]
\centering
\includegraphics[width=0.635\columnwidth, keepaspectratio]{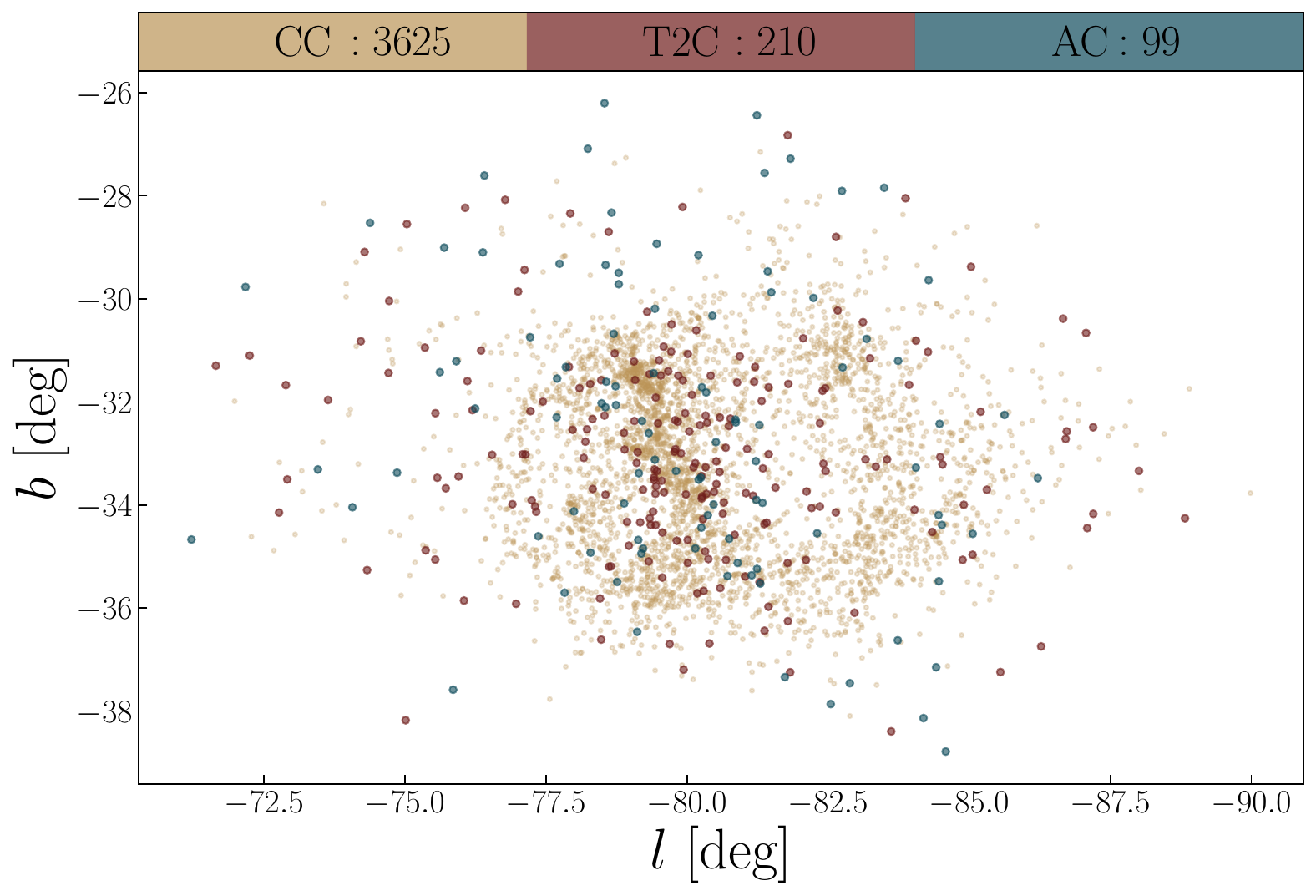}
\caption{On-sky distribution (in Galactic coordinates) of LMC CCs, ACEPs, and T2Cs in our caLMC data. 
%\bl{TBD keep the legend inside, and remove the title}.
} 
\label{Fig: lmc_lwrap_b_map}
\end{figure}

Fig.~\ref{Fig: lmc_lwrap_b_map} shows the on-sky distribution of the various classes of LMC Cepheids in caLMC data. As expected, CCs are mainly located across the LMC disk/bar while T2Cs and ACEPs display a broader radial distribution. 

\par
\begin{figure}[!htbp]
\centering
\includegraphics[width=0.69\columnwidth, keepaspectratio]{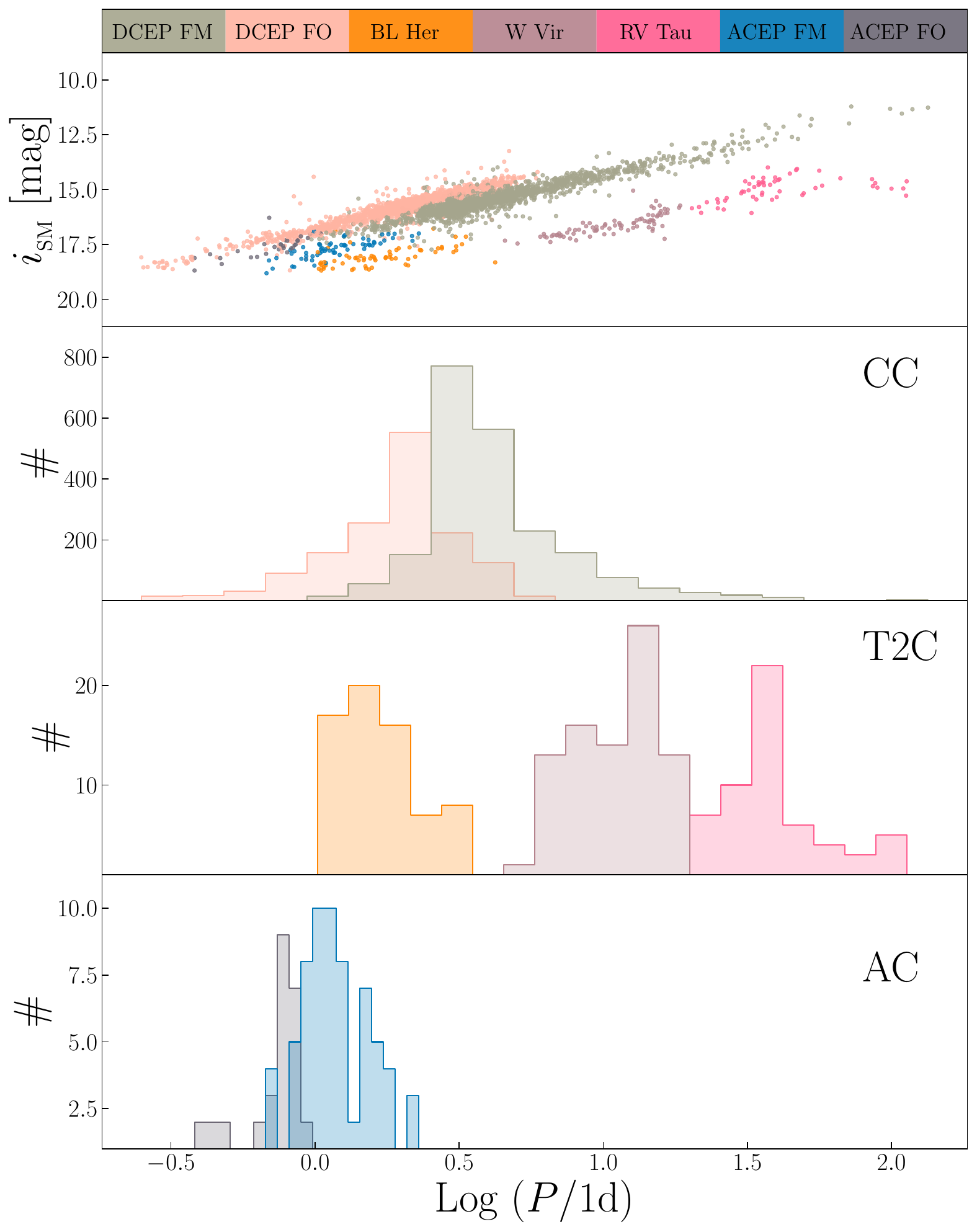}
\caption{Distribution of caLMC calibrating Cepheids per classes (color-coded as per the top bar) and period bins. The upper panel, showing the distribution of the stars in the period-$i$ band luminosity space, confirms that they are properly classified and follow different PL relations. 
%\bl{TBD keep the legend inside (top left, 1st panel only), and remove the title}
} 
\label{Fig: lmc_PL_per_dist}
\end{figure}

 The top panel of Fig.~\ref{Fig: lmc_PL_per_dist} shows the $i$-band PL relations of Cepheids in the caLMC sample, exemplifying their PL relations. We note that no star seems to be wrongly classified (based on its location in this plot). The bottom panels of Fig.~\ref{Fig: lmc_PL_per_dist} display the period distributions of Cepheids in various classes, confirming that ACEPs overlap with BL Her stars and short-period DCEP FMs and DCEP FOs, while the long-period T2Cs of classes W Vir and RV Tau overlap with DCEP FM stars in terms of periods.\\

\begin{figure}[!htbp]
\centering
\includegraphics[width=0.6\columnwidth, keepaspectratio]{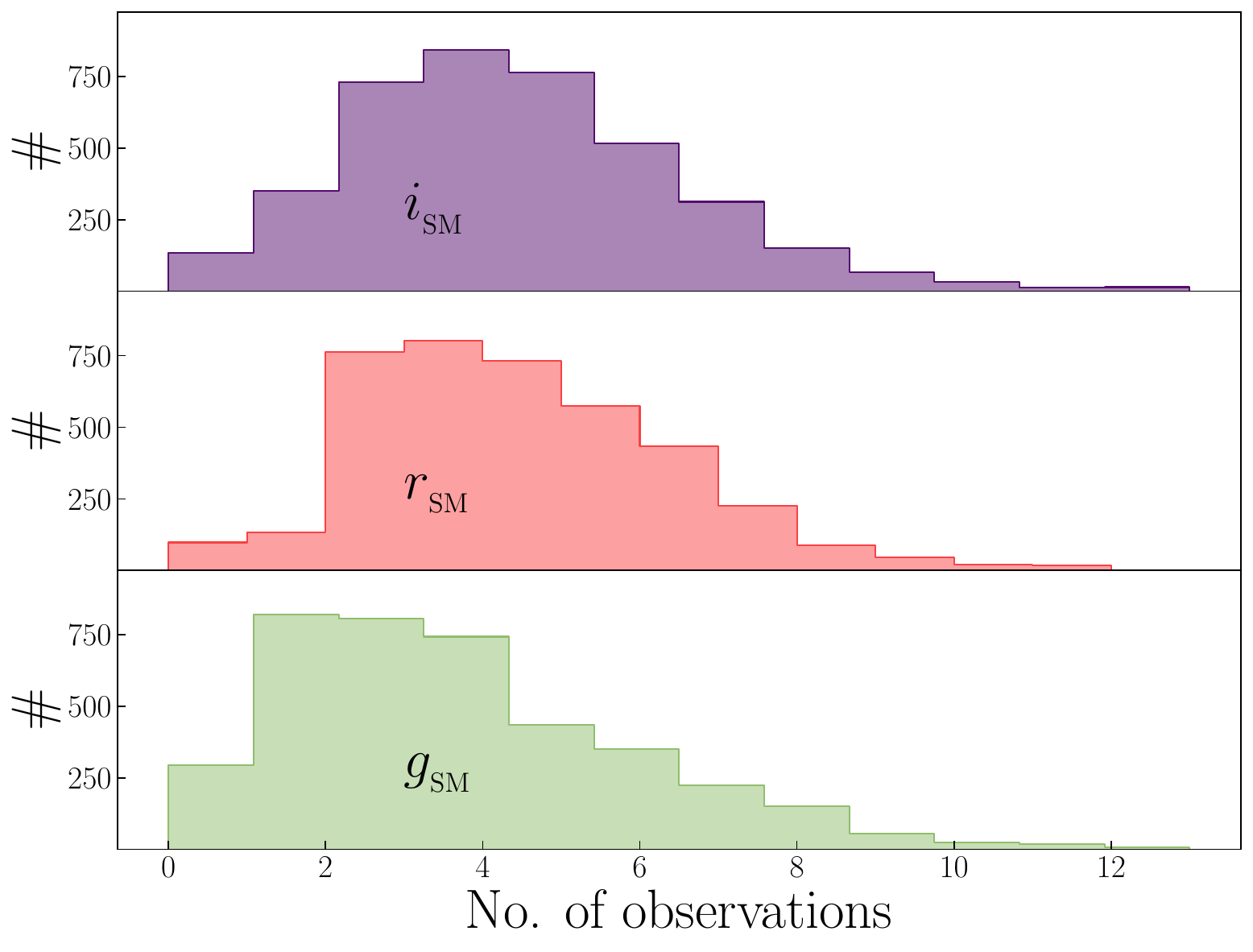}
\caption{Distribution of the number of observations per Cepheid (all classes combined) in our caLMC catalog (see Table~\ref{Table: Generation of caLMC data})}
\label{Fig: lmc_ngood}
\end{figure}

\par Fig.~\ref{Fig: lmc_ngood} indicates the number of good observations in each band for Cepheids in caLMC. We note that the number of good observations decreases from the $i$ to the $g$ band, and, more importantly, that the number of observations is very low, ranging from a few to a maximum of $\sim$ 10 points in the light curve. Such a very small number indicates that we would be unable to classify the Cepheids, or even identify them as Cepheids, solely based on the SMSS DR2. Only the variable nature of these stars could have been noticed. In comparison, OGLE and Gaia light curves contain several 10s to several 100s points in individual Cepheids light curves.  Such a small number of points in the light curves has severe consequences. Indeed, PL relations use the average magnitude of a variable star as input, which is obtained via a Fourier fit to the light curve. Obviously, increasing the number of data points in the light curves helps populate various pulsation phases and improves the quality of the fit, hence the determination of the average magnitude. When the number of points is large and they are well distributed over the period, it is even possible to recover accurate mean magnitudes via a simple arithmetic mean of all observed magnitudes. Even if light curve templates were available in the $gri$ bands (they are not), fitting so sparsely populated light curves would be a challenge. We are left with averaging a small number of apparent magnitudes as the only option, which leads to large uncertainties. However, they cannot exceed half of the peak-to-peak magnitude excursion, in the worst cases where all available observations would have been performed at the maximum (respectively, minimum) brightness. 

\par Since theoretical estimates of $gri$-band amplitudes have not been computed, we derived them empirically. By comparing the $gri$-band photometry of Zwicky Transient Facility Cepheids in the Milky Way (c.f. Sect.~2.3 of Lala et al.) with that of the OGLE-IV stars, we determined the amplitude ratios Amp$_{gri}$/Amp$_{I}$. Using these ratios and the $I$-band amplitudes of stars in the caLMC sample, we derived an estimate of Amp$_{gri}$ for all these stars. To compute the effect of SkyMapper's random sampling on our stars, we computed the uncertainty on mean magnitude, $\sigma_{\text{fit}}$, as ${\frac{\text{Amp}_{gri}}{\sqrt{12 N}}}$, where, $N$ is the number of observations in the SMSS DR2 light curve for a given star. The factor 12 in the denominator arises under the assumption that the magntiude of a star follows a uniform distribution bounded by its amplitude (\citealp{madore2005_NonuniformSamplingPeriodic} and Lala et al.). On average, this $\sigma_{\text{fit}}$ amounts to be \about{0.06}\,mag in the $g$ band, \about{0.04}\,mag in the $r$ band, and \about{0.03}\,mag in the $i$ band. For every star, we added its corresponding $\sigma_{\text{fit}}$ values in quadrature with the photometric uncertainty associated with its light curve. These uncertainties contribute to the error budget of our analysis as statistical uncertainties and are discussed further in Sect.~\ref{M31_result}.

\begin{table}[!htbp]
\centering
\caption{Number of Cepheids of different classes remaining after filtering SMSS DR\,2 photometry according to their survey flags.}
\label{Table: photometric filtering scheme}
\renewcommand{\arraystretch}{1.3} % adds vertical spacing between rows
\begin{tabular}{lcccc}
\toprule\toprule
\textbf{Class} & \textbf{$g$} & \textbf{$r$} & \textbf{$i$} & \textbf{$W_{ri}$} \\ 
\midrule
DCEP FM      & 995 & 885 & 842 & 573 \\
DCEP FO      & 738 & 694 & 655 & 478 \\
T2C: BL Her  & 19  & 21  & 37  & 17  \\
T2C: W Vir   & 48  & 47  & 46  & 33  \\
T2C: RV Tau  & 26  & 23  & 24  & 14  \\
ACEP FM      & 21  & 27  & 27  & 20  \\
ACEP FO      & 11  & 13  & 16  & 11  \\ 
\bottomrule
\end{tabular}
\end{table}

\par We also filter the photometric data based on SMSS DR2 flags \citep{onken2019_SkyMapperSouthernSurvey}. First, SMSS DR2 provides a stellarity index ($class\_star$) varying between 0 (no star) and 1 (a star). For the calibration of our PL/PW relations, we retained only those stars in our caLMC catalog that have a stellarity index $>$ 0.6 in order to refrain from non-stellar contaminants. The $flags\_psf$ flags contain bits to indicate whether a neighbor is expected to bias the photometry by $>$ 1\% in each band. We then remove the corresponding stars with bits 2 (for $i$ band), 4 (for $r$ band), and 8 (for $g$ band) and their bit number combinations. Further down this line, there are filter-specific flags columns ($x\_flags$ \& $x\_nimaflags$; where $x$ indicates a filter) which indicate source extractor warnings about saturation, close neighbors, edge-of-CCD effects as well as warning for bad pixel, cross-talk, and cosmic rays. Therefore, we only use stars for which $x\_flags$ and $x\_nimaflags$ are 0. Finally, we retain only the stars that have been observed at least twice ($x\_ngood$$\geq$2) in the $gri$ bands. Table~\ref{Table: photometric filtering scheme} lists the number of stars remaining in each band after the filtering of SMSS DR2 data.

\newcommand{\twoWabs}{W^{\mathrm{abs}}_{12}}
\newcommand{\twoWapp}{w^{\mathrm{app}}_{12}}
\newcommand{\twoW}{W_{12}}

\newcommand{\threeWapp}{w^{\mathrm{app}}_{123}}
\newcommand{\threeW}{W_{123}}

\newcommand{\Wir}{W^{\mathrm{abs}}_{ir}}
\newcommand{\wir}{w^{\mathrm{app}}_{ir}}

\newcommand{\wri}{w^{\mathrm{app}}_{ri}}

\newcommand{\Wband}[1]{W^{\mathrm{abs}}_{#1}}

\subsection{Method: Derivation of PL/PW relations}
\label{LMC_method}

\subsubsection{PL/PW relations}

\par With such data at hands, we want to derive the coefficients of the PL and PW relations which can be written as shown in  Eqn.~\ref{EqPL} and \ref{EqPW}, respectively:
\begin{equation}
\hspace{1cm}
    M = \alpha + \beta \times (\mathrm{log}P - \mathrm{log}P_{0}),
    \label{EqPL}
\end{equation}

\begin{equation}
\hspace{1cm}
     \twoWabs = \alpha + \beta \times  (\mathrm{log}P - \mathrm{log}P_{0}),
    \label{EqPW}
\end{equation}

where $M$ is the absolute magnitude, $\twoWabs$ is the absolute Wesenheit index, $P$ is the period of the pulsating star, $\alpha$ is the intercept and $\beta$ is the slope of the PL/PW relation. The term $\mathrm{log}P_{0}$ is the mean logarithmic period of the Cepheid variable subsamples based on different classes. We particularly subtract $\mathrm{log}P_{0}$ from $\mathrm{log}P$ in order to minimize the correlation between $\alpha$ and $\beta$. The true $\mathrm{log}P_{0}$ for DCEP FM and T2C subsamples in caLMC data are $\sim$ 0.7\,d and $\sim$ 1.2\,d, respectively. Hence, for simplicity and generality, we adopt $\mathrm{log}P_{0} = 1$ in Eqns \ref{EqPL} and \ref{EqPW} for all the Cepheids subsamples in caLMC. In order to estimate $\twoWabs$, we first need to derive the apparent Wesenheit index $\twoWapp$, which is described in  Eqn.~\ref{EqW12}:

\begin{equation}
\hspace{1cm}
    \twoWapp = \lambda^{\mathrm{mag}}_{1} - R_{12} \times ( \lambda^{\mathrm{mag}}_{1} - \lambda^{\mathrm{mag}}_{2} ),
    \label{EqW12} 
\end{equation}

where $\lambda_{1}$ < $\lambda_{2}$. $R_{12}$ is the total-to-selective extinction ratio which is given by:

\begin{equation}
\hspace{1cm}
    %R_{12} = \frac{A_{\lambda_{1}}}{A_{\lambda_{1}} - A_{\lambda_{2}}},
    R_{12} = \frac{A_{\lambda_{1}}}{E(\lambda_{1} - \lambda_{2})},
    \label{EqR12}
\end{equation}

where $A_{\lambda_{1}}$ is the interstellar extinction in the waveband $\lambda_{1}$, and $E(\lambda_{1} - \lambda_{2})$ is the reddening or color excess. This formulation of $\twoWapp$ in Eqn.~\ref{EqW12} provides the star's pseudo-magnitudes which are reddening-free in nature \citep{ madore1982_PeriodluminosityRelationIV}. Beyond uncertainties on its true value, the potential variability of \textit{R} depending on the direction of the photometric observation is the only way Wesenheit-indices ($\twoW$) can depend on reddening. Thus, PW relations provide more accurate results than PL relations by minimizing the effects of extinction.

\par Based on empirical evidence, \cite{riess2011_SolutionDeterminationHubble} suggested using a three-band PW, for which the Wesenheit index is estimated as follows:

\begin{equation}
\hspace{1cm}
    \hspace{1cm}
    \threeWapp = \lambda^{\mathrm{mag}}_{1} + R_{123} \times ( \lambda^{\mathrm{mag}}_{3} - \lambda^{\mathrm{mag}}_{2} ),
    \label{EqW123} 
\end{equation} 

where $\lambda_{1}$ > $\lambda_{2}$ > $\lambda_{3}$ and $R_{123}$ is the ratio between the selective absorption in the $\lambda_{1}$ band and the color excess $E(\lambda_{2}$ - $\lambda_{3})$. 

According to this paper, the benefits of this $\threeW$ PW relation are twofold: (i) it usually has a small intrinsic dispersion compared to the two-band PW relation; (ii) When the same magnitude is used in the color term of $\threeWapp$, the resultant PW is less prone to systematics. By adopting the optical WFC3-UVIS (\textit{F555W} and \textit{F814W}) bands and the NIR WFC3-IR (\textit{F160W}) band of HST, \cite{li2021_Sub2DistanceM31} used this approach to deliver the most precise distance of M31 ($\mu$=24.407 $\pm$ 0.032 mag) to date. In Sect.~\ref{M31_result}, we use this value as a baseline result to compare our multiple estimates of the M31 distance.\\

\par For PL relations, we correct for the interstellar extinction A$_{y}$ in the individual bands $gri$ as follows:
\begin{equation}
\hspace{1cm}
A_{y}= \frac{R_{y}}{E(B-V)}, \quad \mathrm{with} \hspace{0.15cm} y \in \{g, r, i\};
\end{equation}
where $A_{y}$ is the interstellar extinction in the $y$ band, and $R_{y}$, its extinction coefficient, is taken from the SkyMapper website\footnote{\url{https://skymapper.anu.edu.au/filter-transformations/}}. $R_{y}$ is estimated assuming the reddening law by \cite{fitzpatrick1999_CorrectingEffectsInterstellar}; and $E(B-V)$, the color excess, is converted from the $E(V-I)$ reddening map tabulated by \cite{skowron2021_OGLEingMagellanicSystem} using the conversion factor of 1.237 provided in their  Eqn.~11. As will be explained in Sect.~\ref{LMC_result_subresult}, we do not consider a three-band PW relation and we estimate only $W_{ri}$ among the three potential two-band Wesenheit indices. We form $R_{ri}$ (see Eqn.~\ref{EqR12}) as:
\begin{equation}
\hspace{1cm}
R_{ri}= \frac{A_{r}}{E(r - i)} = \frac{R_{r}}{R_{r}-R_{i}}=3.269
%= \frac{A_{r}}{A_{r}-A_{i}}
\end{equation}
where $R_{ri}$ is the selective-to-total extinction ratio.\\

\par Finally, we calibrate the extinction corrected \textit{gri} PL and the reddening-free $W_{ri}$ PW relation for CCs, T2Cs, and ACs through the 1\% precise distance estimate of the LMC ($\mu$ = 18.477 $\pm$ 0.026 mag) obtained by \citet{pietrzynski2019_DistanceLargeMagellanic} using eclipsing binary systems.

\subsubsection{Bayesian robust regression}

\par Least-squares regression has been until today the traditional approach to fit PL/PW relations. It is simple and straightforward, which makes it computationally cheap. However, it is not very robust when the data sets do not follow standard assumptions or contain outliers.
\par Here, we compute accurate and precise PL/PW relations through a Bayesian robust regression (hereafter, BRR) model developed with the Python package \texttt{pymc3} \citep{salvatier2016_PyMC3PythonProbabilistic}. It delivers a probabilistic programming framework for carrying out Bayesian modeling and visualization. Our BRR model is defined as follows:

\begin{equation}
\hspace{1cm}
    M \text{ or } \twoWabs \sim \mathcal{T} ( \alpha + \beta \times \mathrm{log}_{10}\left\{ \frac{P}{10 \text{ d }} \right\} , \sigma^2 , \nu ),
    \label{EqBRR}
\end{equation}

where $\alpha$ and $\beta$ are the prior probability distributions for the intercept and the slope of the linear regression model. We assumed that these two priors follow a normal distribution with a mean of 0 and a standard deviation of 10, hence we chose very uninformative priors.
As is commonly done in Bayesian inference applications, we utilized a half-normal distribution as the prior probability distribution for the intrinsic scatter ($\sigma$) of the PL/PW relations.
\par The key aspect of our analysis is that we assume that the likelihood of the model follows a Student's $t$-distribution, a generalization of the normal distribution with prominent wings that are constrained by the additional parameter $\nu$, for which we adopted a $\Gamma$ distribution \citep{juarez2010_ModelBasedClusteringNonGaussian}. Thanks to the extended wings of the Student's $t$-distribution, it is possible to attribute a very low weight to outliers. In contrast, the narrow wings of the distribution tend to either fully exclude or attribute a high weight to outliers, thus biasing towards them the derived relations.
\par
\begin{figure}[!htbp]
\centering
\includegraphics[width=0.7\columnwidth, keepaspectratio]{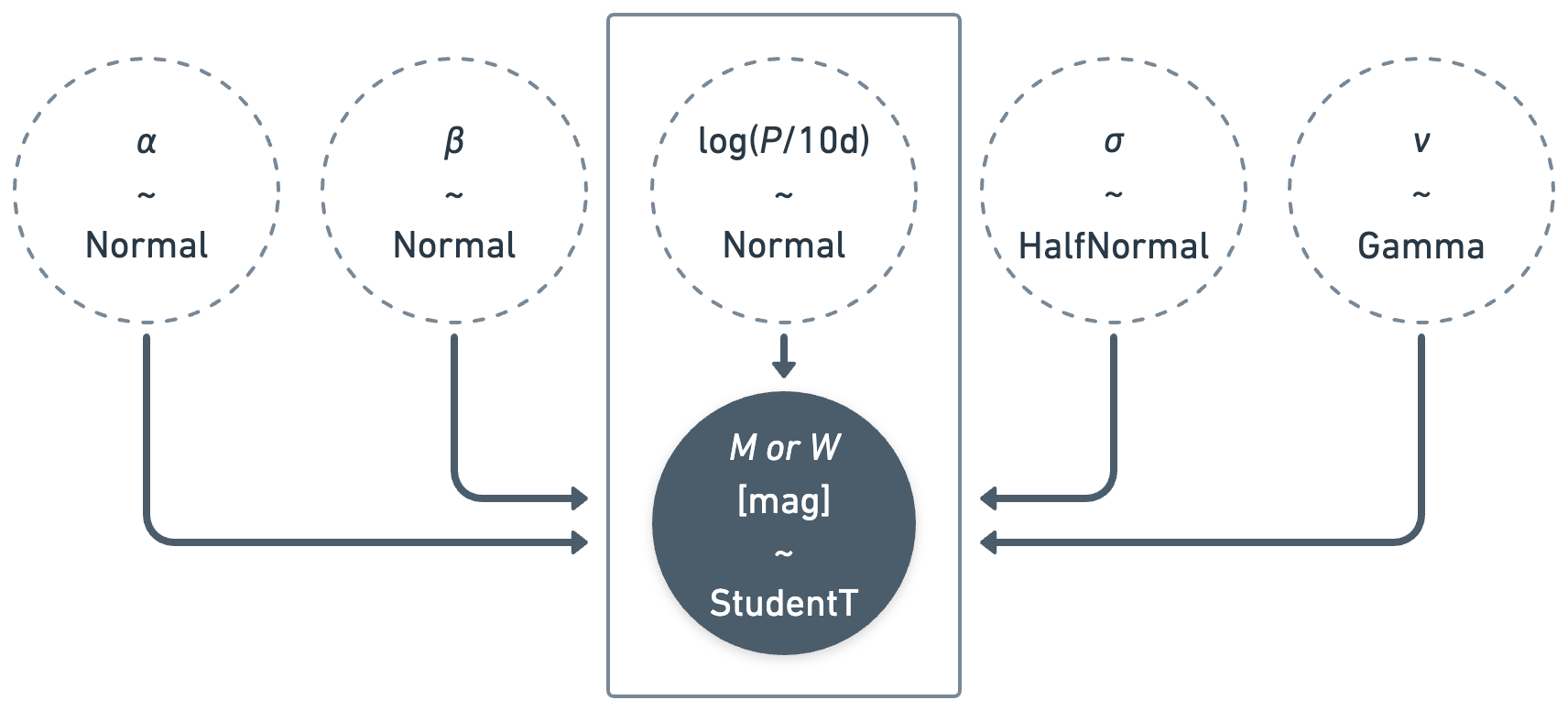}
\caption{Graphical description of the Bayesian robust regression (BRR) model used to determine PL/PW relations. The important feature of the model is that its likelihood is assumed to follow a student-t distribution.} 
\label{Fig: model}
\end{figure}

 The mean of the likelihood distribution is the form of \textit{M} or $\twoWabs$ shown in Eqns.~\ref{EqPL} or \ref{EqPW}, while its standard deviation is the sum in quadrature of $\sigma$ and the uncertainties on \textit{M} or $\twoWabs$. The chart summarizing the BRR model is shown in Fig.~\ref{Fig: model}.

\par Should the likelihood of the model be replaced with a normal distribution, then the model would become a Bayesian least squares (hereafter, BLS) model. Such a BLS fit is already an improvement to the ordinary least squares (hereafter, OLS) approach that has been used in most of the previous studies because the BLS method offers a better treatment of the uncertainties than the OLS.

\subsection{Results: Estimation of PL/PW relation}
\label{LMC_result}

\subsubsection{Results}
\label{LMC_result_subresult}

\par PL and PW relations for CCs, T2Cs, and ACEPs are calibrated on LMC Cepheids using the Bayesian probabilistic method mentioned in Sect. \ref{LMC_method}. Thanks to this approach, we are able to not only provide the slope and intercept and their uncertainties but also the intrinsic dispersion (and its uncertainty) of these individual PL/PW relations. All these values are tabulated in Sect.~\ref{LMC_compare} along with the portrayal of these relations.

\par For the CCs and ACEPs, we derived separate laws for stars pulsating in the fundamental or the first overtone mode. For the T2Cs, we derive laws for individual classes, as well as for the traditional combinations BL Her + W Vir and BL Her + W Vir + RV Tau. It turns out that the \citetalias{kodric2018_CepheidsM31PAndromeda} sample is deprived of BL Her stars, therefore we also determine PL/PW relations by combining W Vir + RV Tau stars. We note in passing that PL/PW relations for CCs are derived using several 1000s of stars, while those for T2Cs and ACEPs rely only on a few tens of stars (see Table~\ref{Table: photometric filtering scheme}). However, our LMC sample of T2Cs and ACEPs remains larger than any other sample in the $gri$ bands in the literature. 

\par The intrinsic dispersion of the PL/PW relations usually decreases when moving toward a longer wavelength, because of the larger intrinsic spread of the instability strip in the bluer bands \citep{madore1991_CepheidDistanceScale}. However, the dispersion of our $g$-band PL relation remains notably too high with respect to that in the $r,i$ bands. The reasons are twofold: first, the \textit{g} band photometry is affected by larger amounts of interstellar extinction than the \textit{r} and \textit{i} bands; second, and more importantly, the luminosity amplitude in the \textit{g} band is larger, leading to larger errors when computing the average magnitude from only a few measurements at random pulsation phases (see Sect.~\ref{LMC_data}). For this reason, we do not compute a three-band PW relation with all the $gri$ bands and limit ourselves to a single two-band PW relation using the \textit{r} and \textit{i} bands. 
\par For the same reason (accuracy of $r$ and $i$ mean magnitudes), the dispersion of the $W_{ri}$ PW relation remains slightly larger than for the individual PL relations in the $r$ and $i$ bands. \\

\subsubsection{Validation using the distance to the LMC}
\label{LMC_result_validation}

\par In order to validate the BRR method, we perform a Monte Carlo analysis and verify that we retrieve the proper distance to the LMC using the PL/PW relations we derived. First, since we derived PL/PW relations of various classes of Cepheid variables, we create a covariance matrix between the uncertainties on the slope ($\beta$) and on the intercept ($\alpha$) we derived and the uncertainties on the log$P$ for each group of variables in our caLMC data set. We are then able to create a single multivariate distribution taking into account $\alpha$, $\beta$, and their uncertainties (from the BRR fit), and log$P$ and their uncertainties from observational data. For every single star in caLMC, we then draw $10^{5}$ random samples from this multivariate distribution. The mean values of each parameter consequently provide us with the absolute magnitude/Wesenheit of this star. In turn, those are converted into individual stellar distances $\mu_{x,i}$ ; where $x$ denotes the class $x$ of Cepheid variables, and $i$ its associated variable star.\\

\par Finally, by grouping individual stellar distances by class(es) of variable stars, we compute the median distance modulus ($\tilde{\mu}_{LMC,x}$) of the LMC as given by this class. Since T2Cs and ACs are located on average at larger LMC radial distances \citep[and likely not uniformly distributed,][]{iwanek2018_ThreeDimensionalDistributionsType}, and since we are not performing geometric corrections, selecting the median over the mean helps to minimize the effect of the probable skewness of their spatial distribution. The uncertainty on this median distance  modulus ($\tilde{\mu}_{err}$) is estimated (for a given class of Cepheid variables) by:

\begin{equation}
\hspace{1cm}
    \tilde{\mu}_{err} = \sqrt{\frac{\pi}{2}} \cdot \frac{\sigma_{\mu,x}}{\sqrt{N_{x}}}
    \label{uncertainty on median distance}
\end{equation}

where $\sigma_{\mu,x}$ is the standard deviation of all the stellar distance moduli computed for the class $x$ of Cepheid variables and N$_{x}$ the number of stars that belong to the class $x$. We further estimate the statistical uncertainty on the distance modulus ($\mu_{err}^{stat}$) by:

\begin{equation}
\hspace{1cm}
    \mu_{err}^{stat} = \frac{\sqrt{\sum{( \sigma_{m,x,y,i}^2 + \sigma_{A,x,y,i}^2 + \beta_{sd,x,y}^2}} )}{N_{x}}
    \label{statistical uncertainty on distance modulus}
\end{equation}

where $\sigma_{m,x,y,i}$ and $\sigma_{A,x,y,i}$ are the uncertainty on the mean apparent magnitude $m$ and on the extinction $A$ for the star $i$ of class $x$ in the band $y$, respectively. They are estimated from the standard deviation of the $10^{5}$ draws from their corresponding random normal distributions. Note that $\sigma_{m,x,y,i}$ includes both the photometric uncertainty and the $\sigma_{\text{fit}}$ derived in Sect.~\ref{LMC_data}. Further, $\beta_{sd,x,y}$ is the uncertainty on the derived slope $\beta$ (coefficient of Log$P$) in the PL/PW relations of class $x$ in the band $y$ from the BRR model described in Sect.~\ref{LMC_method}. Using the $\beta$ and $\beta_{sd}$ values for the various classes of Cepheid variables that are listed in Sect.~\ref{LMC_compare}, we similarly draw $10^5$ samples and take their standard deviation to determine $\beta_{sd,x,y}$. In order to estimate the systematic uncertainty on the distance modulus ($\mu_{err}^{sys}$), we follow the following formalism:

\begin{equation}
\hspace{1cm}
    \mu_{err}^{sys} = \sqrt{\sigma_{\mu,cal}^2 + \alpha_{sd,x,y}^2 + \beta_{sd,x,y}^2}
    \label{systematic uncertainty on distance modulus}
\end{equation}

where $\sigma_{\mu,cal}$ = 0.026 mag is the total uncertainty on the 1\% precise geometric distance of LMC that we used to calibrate our PL/PW relations in Sect. \ref{LMC_method}. Here, $\beta_{sd,x,y}$ is the uncertainty on the coefficient of Log$P_{0}$ (where Log$P_{0}$ = 1) and $\alpha_{sd,x,y}$ is the uncertainty on the intercept $\alpha$ in the PL/PW relations, similarly estimated as $\beta_{sd,x,y}$.\\

\par From these resultant quantities in  Eqn.~\ref{uncertainty on median distance}, \ref{statistical uncertainty on distance modulus}, and \ref{systematic uncertainty on distance modulus}, it is possible to derive the total uncertainty ($\mu_{err}^{tot}$) on the median LMC distance modulus ($\tilde{\mu}_{LMC,x}$) linked with the various classes of Cepheid variables. This is done by summing in quadrature the uncertainty on the median distance modulus  $\tilde{\mu}_{err}$ from  Eqn.~\ref{uncertainty on median distance}, the statistical uncertainty ($\mu_{err}^{stat}$) from  Eqn.~\ref{statistical uncertainty on distance modulus} and the systematic uncertainty ($\mu_{err}^{sys}$) from  Eqn.~\ref{systematic uncertainty on distance modulus}:

\begin{equation}
\hspace{1cm}
    \mu_{err}^{tot} = \sqrt{ {\tilde{\mu}_{err}}^{2} + {\mu_{err}^{stat}}^{2} + {\mu_{err}^{sys}}^{2}  }
    \label{total unceratinty}
\end{equation}

%--------------------------------------------------------------------------------------------------------------------

\begin{table}[!htbp]
\centering
\caption{Summary of the total error budget for various median distance determinations of the LMC.}
\label{tab: LMC distance error budget}
\begin{tabular}{lll}
\toprule\toprule
\multicolumn{3}{c}{$\mu_{\mathrm{err}}^{\mathrm{tot}}$ on the LMC distance modulus $\tilde{\mu}_{\mathrm{LMC},x}$} \\
\midrule
\textbf{Uncertainty type} & \textbf{Contributing term} & \textbf{Description} \\

\midrule

\begin{tabular}[c]{@{}l@{}}Uncertainty on median \\ distance modulus \\ ($\tilde{\mu}_{\mathrm{err}}$)\end{tabular} 
& $\sigma_{\mu,x}$ 
& \begin{tabular}[c]{@{}l@{}}Standard deviation of all stellar distance moduli \\ ($\sigma_{\mu}$) in class $x$\end{tabular} \\[2.5mm]

\midrule

\multirow{3}{*}{\begin{tabular}[c]{@{}l@{}}Statistical uncertainty \\ ($\mu_{\mathrm{err}}^{\mathrm{stat}}$ per star $i$ \\ of class $x$ in band $y$)\end{tabular}}
& $\sigma_{m,x,y}$ & Uncertainty on mean apparent magnitude $m$ \\[2mm]
& $\sigma_{A,x,y}$ & Uncertainty on interstellar extinction $A$ \\[2mm]
& $\beta_{sd,x,y}$ & \begin{tabular}[c]{@{}l@{}}Uncertainty on slope $\beta$ (coefficient of $\log P$) \\ of PL/PW relations\end{tabular} \\[2.5mm]

\midrule

\multirow{3}{*}{\begin{tabular}[c]{@{}l@{}}Systematic uncertainty \\ ($\mu_{\mathrm{err}}^{\mathrm{sys}}$)\end{tabular}}
& $\sigma_{\mu,cal}$ & \begin{tabular}[c]{@{}l@{}}Uncertainty on 1\% precise LMC calibration \\ distance modulus\end{tabular} \\[3.5mm]
& $\alpha_{sd,x,y}$ & Uncertainty on intercept $\alpha$ of PL/PW relations \\[2mm]
& $\beta_{sd,x,y}$ & \begin{tabular}[c]{@{}l@{}}Uncertainty on slope $\beta$ (coefficient of $\log P_0$, \\ where $\log P_0 = 1$) of PL/PW relations\end{tabular} \\

\bottomrule
\end{tabular}
\end{table}

%--------------------------------------------------------------------------------------------------------------------

\par Table \ref{tab: LMC distance error budget} shows the total error budget and covers all these contribution terms with their description that makes up the $ \mu_{err}^{tot}$ on the median distance estimates of LMC $\tilde{\mu}_{LMC,x}$. Moreover, Table \ref{tab:LMC Distance estimates} gives these median distance estimates for all the distinct classes of Cepheid variables available in the caLMC data. Along with this, we provide the estimated $\tilde{\mu}_{err}$, $\mu_{err}^{stat}$, $\mu_{err}^{sys}$, and $\mu_{err}^{tot}$. In addition to this, the last column of Table \ref{tab:LMC Distance estimates} quantifies the accuracy of our estimated median distances w.r.t. the 1\% precise geometric distance of LMC ($\mu$ = 18.477 $\pm$ 0.026 mag) by \cite{pietrzynski2019_DistanceLargeMagellanic}. It is evident from this column that the vast majority of these LMC distance estimates are > 99\% accurate; indicating the robustness of the associated PL/PW relations.

\par However, there are a few cases for which the accuracy is less than 99\%. For instance, in the case of distance estimates based on the $gri$ bands PL relations of the individual classes of T2Cs (see Sect.~\ref{t2c_relation_compare}), the accuracy in the $g$ band is lower. This might be due to the fact that T2Cs are $\sim$ 1.5 mag intrinsically fainter than CCs, and therefore would require longer exposures to reach the same signal/noise (S/N), especially in the bluer wavelengths such as the $g$ band. We also notice the same effect in the accuracy of global T2Cs PL relations in the $g$ band.

\par Further, we note that even though the accuracy of distances estimated using ACEP FM and ACEP FO stars is > 98\% (except the one in $W_{ri}$ index for ACEP FO), the total fractional errors on median distance moduli ($\mu_{err}^{tot}$ / $\tilde{\mu}_{LMC,x}$) range from $\sim$ 5\% to 12\% in the $g$ band and the $W_{ri}$ index for ACEP FM and from $\sim$ 5\% to 18\% in the $gr$ bands and the $W_{ri}$ index. This is because the slopes and intercepts of the corresponding PL/PW relations for these stars possess extremely high uncertainties (see Sect.~\ref{ac_relation_compare}). Hence, we caution against the usage of these PL/PW relations for ACEP FM and ACEP FO stars.\\

\par
\begin{figure}[!htbp]
\centering
\includegraphics[width=0.78\columnwidth, keepaspectratio]{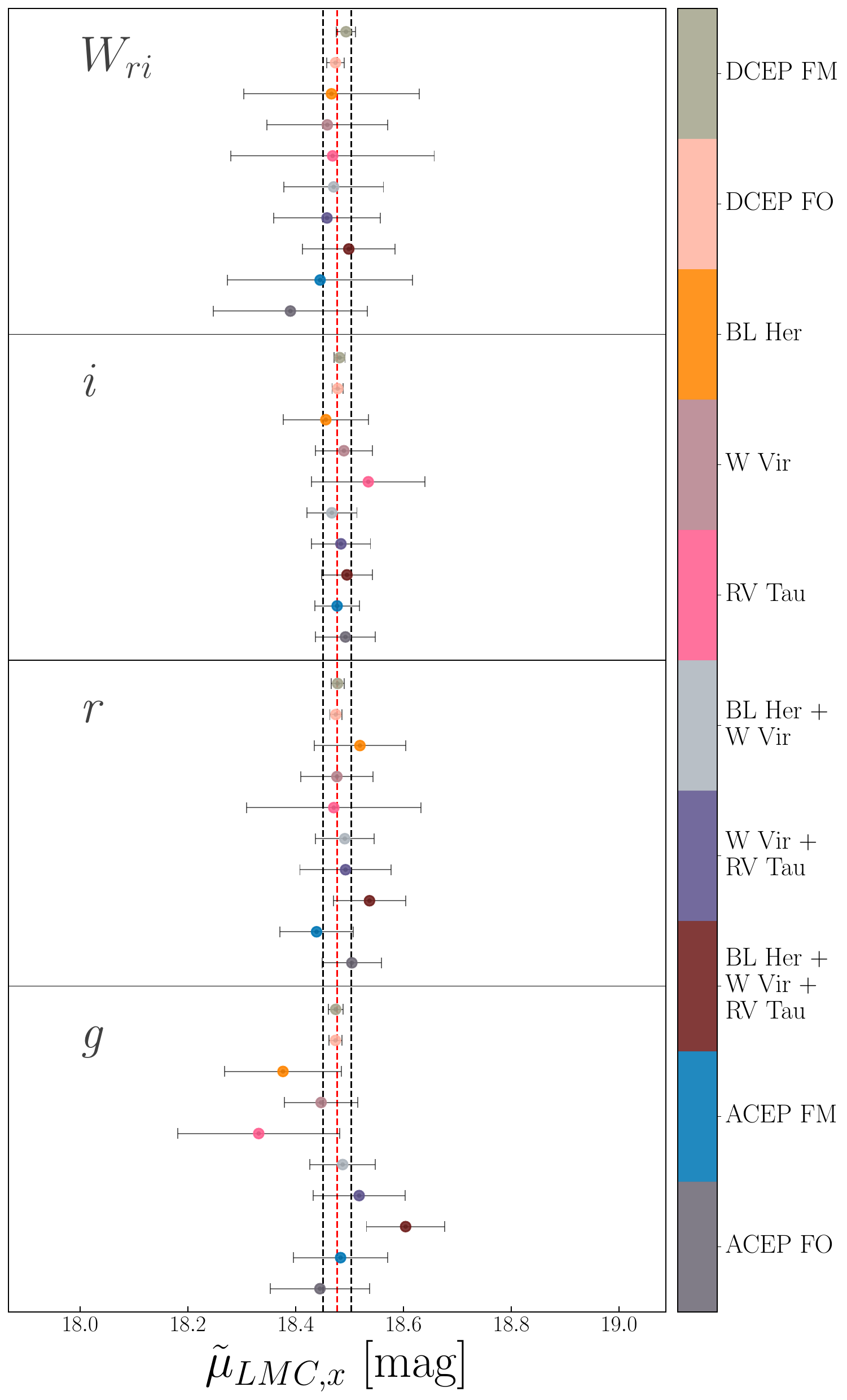}
\caption{Median value of the LMC distance modulus (colored dots) and the uncertainty on this median value $\tilde{\mu}_{err}$ (horizontal thin gray lines) for different classes of pulsating stars (in different colors) and different PL/PW relations (in different panels). The vertical thin red line marks the LMC canonical distance established by \cite{pietrzynski2019_DistanceLargeMagellanic} and the vertical black lines the total uncertainty on this distance.}

\label{Fig: lmc_result_boxplot}
\end{figure}

 The remarkable agreement between these distance moduli retrieved for the different classes of Cepheid variables can also be noticed in Fig.~\ref{Fig: lmc_result_boxplot}; where for comparison we have also shown the 1\% precise LMC distance. The precision of our estimates is affected by both the large photometric uncertainties of SkyMapper and the fewer number of observations in SMSS DR2 (as discussed in Sect.~\ref{LMC_data}). Nevertheless, the overall robust concurrence validates our PL/PW relations and allows us to apply them to other stellar systems, and in particular, to M31. Before doing this (see Sect.~\ref{M31}), we compare our PL/PW relations with previous studies in the $gri$ bands.

%--------------------------------------------------------------------------------------------------------------------

%--------------------------------------------------------------------------------------------------------------------
\begin{center}
\begin{table*}[!htbp]
%\begin{adjustbox}{width=0.992\columnwidth, keepaspectratio}
\centering
\caption{LMC distance estimates for different classes ($x$) of Cepheids in different bands. The table displays the median distance modulus $\tilde{\mu}_{LMC,x}$, its statistical uncertainty $\mu_{err}^{stat}$, its systematic uncertainty $\tilde{\mu}_{err}$, and its total uncertainty $\mu_{err}^{tot}$. The last column displays the accuracy (in \%) of $\tilde{\mu}_{LMC,x}$ with respect to the 1\% precise distance of the LMC ($\mu$ = 18.477 $\pm$ 0.026\,mag) by \citet{pietrzynski2019_DistanceLargeMagellanic}.}
\label{tab:LMC Distance estimates}
\begin{tabular}{cccccccc}
\toprule\toprule
\textbf{Class ($x$)} &
  \textbf{Filter} &
  \textbf{$\tilde{\mu}_{LMC,x}$ [mag]} &
  \textbf{$\tilde{\mu}_{err}$ [mag]} &
  \textbf{$\mu_{err}^{stat}$ [mag]} &
  \textbf{$\mu_{err}^{sys}$ [mag]} &
  \textbf{$\mu_{err}^{tot}$ [mag]} &
  \textbf{Accuracy [\%]} \\ 
\midrule 
\multirow{4}{*}{DCEP FM}                                                  & $g$      & 18.474 & 0.013 & 0.009 & 0.053 & 0.056 & 99.88 \\
                                                                          & $r$      & 18.478 & 0.012 & 0.007 & 0.050 & 0.052 & 99.96 \\
                                                                          & $i$      & 18.482 & 0.010 & 0.005 & 0.042 & 0.043 & 99.78 \\
                                                                          & $W_{ri}$ & 18.494 & 0.018 & 0.012 & 0.067 & 0.071 & 99.23 \\
\midrule
\multirow{4}{*}{DCEP FO}                                                  & $g$      & 18.474 & 0.012 & 0.010 & 0.059 & 0.061 & 99.86 \\
                                                                          & $r$      & 18.475 & 0.011 & 0.007 & 0.049 & 0.051 & 99.89 \\
                                                                          & $i$      & 18.478 & 0.010 & 0.005 & 0.043 & 0.044 & 99.95 \\
                                                                          & $W_{ri}$ & 18.474 & 0.016 & 0.008 & 0.062 & 0.064 & 99.86 \\
\midrule
\multirow{4}{*}{BL Her}                                                   & $g$      & 18.377 & 0.108 & 0.173 & 0.859 & 0.883 & 95.48 \\
                                                                          & $r$      & 18.519 & 0.085 & 0.113 & 0.557 & 0.575 & 98.03 \\
                                                                          & $i$      & 18.456 & 0.079 & 0.064 & 0.460 & 0.471 & 99.04 \\
                                                                          & $W_{ri}$ & 18.467 & 0.163 & 0.219 & 0.948 & 0.986 & 99.52 \\
\midrule
\multirow{4}{*}{W Vir}                                                    & $g$      & 18.447 & 0.068 & 0.057 & 0.302 & 0.315 & 98.64 \\
                                                                          & $r$      & 18.477 & 0.067 & 0.047 & 0.255 & 0.268 & 99.99 \\
                                                                          & $i$      & 18.490 & 0.053 & 0.045 & 0.263 & 0.272 & 99.42 \\
                                                                          & $W_{ri}$ & 18.459 & 0.112 & 0.096 & 0.422 & 0.447 & 99.17 \\
\midrule
\multirow{4}{*}{RV Tau}                                                   & $g$      & 18.332 & 0.150 & 0.139 & 0.792 & 0.818 & 93.52 \\
                                                                          & $r$      & 18.471 & 0.162 & 0.144 & 0.797 & 0.826 & 99.72 \\
                                                                          & $i$      & 18.535 & 0.105 & 0.099 & 0.552 & 0.570 & 97.30 \\
                                                                          & $W_{ri}$ & 18.469 & 0.189 & 0.244 & 1.070 & 1.114 & 99.62 \\
\midrule
\multirow{4}{*}{\begin{tabular}[c]{@{}c@{}}BL Her +\\ W Vir\end{tabular}} & $g$      & 18.487 & 0.061 & 0.037 & 0.136 & 0.154 & 99.52 \\
                                                                          & $r$      & 18.491 & 0.055 & 0.030 & 0.109 & 0.125 & 99.34 \\
                                                                          & $i$      & 18.467 & 0.046 & 0.019 & 0.089 & 0.102 & 99.55 \\
                                                                          & $W_{ri}$ & 18.471 & 0.092 & 0.061 & 0.154 & 0.190 & 99.71 \\
\midrule
\multirow{4}{*}{\begin{tabular}[c]{@{}c@{}}W Vir +\\ RV Tau\end{tabular}} & $g$      & 18.518 & 0.085 & 0.052 & 0.380 & 0.393 & 98.10 \\
                                                                          & $r$      & 18.493 & 0.085 & 0.041 & 0.285 & 0.300 & 99.28 \\
                                                                          & $i$      & 18.484 & 0.055 & 0.025 & 0.151 & 0.163 & 99.67 \\
                                                                          & $W_{ri}$ & 18.458 & 0.099 & 0.054 & 0.181 & 0.213 & 99.14 \\
\midrule
\multirow{4}{*}{\begin{tabular}[c]{@{}c@{}}BL Her +\\ W Vir +\\ RV Tau\end{tabular}} &
  $g$ &
  18.604 &
  0.073 &
  0.030 &
  0.140 &
  0.161 &
  93.97 \\
                                                                          & $r$      & 18.537 & 0.067 & 0.025 & 0.121 & 0.141 & 97.19 \\
                                                                          & $i$      & 18.495 & 0.047 & 0.016 & 0.077 & 0.091 & 99.16 \\
                                                                          & $W_{ri}$ & 18.499 & 0.086 & 0.049 & 0.132 & 0.165 & 99.00 \\
\midrule
\multirow{4}{*}{ACEP FM}                                                  & $g$      & 18.483 & 0.087 & 0.171 & 0.980 & 0.998 & 99.70 \\
                                                                          & $r$      & 18.439 & 0.068 & 0.114 & 0.743 & 0.754 & 98.26 \\
                                                                          & $i$      & 18.477 & 0.042 & 0.074 & 0.482 & 0.489 & 100.00 \\
                                                                          & $W_{ri}$ & 18.445 & 0.172 & 0.403 & 2.372 & 2.412 & 98.56 \\
\midrule
\multirow{4}{*}{ACEP FO}                                                  & $g$      & 18.445 & 0.092 & 0.501 & 2.433 & 2.486 & 98.55 \\
                                                                          & $r$      & 18.504 & 0.055 & 0.192 & 0.990 & 1.010 & 98.73 \\
                                                                          & $i$      & 18.493 & 0.056 & 0.128 & 0.747 & 0.760 & 99.28 \\
                                                                          & $W_{ri}$ & 18.390 & 0.143 & 0.684 & 3.315 & 3.388 & 96.09 \\ 
\bottomrule                                                                          
\end{tabular}
%\end{adjustbox}
\end{table*}
\end{center}

%--------------------------------------------------------------------------------------------------------------------

%\FloatBarrier
\subsection{Comparison with PL/PW relation in previous studies}
\label{LMC_compare}

\subsubsection{CCs PL/PW relation comparison}
\label{cc_relation_compare}

\par The first study which touched upon the LMC-based PL relations of DCEP FM in the Sloan photometric system is by \cite{ngeow2007_SemiempiricalCepheidPeriodLuminosity}. They derived the semi-empirical $ugriz$ bands PL relations by combining the observed $BVI$ mean magnitudes of DCEP FM with their theoretical bolometric corrections. For the sanity check, they further used publicly available Johnson-Sloan photometric transformations to estimate empirical $gr$ bands PL relations and found a good agreement with the semi-empirical one. Later, \cite{dicriscienzo2013_PredictedPropertiesGalactic} presented the first theoretical PL/PW relations for DCEP FM and DCEP FO stars in these SDSS bands. This study exploits non-linear convective pulsation models that account for the different chemical compositions of these stars in the  Milky Way, LMC, and SMC and provides theoretical PL/PW relations in the $ugriz$ bands by using the bolometric corrections and color-temperature transformations based on these updated model atmospheres. They report that the deviation in their PL slopes from the ones presented by \cite{ngeow2007_SemiempiricalCepheidPeriodLuminosity} decreases from $\sim$ 13\% in the $u$ band to $\sim$ 3\% in the $z$ band for DCEP FM stars with a period shorter than 100\,d in the LMC ($Z=0.008$). In order to supplement these semi-empirical and theoretical studies, \cite{hoffmann2015_CepheidVariablesMaserhost} additionally provide an empirical calibration of PL relations of DCEP FM stars located in the LMC using the models of \cite{dicriscienzo2013_PredictedPropertiesGalactic}. Further, by applying these synthetic PL relations to their DCEP FM sample of the NGC 4258 galaxy, they obtained its distance modulus, which is in good agreement with the maser distance to this galaxy obtained by \cite{humphreys2013_NewGeometricDistance}.

\par Despite these semi-empirical and theoretical studies, the Sloan filters have rarely been used for Cepheid photometry and for the subsequent derivation of their PL/PW relations.
Recently, using Gaia DR3 parallaxes, \cite{narloch2023_PeriodLuminosityRelationsGalactic} provided the first observational calibration of $gri$ bands absolute ($L_{\lambda}$) as well as astrometry-based luminosity ($ABL_{\lambda}$) PL/PW relations for DCEP FM stars in the Milky Way (MW). Since \citetalias{kodric2018_CepheidsM31PAndromeda} not only provides the largest $gri$ bands sample of Cepheid located in M31 (see sect. \ref{M31_data}) but also estimates the {\it apparent} PL/PW relations for DCEP FM and DCEP FO stars in these bands, \cite{narloch2023_PeriodLuminosityRelationsGalactic} compared their results with the \citetalias{kodric2018_CepheidsM31PAndromeda} PL/PW and report an overall good agreement in the $r$ and $i$ bands.\\

\par
\begin{figure}[!htbp]
\centering
\includegraphics[width=0.78\columnwidth, keepaspectratio]{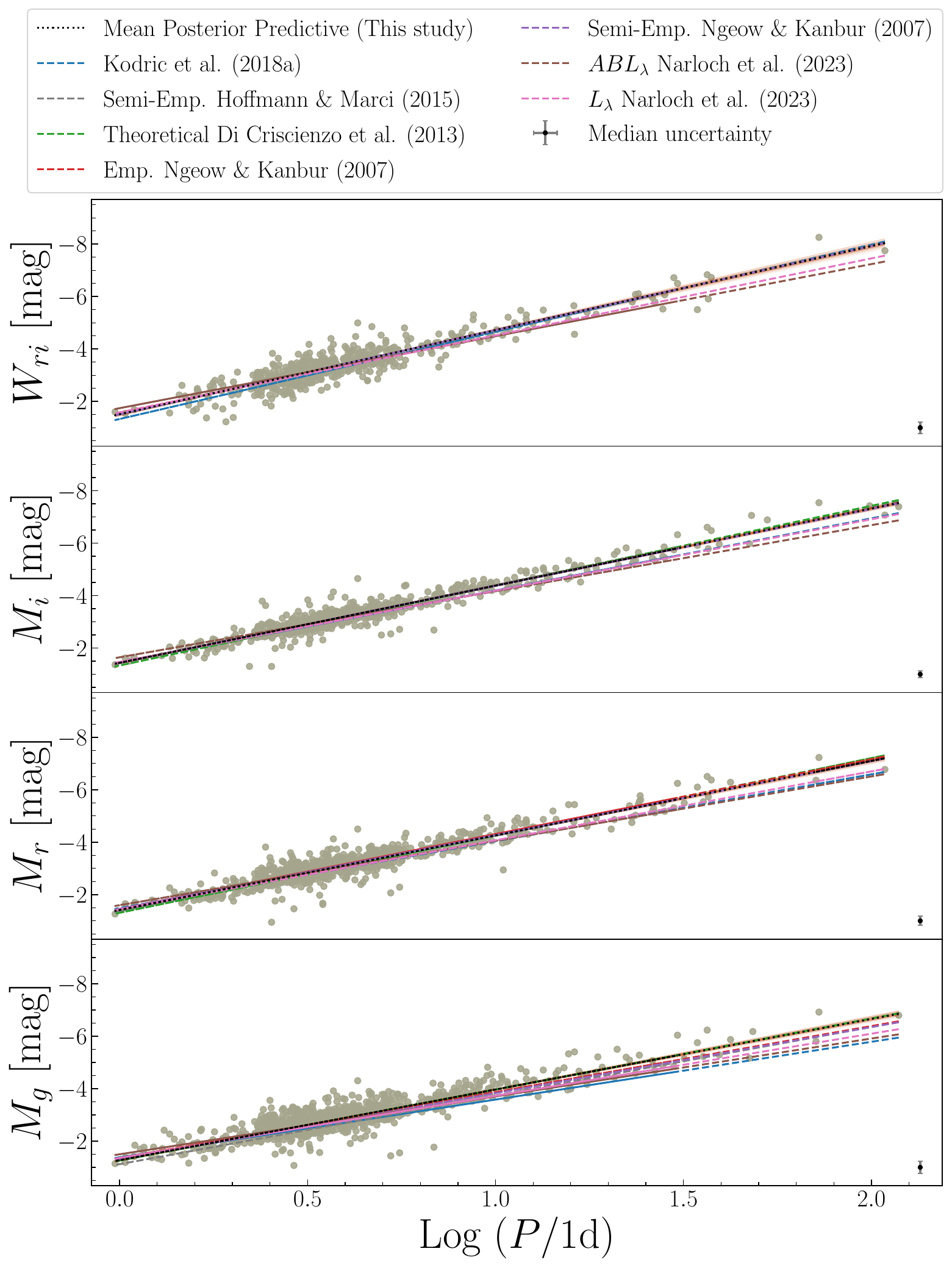}
\caption{Comparison of theoretical (\citealp{dicriscienzo2013_PredictedPropertiesGalactic}), semi-empirical (\citealp{ngeow2007_SemiempiricalCepheidPeriodLuminosity, hoffmann2015_CepheidVariablesMaserhost}), and empirical (\citealp{ngeow2007_SemiempiricalCepheidPeriodLuminosity}) PL/PW relations. We note that all these relations have been derived in the common metallicity regime of the LMC (\citealp{ngeow2007_SemiempiricalCepheidPeriodLuminosity, dicriscienzo2013_PredictedPropertiesGalactic, hoffmann2015_CepheidVariablesMaserhost}), the MW (\citealp{narloch2023_PeriodLuminosityRelationsGalactic}), and M31 (\citealp{kodric2018_CepheidsM31PAndromeda}). For the PL relations derived in this study, we show the mean value of the posterior predictive (in black) and 200 random draws of the posterior predictives within the 94\% credible interval (in orange). The cross in the bottom right corner represents typical (median) uncertainties on $M$ or $W$ and Log($P$/1d). Note that the errors on periods are vanishing on the logarithmic scale. 
%Same as Fig.~\ref{Fig: F_lmc_PL_pp_study_comp_gband} for DCEP FM PL relations in the $gri$ bands and PW$_{ri}$ relations.  
% \vp{Add M33 slope values for DCEP FM by Adair and lee 2023}
}
\label{Fig: F_lmc_PL_pp_study_comp}
\end{figure}

\begin{figure}[!htbp]
\centering
\includegraphics[width=0.7\columnwidth, keepaspectratio]{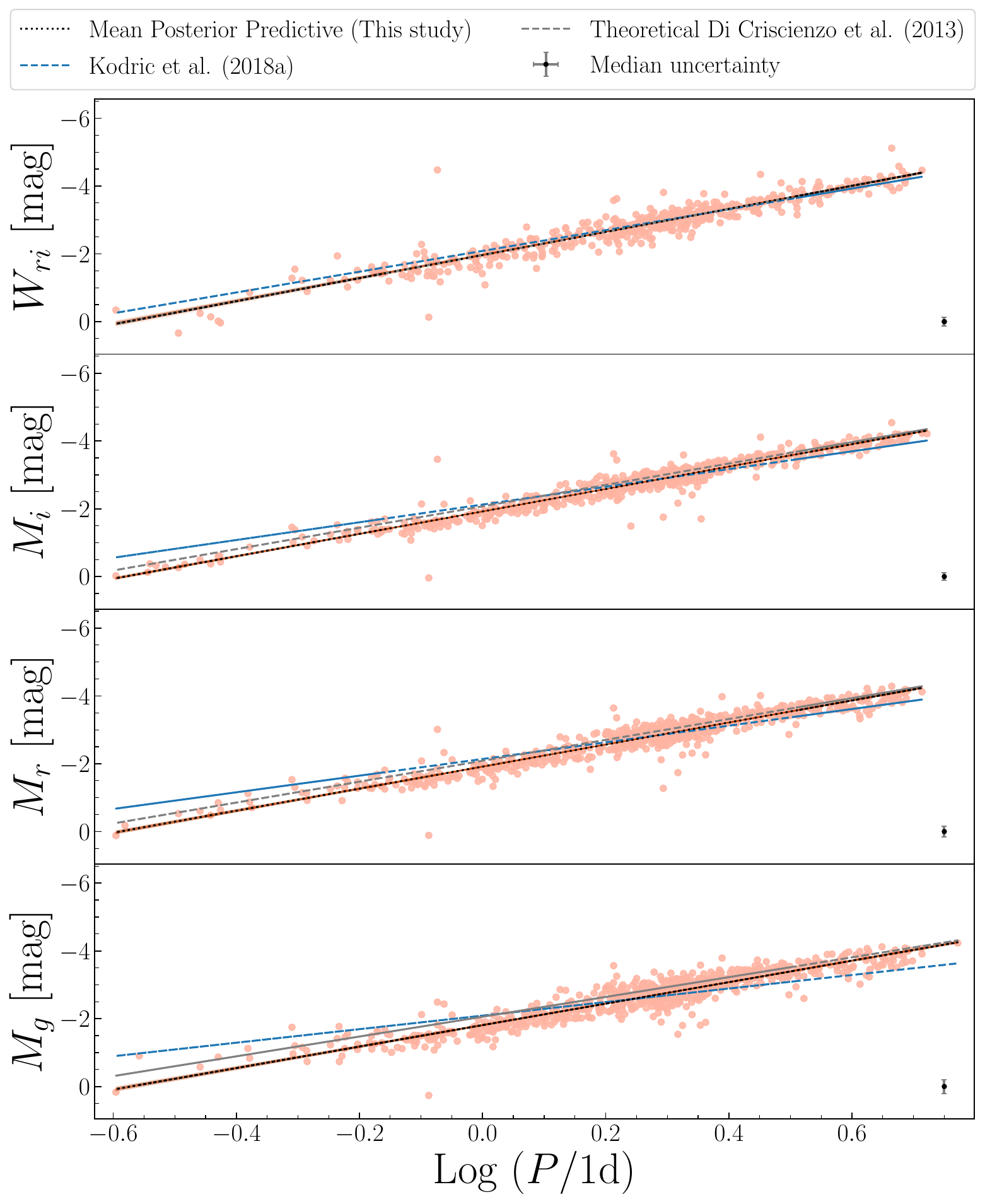}
\caption{Same as Fig.~\ref{Fig: F_lmc_PL_pp_study_comp} for DCEP FO PL relations in the $gri$ bands and PW$_{ri}$ relations. 
%\vp{DOUBLE CHECK}
} 
\label{Fig: FO_lmc_PL_pp_study_comp}
\end{figure}

\begin{figure*}[!htbp]
\centering
\includegraphics[width=0.992\textwidth, keepaspectratio]{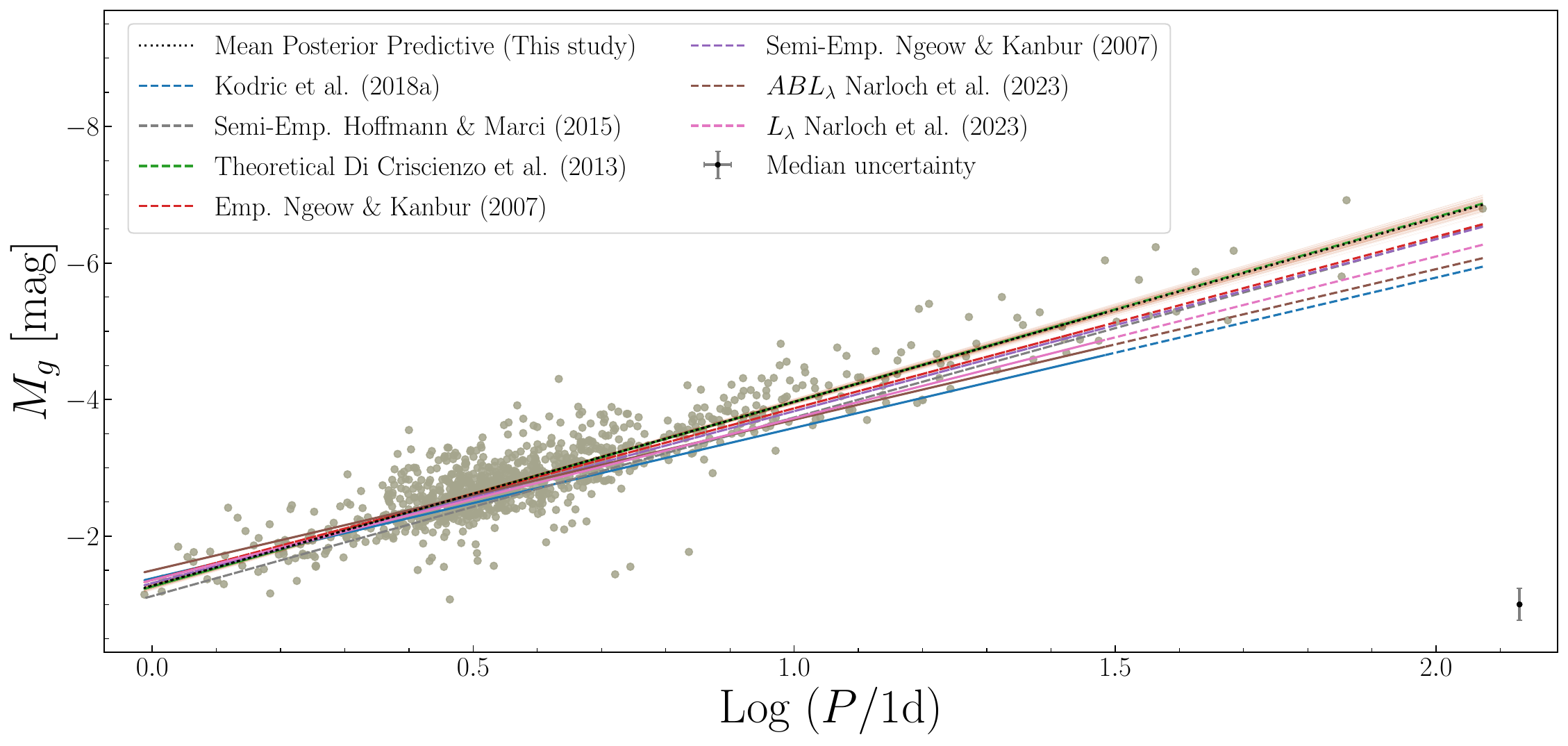}
%\captionsetup{justification=centering}
\caption{Same as Fig.~\ref{Fig: F_lmc_PL_pp_study_comp} for DCEP FM PL relations in the $g$ band.}
\label{Fig: F_lmc_PL_pp_study_comp_gband}
\end{figure*}

\begin{figure}[!htbp]
\centering
\includegraphics[width=0.6\columnwidth, keepaspectratio]{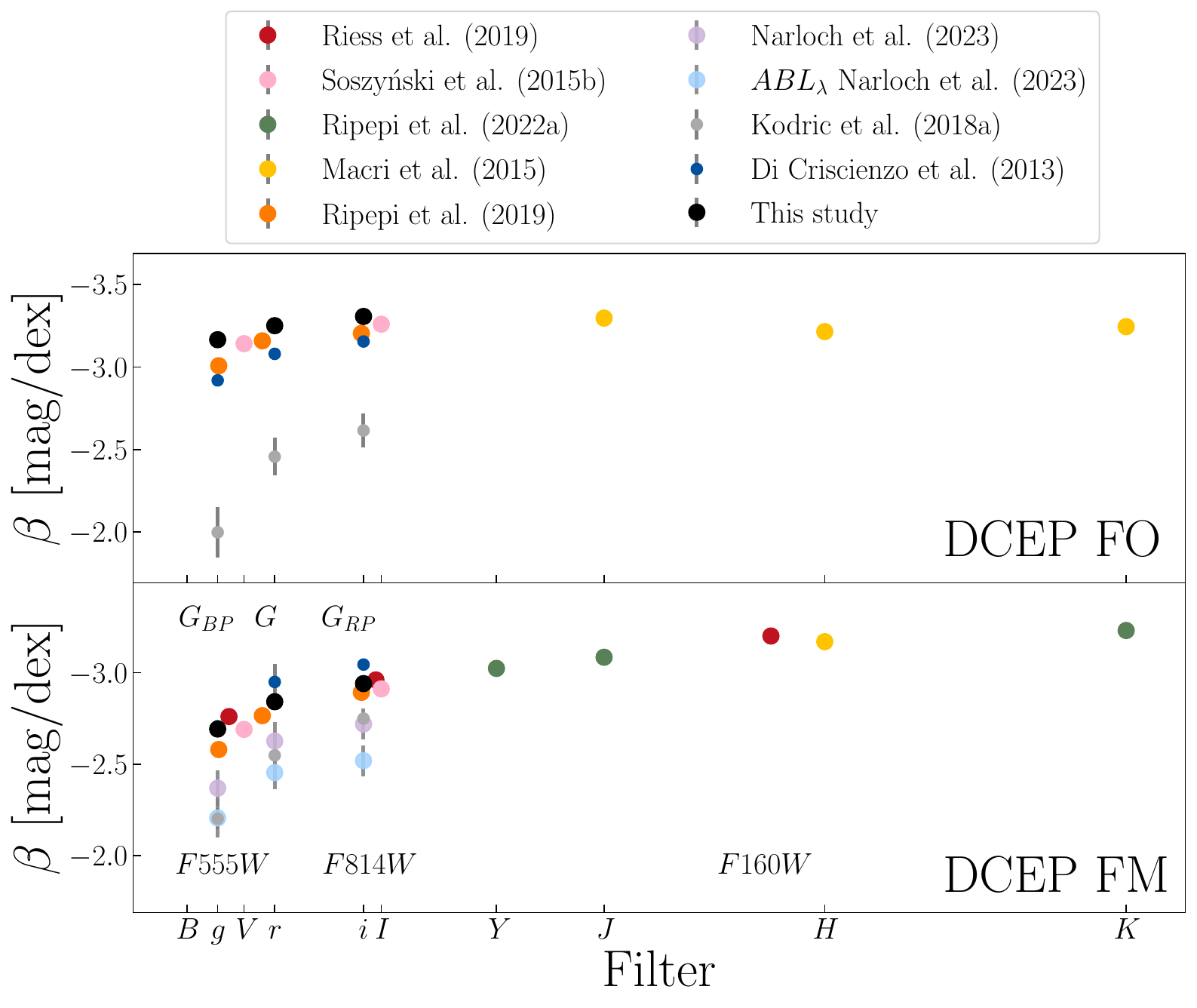}
\caption{Comparison of the slopes of Period luminosity relations obtained in various optical and near-infrared photometric bands for fundamental mode {\it (lower panel)}
and first overtone {\it (upper panel)} classical Cepheids. Note that the y-axis is reversed, showing steeper slopes above shallower slopes. The studies included in the comparison are those of \cite{
dicriscienzo2013_PredictedPropertiesGalactic,
macri2015_LargeMagellanicCloud,
soszynski2015_OGLECollectionVariablea,
kodric2018_CepheidsM31PAndromeda,
riess2019_LargeMagellanicCloud,
ripepi2019_ReclassificationCepheidsGaia,
ripepi2022_VMCSurveyXLVIII} and \cite{narloch2023_PeriodLuminosityRelationsGalactic}.
}
\label{Fig: CC_gri_slope_comparison}
\end{figure}

 In this study, we derive the first observational-based $gri$ bands PL relations and $W_{ri}$ PW relations for DCEP FM and DCEP FO stars located in the LMC. These relations are listed in Table~\ref{tab: CC PLPW relations}. Along with their posterior predictive samples under the 94\% credible interval\footnote{The credible interval is a Bayesian analog of the confidence interval. Here, a 94\% credible interval means that there is a 94\% probability that our BRR parameter values lie within this interval.}, Fig.~\ref{Fig: F_lmc_PL_pp_study_comp} and Fig.~\ref{Fig: FO_lmc_PL_pp_study_comp} show our mean posterior predictive PL/PW relations compared with relations from the aforementioned studies for DCEP FM and DCEP FO stars, respectively.
DCEP FM stars tightly follow PL relations and their 94\% credible interval is very narrow, especially when compared to those of T2Cs (see figures in Sect.~\ref{t2c_relation_compare}). Moreover, the overall alignment of our derived PL relations for these stars with various theoretical/empirical/semi-empirical studies is so exemplary that most of their PL relations fall in this interval area in Fig.~\ref{Fig: F_lmc_PL_pp_study_comp}. In particular, we notice that our $gri$ bands relations are in excellent agreement with those derived by \cite{dicriscienzo2013_PredictedPropertiesGalactic} in the LMC. Since \cite{dicriscienzo2013_PredictedPropertiesGalactic} provide only the slopes of their relations, we followed the approach adopted by \citet{narloch2023_PeriodLuminosityRelationsGalactic} and combined their slopes with our intercepts. Although this does not enable a true comparison between both studies, we note that the agreement between their and our study is exquisite both for the $r$ and $i$ bands PL relations (Fig.~\ref{Fig: F_lmc_PL_pp_study_comp}) and remains excellent even in the case of our relatively less accurate $g$ band PL relation (Fig.~~\ref{Fig: F_lmc_PL_pp_study_comp_gband}). Even though we witness a small deviation in our $g$ and $i$ bands mean posterior predictive PL relations, there is an remarkable agreement in the $r$ band when comparing the relations with the empirical/semi-empirical study of \cite{ngeow2007_SemiempiricalCepheidPeriodLuminosity}. By comparing the relations from \cite{hoffmann2015_CepheidVariablesMaserhost}, we found flawless conformity only in the $r$ and $i$ bands. In Fig.~\ref{Fig: F_lmc_PL_pp_study_comp}, we also note while comparing observation-based PL/PW relationships for DCEP FM stars in the MW and M31 that our LMC-based PL/PW relations are steeper in all the $gri$ bands and $W_{ri}$ index $wrt$ those derived by \cite{narloch2023_PeriodLuminosityRelationsGalactic}\footnote{Note that \cite{narloch2023_PeriodLuminosityRelationsGalactic} provides two different $L_{\lambda}$ or $ABL_{\lambda}$ PL/PW relations using two corresponding extinction correction approach from \cite{green2019_3DDustMap} and \cite{fitzpatrick1999_CorrectingEffectsInterstellar}. Since we used the \cite{fitzpatrick1999_CorrectingEffectsInterstellar} approach in our study, we only show the comparison of our derived relations with the ones associated with \cite{fitzpatrick1999_CorrectingEffectsInterstellar} in Fig. \ref{Fig: F_lmc_PL_pp_study_comp}, \ref{Fig: F_lmc_PL_pp_study_comp_gband}, \ref{Fig: CC_gri_slope_comparison}, and  from this study.} and \citetalias{kodric2018_CepheidsM31PAndromeda}.\\

\par On the other hand, the agreement is relatively poor when comparing the PL relations of DCEP FO stars to the ones predicted by \cite{dicriscienzo2013_PredictedPropertiesGalactic} (Fig.~\ref{Fig: FO_lmc_PL_pp_study_comp}). However, one can notice that the difference in our intercept with this study decreases while moving from the $g$ band to the $i$ band. In addition, the comparison of our relations with the ones derived by \citetalias{kodric2018_CepheidsM31PAndromeda} using the actual photometric data of M31 reveals that there is a relatively good agreement for the $W_{ri}$ PW relation but not for the $gri$ PL relations.\\

\par We further compare our $gri$ bands slope estimates with other LMC-based slope estimates in the different photometric systems for these classes of CCs. This comparison is shown in Fig.~\ref{Fig: CC_gri_slope_comparison} and includes the literature slope estimates in HST filters \citep{riess2019_LargeMagellanicCloud}, OGLE $VI$ bands \citep{soszynski2015_OGLECollectionVariablea}, VMC $YJK$ bands \citep{ripepi2022_VMCSurveyXLVIII}, NIR synoptic survey $JHK$ bands \citep{macri2015_LargeMagellanicCloud} and Gaia filters \citep{ripepi2019_ReclassificationCepheidsGaia}. It is evident that our derived slopes for these pulsating stars follow the general trend of steeper slopes while moving towards a longer wavelength (\citealp{bono2010_InsightsCepheidDistance} for DCEP FM and \citealp{baraffe2001_PeriodMagnitudeRelationships} for DCEP FO). Moreover, we also include the $gri$ bands slopes estimated by \cite{narloch2023_PeriodLuminosityRelationsGalactic} (only for DCEP FM) and \citetalias{kodric2018_CepheidsM31PAndromeda} (for both) in Fig.~\ref{Fig: CC_gri_slope_comparison} in order to compare them with our slope estimate. \cite{narloch2023_PeriodLuminosityRelationsGalactic} quote the good agreement of their slopes with the ones from \citetalias{kodric2018_CepheidsM31PAndromeda} for DCEP FM stars (also visible from Fig.~\ref{Fig: CC_gri_slope_comparison}). However, these slope values are significantly shallower than ours in all the $gri$ bands for both classes of CCs. One possible reason to explain this significant discrepancy is the metallicity dependence of the slope which we will later discuss in sect. \ref{discussion}.

%--------------------------------------------------------------------------------------------------------------------
%\begin{center}
\begin{table}[!htbp]
%\begin{adjustbox}{width=0.992\columnwidth, keepaspectratio}
\centering
\caption{Period Luminosity and Period Wesenheit relations for fundamental mode (DCEP FM) and first-overtone (DCEP FO) classical Cepheids located in the LMC derived using a Bayesian robust regression. The table lists the means and standard deviations of the posterior distributions of the model parameters.}
\label{tab: CC PLPW relations}
\begin{tabular}{lccccccc}
\toprule\toprule
\textbf{CC Class} &
  \textbf{Filter} &
  \textbf{$\alpha_{mean}$} &
  \textbf{$\alpha_{sd}$} &
  \textbf{$\beta_{mean}$} &
  \begin{tabular}[c]{@{}c@{}}\textbf{$\beta_{sd}$}\end{tabular} &
  $\sigma_{mean}$ &
  \begin{tabular}[c]{@{}c@{}}\textbf{$\sigma_{sd}$}\end{tabular} \\ 
\midrule
\multirow{4}{*}{DCEP FM} & $g$   & -3.968 & 0.020 & -2.692 & 0.042 & 0.167 & 0.016 \\
                         & $r$   & -4.255 & 0.018 & -2.841 & 0.038 & 0.129 & 0.013 \\
                         & $i$   & -4.378 & 0.014 & -2.940 & 0.029 & 0.092 & 0.010 \\
                         & $W_{ri}$ & -4.720 & 0.026 & -3.206 & 0.056 & 0.203 & 0.017 \\
\midrule
\multirow{4}{*}{DCEP FO} & $g$   & -4.977 & 0.032 & -3.166 & 0.042 & 0.080 & 0.023 \\
                         & $r$   & -5.167 & 0.025 & -3.251 & 0.033 & 0.056 & 0.020 \\
                         & $i$   & -5.228 & 0.021 & -3.307 & 0.026 & 0.074 & 0.009 \\
                         & $W_{ri}$ & -5.368 & 0.033 & -3.404 & 0.045 & 0.120 & 0.012 \\ 
\bottomrule
                         %\cline{2-8} 
\end{tabular}
%\end{adjustbox}
\end{table}
%\end{center}

%--------------------------------------------------------------------------------------------------------------------

\subsubsection{T2Cs PL/PW relation comparison}
\label{t2c_relation_compare}

\par To our knowledge, the first T2C's \textit{gri} bands PL/PW relations were calculated by \cite{ngeow2022_ZwickyTransientFacility} using a small sample of stars located in Milky Way globular clusters. Some stars tabulated as T2Cs in the literature were not retained by \cite{ngeow2022_ZwickyTransientFacility} as they suspected a wrong classification, and the periods of the stars in their final sample were verified and adjusted if needed. This data set combines (a) the ZTF DR10 photometric mean magnitudes in the $gr$ bands for 37 T2Cs; (b) the ZTF DR10 photometric mean magnitudes in the $i$ band for 17 T2Cs; and (c) $gri$ band mean magnitudes for 25 T2Cs in common with \cite{bhardwaj2022_RRLyraeType} obtained by transforming their $BVI$-band mean magnitudes using the relations provided by \cite{tonry2012_PanSTARRS1PhotometricSystem}. 
The PL/PW relations are calibrated against individual accurate GC distances determined by \cite{baumgardt2021_AccurateDistancesGalactic}.
We note that for the estimation of the PL/PW relations, \cite{ngeow2022_ZwickyTransientFacility} used an iterative 3$\sigma$-clipping method in order to exclude the outliers while fitting a linear regression to their data, a method that is not without danger especially when dealing with inhomogeneous photometric data. Moreover, due to the small size of their sample, they did not further separate their T2Cs into three sub-classes (BL Her, W Vir, RV Tau) but computed a common law for their combined sample.\\

\par In contrast, the caLMC data contains a much larger sample of BL Her, W Vir, and RV Tau stars (listed in Table \ref{Table: Generation of caLMC data}). Moreover, they have homogeneous SkyMapper SMSS DR2 photometry \textit{gri} bands. However, the mean magnitudes are formed from a few measurements only, while those of \cite{ngeow2022_ZwickyTransientFacility} are derived from a few tens up to a few hundreds.
Given our lower-quality data, the selection of the BRR fitting method, which minimizes biases towards outliers, especially when compared with traditional fitting and outlier rejection methods is crucial, and it provides a better treatment of the uncertainties by taking the covariances between the parameters into account. Along with the PL/PW relations for the three individual classes and the global relations for all T2Cs, we provide relations for the combination of 2 classes, namely BL Her + W Vir and W Vir + RV Tau. The parameters of these relations are listed in Tables~\ref{tab:T2C class PLPW relation} and \ref{tab: T2C PLPW relations}. As for the CCs, and for the same reasons, the $g$-band PL relations show a large dispersion, and the $r$ and $i$ band PL relations end up being slightly more accurate than the $W_{ri}$ PW relation.

%--------------------------------------------------------------------------------------------------------------------

%\begin{center}
\begin{table}[!htbp]
%\begin{adjustbox}{width=0.992\columnwidth, keepaspectratio}
\centering
\caption{Same as Table~\ref{tab: CC PLPW relations} for individual classes of T2Cs.}
\label{tab:T2C class PLPW relation}
\begin{tabular}{cccccccc}
\toprule\toprule
\textbf{T2C Class} & \textbf{Filter} & \textbf{$\alpha_{mean}$} & \textbf{$\alpha_{sd}$} & \textbf{$\beta_{mean}$} & \textbf{$\beta_{sd}$} & \textbf{$\sigma_{mean}$} & \textbf{$\sigma_{sd}$} \\ 
\midrule
\multirow{4}{*}{BL Her} & $g$      & -1.964 & 0.520 & -1.854 & 0.683 & 0.338 & 0.095 \\
                        & $r$      & -1.892 & 0.328 & -1.712 & 0.450 & 0.188 & 0.075 \\
                        & $i$      & -1.991 & 0.285 & -1.840 & 0.360 & 0.220 & 0.065 \\
                        & $W_{ri}$ & -2.579 & 0.559 & -2.545 & 0.765 & 0.303 & 0.147 \\
\midrule
\multirow{4}{*}{W Vir}  & $g$      & -1.535 & 0.043 & -2.368 & 0.298 & 0.148 & 0.060 \\
                        & $r$      & -1.873 & 0.034 & -2.622 & 0.251 & 0.103 & 0.058 \\
                        & $i$      & -2.010 & 0.037 & -2.532 & 0.259 & 0.167 & 0.055 \\
                        & $W_{ri}$ & -2.361 & 0.058 & -2.286 & 0.417 & 0.202 & 0.066 \\
\midrule
\multirow{4}{*}{RV Tau} & $g$      & -3.630 & 0.432 & 0.746  & 0.663 & 0.559 & 0.119 \\
                        & $r$      & -3.620 & 0.437 & 0.159  & 0.666 & 0.544 & 0.114 \\
                        & $i$      & -2.871 & 0.302 & -1.258 & 0.461 & 0.415 & 0.081 \\
                        & $W_{ri}$ & -2.054 & 0.607 & -3.413 & 0.881 & 0.593 & 0.174 \\ 
\bottomrule
\end{tabular}
%\end{adjustbox}
\end{table}
%\end{center}

%--------------------------------------------------------------------------------------------------------------------

%\begin{center}
\begin{table}[!htbp]
%\begin{adjustbox}{width=0.992\columnwidth, keepaspectratio}
\centering
\caption{Same as Table~\ref{tab: CC PLPW relations} for various combinations of T2Cs.}
\label{tab: T2C PLPW relations}
\begin{tabular}{cccccccc}
\toprule\toprule
\textbf{T2C Class}  & \textbf{Filter} & \textbf{$\alpha_{mean}$} & \textbf{$\alpha_{sd}$} & \textbf{$\beta_{mean}$} & \textbf{$\beta_{sd}$} & \textbf{$\sigma_{mean}$} & \textbf{$\sigma_{sd}$} \\ 
\midrule
\multirow{4}{*}{\begin{tabular}[c]{@{}c@{}}BL Her + \\ W Vir + \\ RV Tau\end{tabular}} & $g$ & -1.764 & 0.060 & -1.811 & 0.124 & 0.480 & 0.049 \\
 &
  $r$ &
  -2.012 &
  0.050 &
  -2.034 &
  0.107 &
  0.365 &
  0.055 \\
 &
  $i$ &
  -2.135 &
  0.037 &
  -2.093 &
  0.062 &
  0.271 &
  0.043 \\
 &
  $W_{ri}$ &
  -2.481 &
  0.057 &
  -2.621 &
  0.116 &
  0.340 &
  0.071 \\
\midrule
\multirow{4}{*}{\begin{tabular}[c]{@{}c@{}}BL Her + \\ W Vir\end{tabular}} &
  $g$ &
  -1.580 &
  0.050 &
  -1.460 &
  0.124 &
  0.245 &
  0.044 \\
 &
  $r$ &
  -1.878 &
  0.039 &
  -1.772 &
  0.098 &
  0.176 &
  0.045 \\
 &
  $i$ &
  -2.006 &
  0.041 &
  -1.890 &
  0.074 &
  0.179 &
  0.040 \\
 &
  $W_{ri}$ &
  -2.366 &
  0.054 &
  -2.259 &
  0.142 &
  0.197 &
  0.057 \\
\midrule
\multirow{4}{*}{\begin{tabular}[c]{@{}c@{}}W Vir + \\ RV Tau\end{tabular}} &
  $g$ &
  -1.622 &
  0.075 &
  -2.450 &
  0.372 &
  0.428 &
  0.108 \\
 &
  $r$ &
  -1.909 &
  0.039 &
  -2.825 &
  0.281 &
  0.213 &
  0.101 \\
 &
  $i$ &
  -2.048 &
  0.048 &
  -2.468 &
  0.141 &
  0.298 &
  0.051 \\
 &
  $W_{ri}$ &
  -2.396 &
  0.066 &
  -2.901 &
  0.166 &
  0.303 &
  0.091 \\
\bottomrule 
\end{tabular}
%\end{adjustbox}
\end{table}
%\end{center}

%--------------------------------------------------------------------------------------------------------------------

\par
\begin{figure}[!htbp]
\centering
\includegraphics[width=0.6\columnwidth, keepaspectratio]{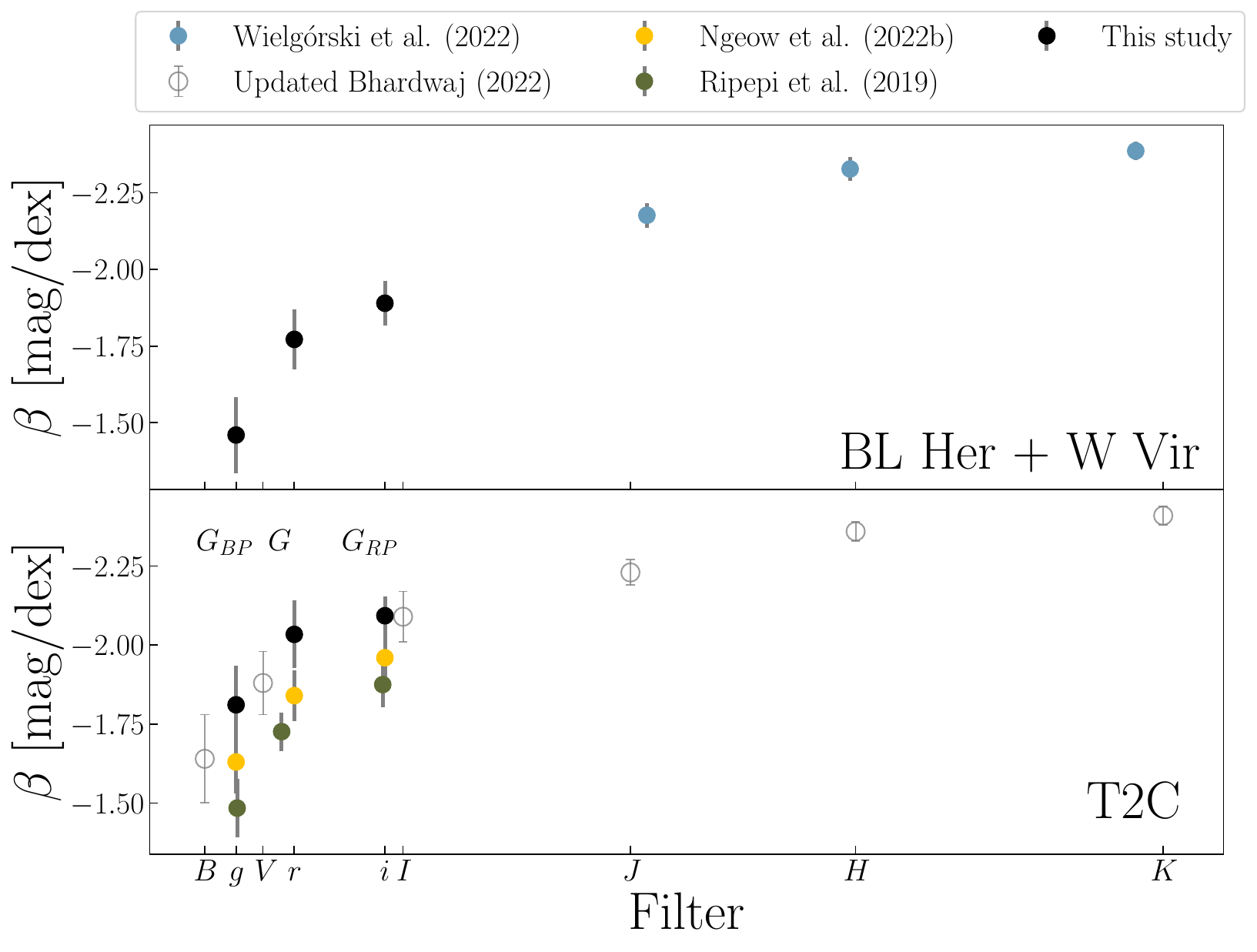}
\caption{Comparison of the slopes of Period luminosity relations obtained in various optical and near-infrared photometric bands for T2Cs {\it (lower panel)}
and the BL Her + W Vir combination {\it (upper panel)}. Note that the y-axis is reversed, showing steeper slopes above shallower slopes. The studies included in the comparison are those of \cite{ripepi2019_ReclassificationCepheidsGaia,wielgorski2022_AbsoluteCalibrationNearinfrared, ngeow2022_ZwickyTransientFacility}, and the updates by \cite{ngeow2022_ZwickyTransientFacility} using accurate, individual globular cluster distances \citep{baumgardt2021_AccurateDistancesGalactic} of the PL relations previously derived by \cite{bhardwaj2022_RRLyraeType}.} 
\label{Fig: T2C_gri_slope_comparison}
\end{figure}

 Given the paucity of PL/PW relations in the literature, we can compare our slopes (in the $gri$ bands) only for relations combining all classes of T2Cs. Fig.~\ref{Fig: T2C_gri_slope_comparison} indicates that they are in line with the \citet{bhardwaj2022_RRLyraeType} slopes in the $BVI$ bands as updated by \citet{ngeow2022_ZwickyTransientFacility}, while those derived by \citet{ngeow2022_ZwickyTransientFacility} reach smaller values and those of \citet{ripepi2019_ReclassificationCepheidsGaia} in the Gaia filters lie even below.

\par
\begin{figure}[!htbp]
\centering
\includegraphics[width=0.6\columnwidth, keepaspectratio]{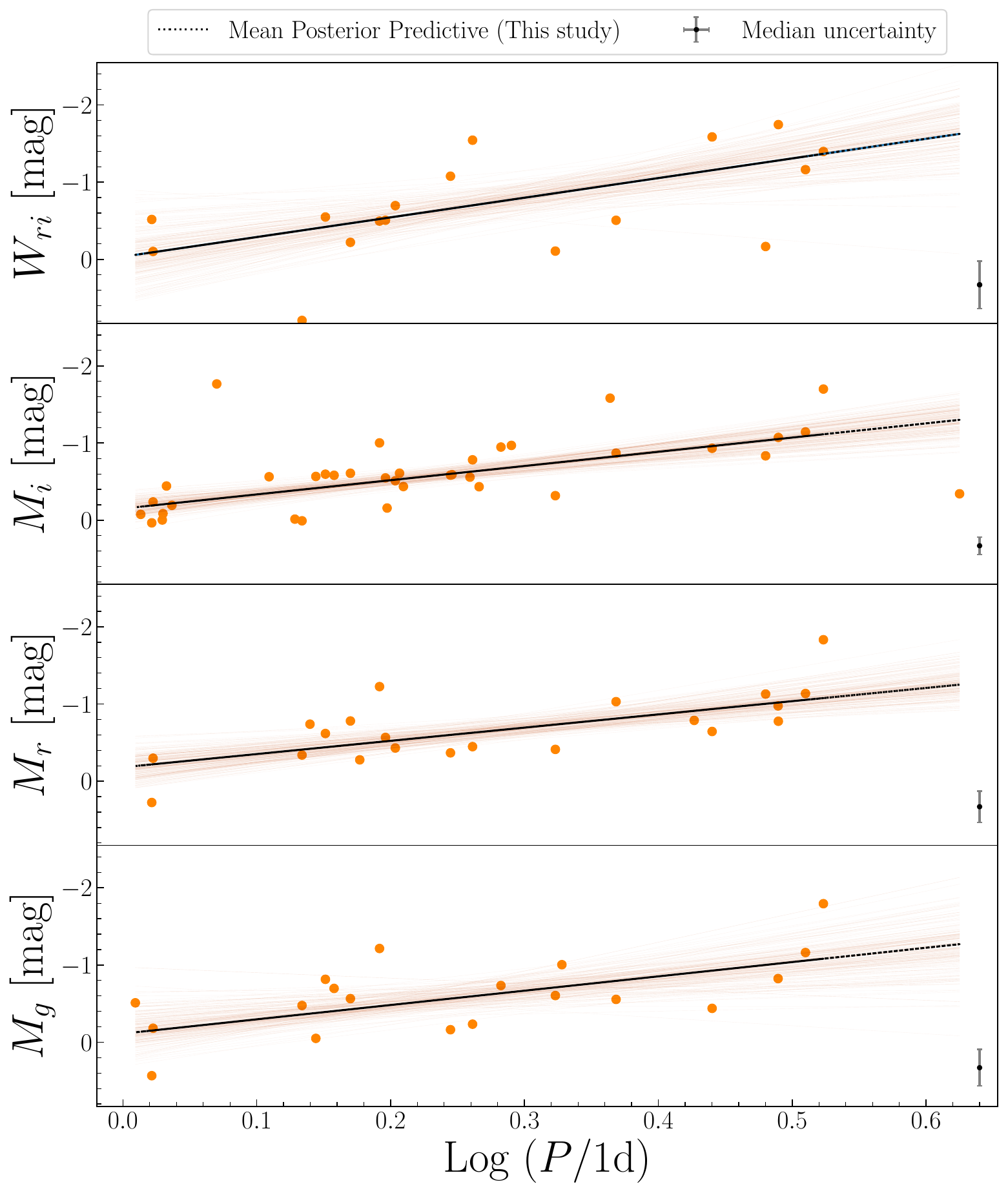}
\caption{Same as Fig.~\ref{Fig: F_lmc_PL_pp_study_comp_gband} for BL Her PL relations in the $gri$ bands and PW$_{ri}$ relations.} 
\label{Fig: BLHER_lmc_PL_pp}
\end{figure}

\begin{figure}[!htbp]
\centering
\includegraphics[width=0.6\columnwidth, keepaspectratio]{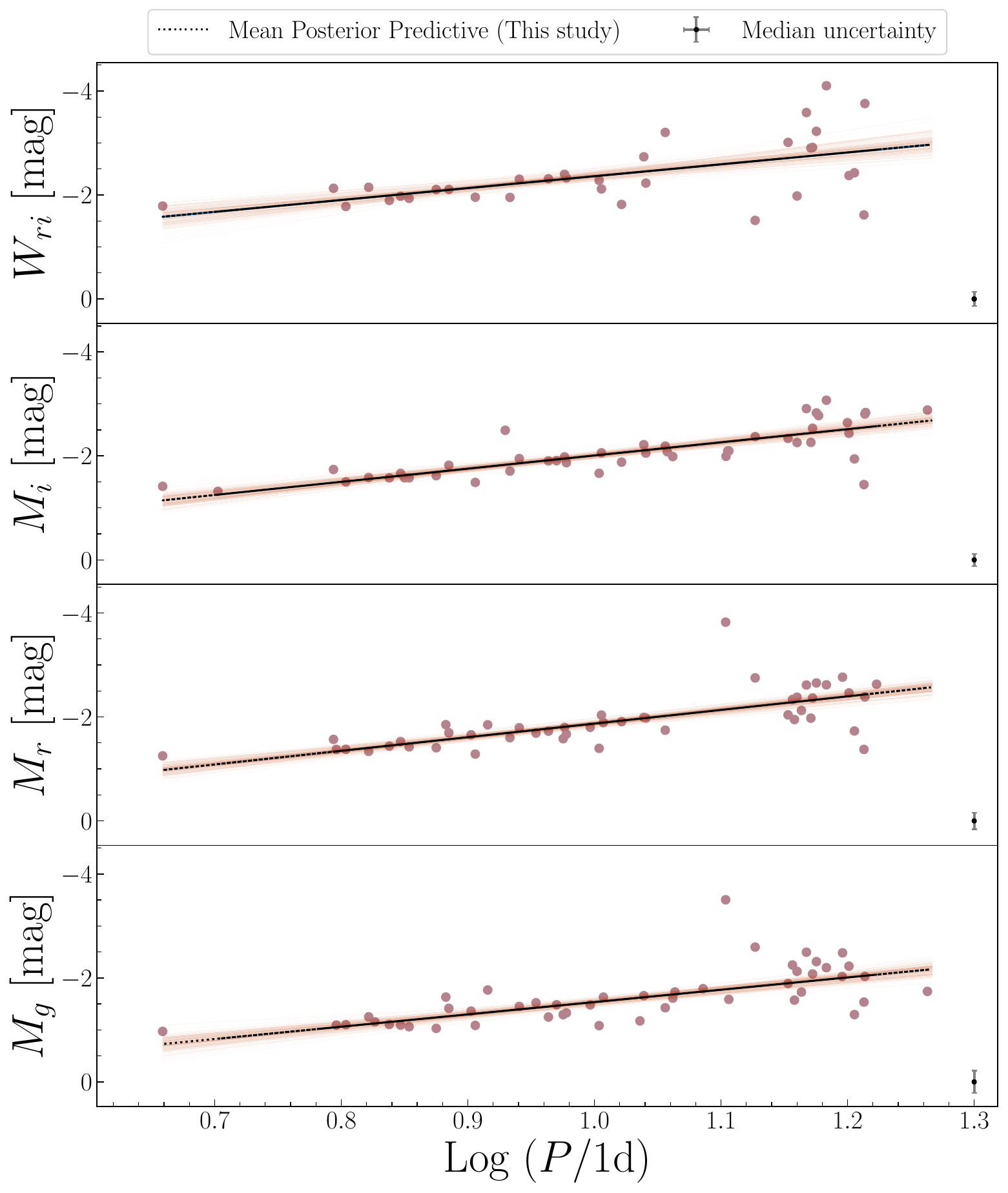}
\caption{Same as Fig.~\ref{Fig: F_lmc_PL_pp_study_comp_gband} for W Vir PL relations in the $gri$ bands and PW$_{ri}$ relations.} 
\label{Fig: WVIR_lmc_PL_pp}
\end{figure}

\begin{figure}[!htbp]
\centering
\includegraphics[width=0.6\columnwidth, keepaspectratio]{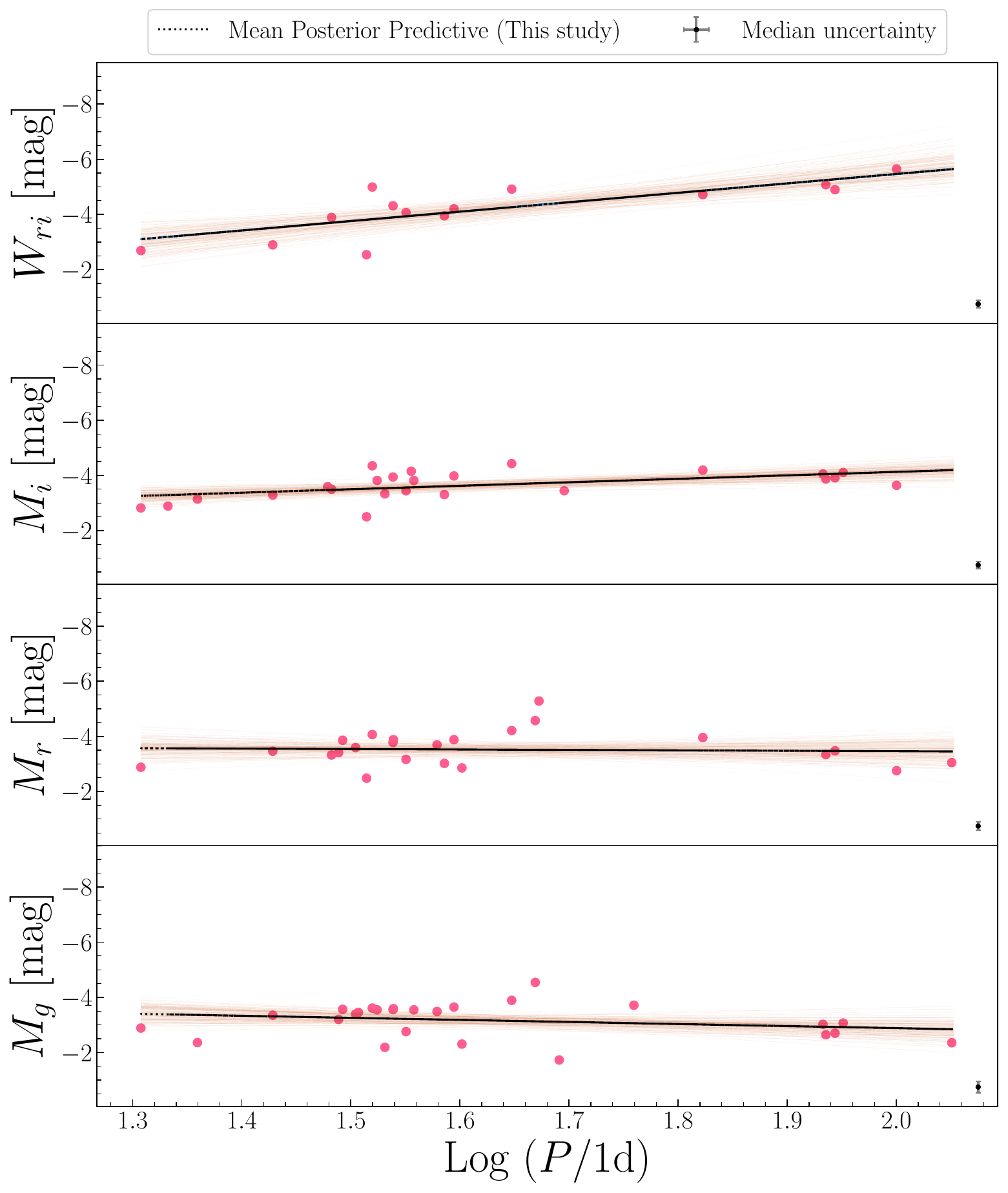}
\caption{Same as Fig.~\ref{Fig: F_lmc_PL_pp_study_comp_gband} for RV Tau PL relations in the $gri$ bands and PW$_{ri}$ relations.}  
\label{Fig: RVTAU_lmc_PL_pp}
\end{figure}

 The 94\% credible intervals of our PL/PW relations for BL Her stars only (Fig.~\ref{Fig: BLHER_lmc_PL_pp}) are broad, reflecting the relatively large dispersion of our data points. The dispersions of the PL/PW relations for W Vir stars only (Fig.~\ref{Fig: WVIR_lmc_PL_pp}) are much smaller, while those for RV Tau stars only (Fig.~\ref{Fig: RVTAU_lmc_PL_pp}) are similar to the BL Her ones. We note in passing that the RV Tau PL relations in the $g$ and $r$ bands have unexpected positive slope which casts some ambiguity on their reliability. This might be potentially related to the undersampling of the light curve, considering RV Tau stars pulsation period can reach up to $\sim$ 100d. Moreover, the intrinsic properties of RV Tau stars may impact their PL relations in the optical bands. This is more extensively discussed in Sect. \ref{Intrinsic properties of RV Tau stars}. 

\par
\begin{figure}[!htbp]
\centering
\includegraphics[width=0.6\columnwidth, keepaspectratio]{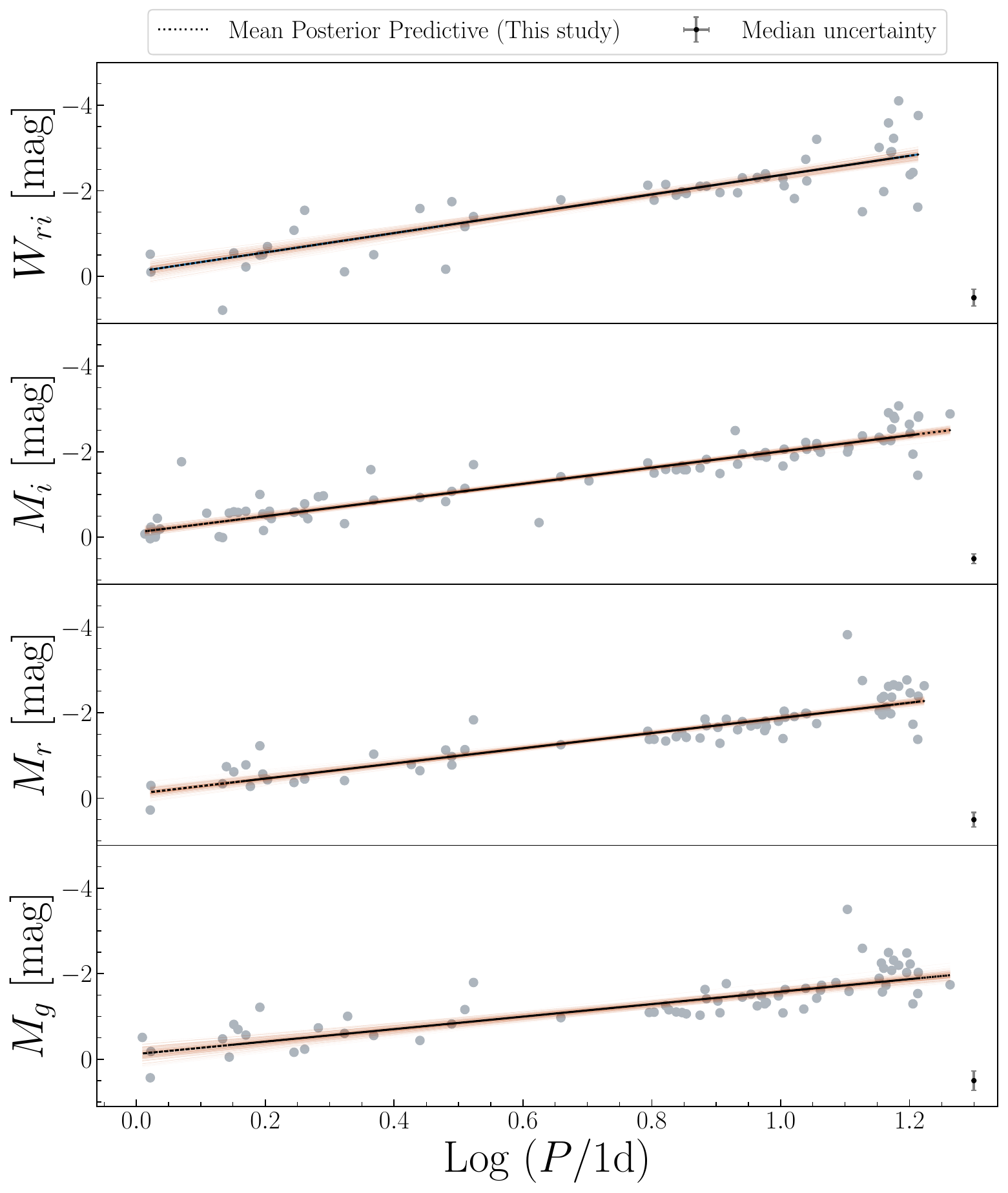}
\caption{Same as Fig.~\ref{Fig: F_lmc_PL_pp_study_comp_gband} for BL Her + W Vir PL relations in the $gri$ bands and PW$_{ri}$ relations.}  
\label{Fig: BLHER_WVIR_lmc_PL_pp}
\end{figure}

\begin{figure}[!htbp]
\centering
\includegraphics[width=0.6\columnwidth, keepaspectratio]{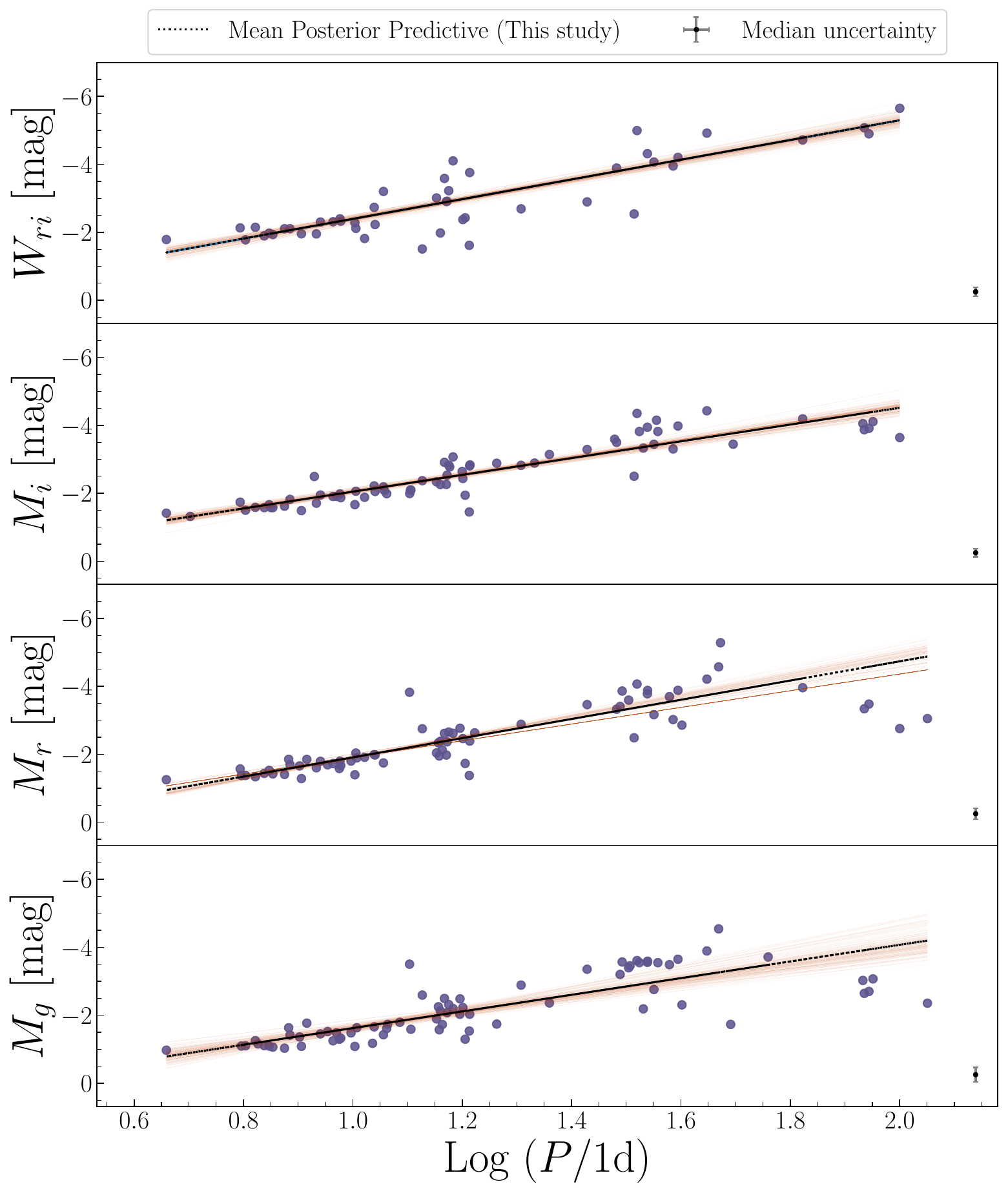}
\caption{Same as Fig.~\ref{Fig: F_lmc_PL_pp_study_comp_gband} for W Vir + RV Tau PL relations in the $gri$ bands and PW$_{ri}$ relations.}  
\label{Fig: WVIR_RVTAU_lmc_PL_pp}
\end{figure}

\begin{figure}[!htbp]
\centering
\includegraphics[width=0.78\columnwidth, keepaspectratio]{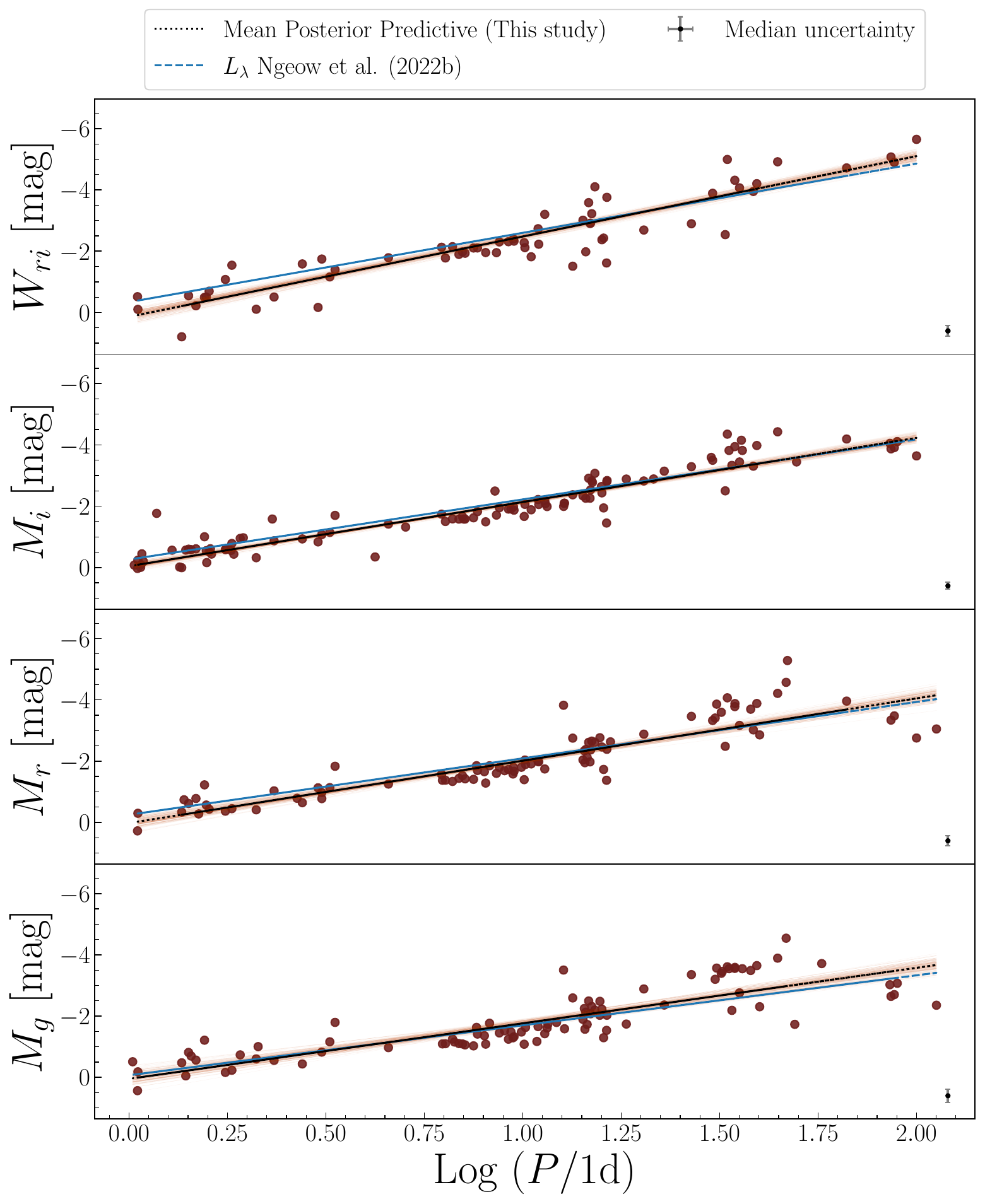}
\caption{Same as Fig.~\ref{Fig: F_lmc_PL_pp_study_comp_gband} for T2Cs (BL Her + W Vir + RV TAU) PL relations in the $gri$ bands and PW$_{ri}$ relations.} 
\label{Fig: T2C_lmc_PL_pp_study_comp}
\end{figure}

 To our knowledge, since we are the first to provide such relations in the $gri$ bands, there is no literature to compare them with. Combining BL Her and W Vir stars (Fig.~\ref{Fig: BLHER_WVIR_lmc_PL_pp}) or W Vir and RV Tau stars (Fig.~\ref{Fig: WVIR_RVTAU_lmc_PL_pp}) makes the relations tighter, but not combining the three subclasses of T2Cs (Fig.~\ref{Fig: T2C_lmc_PL_pp_study_comp}), potentially because of the RV Tau stars (see Sect.~\ref{Intrinsic properties of RV Tau stars}). 

\par In the latter case, we can compare our relations to those of \citet{ngeow2022_ZwickyTransientFacility}, and they agree quite well, the relations of \citet{ngeow2022_ZwickyTransientFacility} lying within or just at the edge of the 94\% credible interval. As noted earlier, the slopes of their relations are systematically (slightly) smaller than ours.

%\FloatBarrier
\subsubsection{ACs PL/PW relation comparison}
\label{ac_relation_compare}

\par When it comes to the PL/PW relations for ACEPs, only a few studies estimated theoretical or observational PL/PW relation in the Johnson-Cousins $BVI$ \citep{nemec1988_VariableStarsUrsa, nemec1994_PeriodLuminosityMetallicityRelationsPulsation,bono1997_EvolutionaryScenarioMetalPoor,pritzl2002_DwarfSpheroidalCompanions,marconi2004_UpdatedPulsationModels,ripepi2014_VMCSurveyVIII,soszynski2015_OGLECollectionVariable,groenewegen2017_PeriodluminosityPeriodradiusRelations, iwanek2018_ThreeDimensionalDistributionsType} or the Gaia $GB_{p}R_{p}$ filters \citep{ripepi2019_ReclassificationCepheidsGaia, ripepi2022_GaiaDR3Specific}. 
\par \cite{ngeow2022_ZwickyTransientFacilitya} recently estimated the first PL/PW relations for ACEP FM stars in the $gri$ bands. Similarly to their study of T2Cs, they used a combination of ZTF $gri$ band data and of $BVI$ photometry transformed into $gri$ mean magnitudes for a very small sample of 5 ACEP FM and 2 ACEP FO (not included in the fit) located in the M92 GC and in the LMC. The PL/PW relations are calibrated using the accurate distances to M92 \citep{baumgardt2021_AccurateDistancesGalactic} and to the LMC \citep{pietrzynski2019_DistanceLargeMagellanic}, and tested against 7 ACEPs in the Crater II dwarf galaxy reported by \cite{vivas2020_DECamViewDiffuse}. For these stars, \cite{ngeow2022_ZwickyTransientFacilitya} identify one as a foreground ACEP or misclassified object, and speculate that three of them are ACEP FMs and the remaining three ACEP FO stars, based on their $g$ band mean magnitudes overlaid on $B$ band PL-relations from \cite{pritzl2002_DwarfSpheroidalCompanions}. After further transforming their PL/PW relations into the $gri$ SDSS system, in which the data of \citet{vivas2020_DECamViewDiffuse} is provided, they obtain $\mu_{g}$ = 20.45 $\pm$ 0.25\,mag, $\mu_{i}$ = 20.55 $\pm$ 0.29\,mag and $\mu_{gi}^{W}$ = 20.54 $\pm$ 0.30\,mag for the distance modulus of Crater II, in agreement with the value ($\mu$ = 20.33 $\pm$ 0.01\,mag) provided by \citet{vivas2020_DECamViewDiffuse} from RR Lyrae theoretical PL relation.\\

\par
\begin{figure}
\centering
\includegraphics[width=0.55\columnwidth, keepaspectratio]{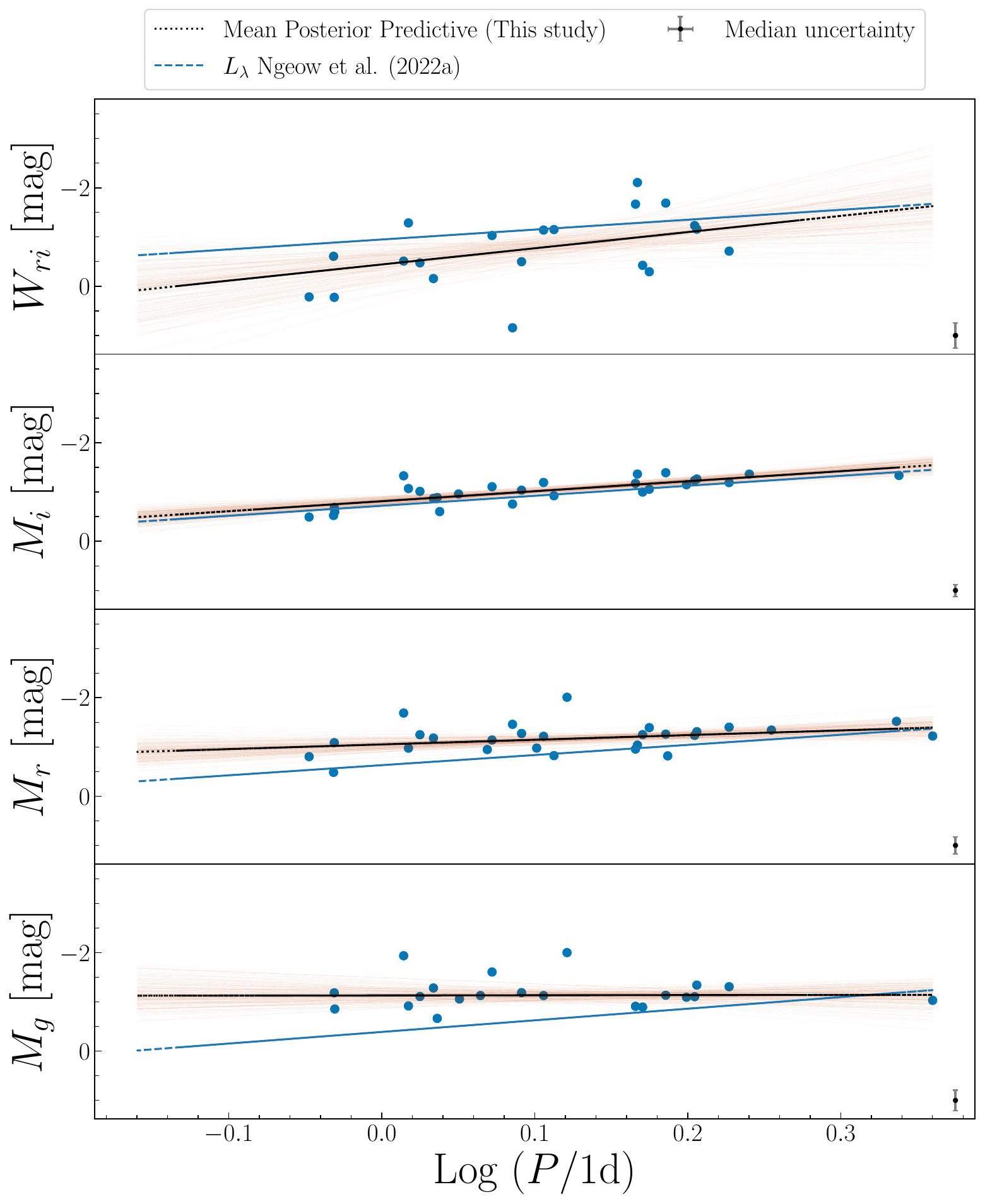}
\caption{{Same as Fig.~\ref{Fig: F_lmc_PL_pp_study_comp_gband} for ACEP FM PL relations in the $gri$ bands and PW$_{ri}$ relations.} } 
\label{Fig: ACEP_F_lmc_PL_pp_study_comp}
\end{figure}

\begin{figure}
\centering
\includegraphics[width=0.55\columnwidth, keepaspectratio]{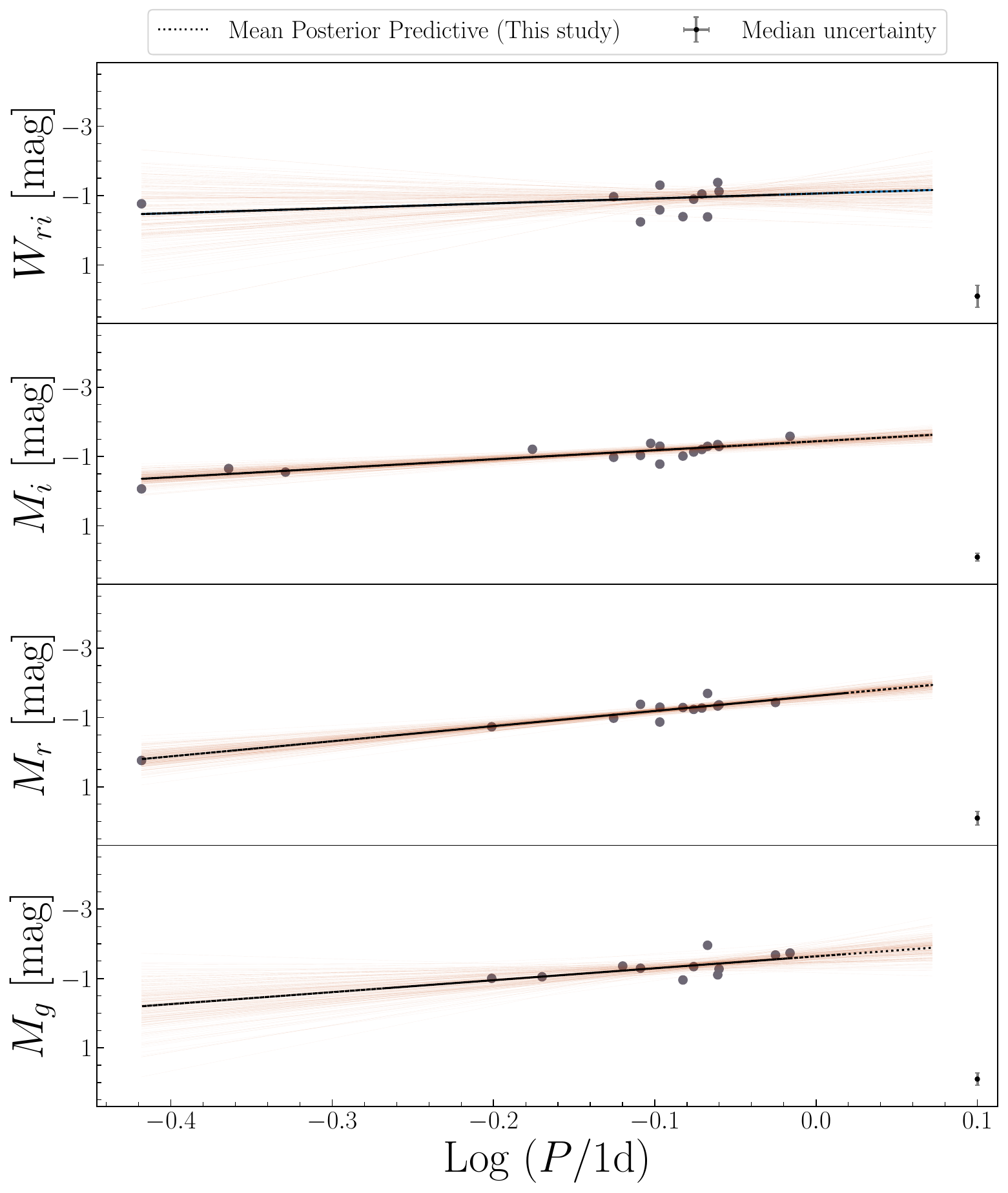}
\caption{Same as Fig.~\ref{Fig: F_lmc_PL_pp_study_comp_gband} for ACEP FO PL relations in the $gri$ bands and PW$_{ri}$ relations.}  
\label{Fig: ACEP_1O_lmc_PL_pp}
\end{figure}

\begin{figure}[!htbp]
\centering
\includegraphics[width=0.6\columnwidth, keepaspectratio]{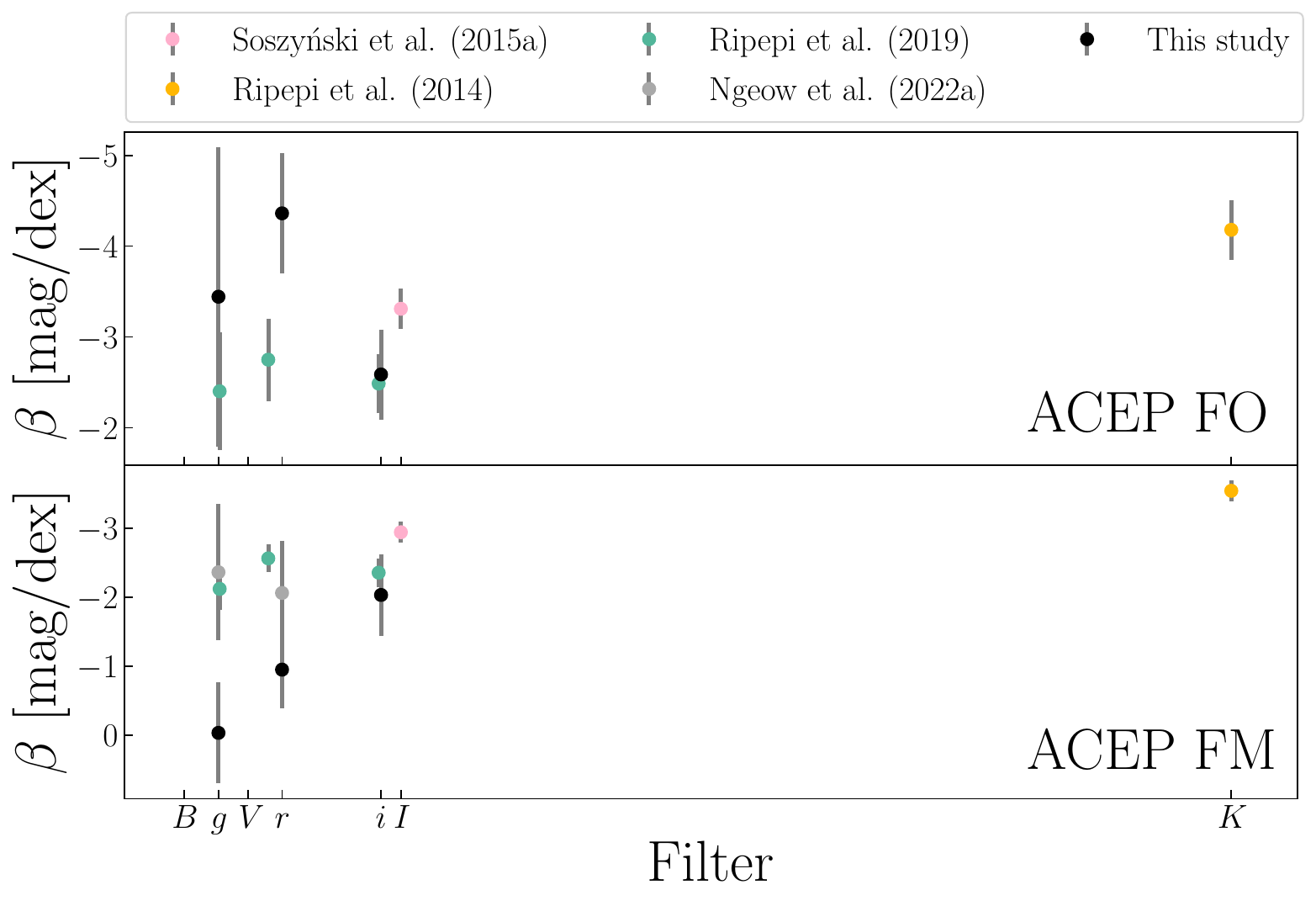}
\caption{Comparison of the slopes of Period luminosity relations obtained in various optical and near-infrared photometric bands for ACEP FMs {\it (lower panel)}
and the ACEP FOs {\it (upper panel)}. Note that the y-axis is reversed, showing steeper slopes above shallower slopes. The studies included in the comparison are those of \cite{ripepi2014_VMCSurveyVIII,
ripepi2019_ReclassificationCepheidsGaia,
soszynski2015_OGLECollectionVariable}
and \cite{ngeow2022_ZwickyTransientFacilitya}} 
\label{Fig: AC_gri_slope_comparison}
\end{figure}

 The parameters of our PL/PW relations for ACEP FMs and ACEP FOs are listed in Table~\ref{tab: AC PLPW relations}. They are derived using a much larger number of stars (see Table \ref{Table: photometric filtering scheme}) with homogeneous SMSS DR2 photometry. They share the same characteristics as the PL/PW relations for CCs and T2Cs, the PL relations in the $r$ and $i$ bands being once again the ones with the smallest intrinsic dispersion thanks to relatively higher-quality data. They are shown in Figs.~\ref{Fig: ACEP_F_lmc_PL_pp_study_comp},\ref{Fig: ACEP_1O_lmc_PL_pp}. As far as the slopes are concerned, Fig.~\ref{Fig: AC_gri_slope_comparison} indicates that our slopes for the ACEP FOs PL relations in the $g$ and $r$ bands might be too large when compared to those obtained in the $BP$ and $G$ Gaia filters by \citep{ripepi2019_ReclassificationCepheidsGaia} and in the VMC $K$ band \citep{ripepi2014_VMCSurveyVIII}. On the other hand, the PL relation in the $i$ band is in excellent agreement with the Gaia $RP$, and does not lie too far from the one derived in the OGLE $I$ band by \citep{soszynski2015_OGLECollectionVariable}, In surprising contrast, our $g$ and $r$ bands slopes for ACEP FMs are much smaller than those obtained by \citet{ripepi2019_ReclassificationCepheidsGaia} in the $BP$ and $G$, but also by \citet{ngeow2022_ZwickyTransientFacilitya} in $g$ and $r$ as well.  In the $i$ band, our slope agrees with the one of \citet{ngeow2022_ZwickyTransientFacilitya}, both being slightly smaller than the one derived in the $I$ band by \citet{soszynski2015_OGLECollectionVariable}, Clearly, PL/PW relations for Anomalous Cepheids deserve further inspection. Similarly, a comparison with the PL/PW relations derived by \cite{ngeow2022_ZwickyTransientFacilitya} indicates that only their PL relation in the $i$ band is in excellent agreement with ours, as it falls well within our 94\% credible interval, while the $W_{ri}$ PW relation lies only slightly outside of it. The PL relations in the $g$ and $r$ bands disagree, which may be due either to inaccurate mean magnitudes in our study or to the very small number of ACEPs in the sample of  \cite{ngeow2022_ZwickyTransientFacilitya}. We note in passing that ACEP FO can easily be confused with DCEP FO, and ACEP FM with DCEP FM or RR$_{ab}$, making their classification from the shape of their light curve particularly difficult when their distance is not known.

%\begin{center}
\begin{table}[!htbp]
%\begin{adjustbox}{width=0.992\columnwidth, keepaspectratio}
\centering
\caption{Same as Table~\ref{tab: CC PLPW relations} for ACEP FMs and ACEP FOs.}
\label{tab: AC PLPW relations}
\begin{tabular}{lccccccc}
\toprule\toprule
\multicolumn{1}{c}{\textbf{AC Class}} & \textbf{Filter} & \textbf{$\alpha_{mean}$} & \textbf{$\alpha_{sd}$} & \textbf{$\beta_{mean}$} & \textbf{$\beta_{sd}$} & \textbf{$\sigma_{mean}$} & \textbf{$\sigma_{sd}$} \\ 
\midrule
\multirow{4}{*}{ACEP FM} & $g$      & -1.165 & 0.647 & -0.033 & 0.735 & 0.210 & 0.089 \\
                         & $r$      & -2.000 & 0.487 & -0.949 & 0.560 & 0.216 & 0.057 \\
                         & $i$      & -2.845 & 0.319 & -2.031 & 0.360 & 0.117 & 0.037 \\
                         & $W_{ri}$ & -3.731 & 1.587 & -3.289 & 1.763 & 0.573 & 0.146 \\
\midrule
\multirow{4}{*}{ACEP FO} & $g$      & -5.083 & 1.788 & -3.443 & 1.650 & 0.189 & 0.111 \\
                         & $r$      & -5.987 & 0.734 & -4.362 & 0.664 & 0.075 & 0.060 \\
                         & $i$      & -4.026 & 0.558 & -2.586 & 0.496 & 0.148 & 0.052 \\
                         & $W_{ri}$ & -2.473 & 2.444 & -1.414 & 2.239 & 0.239 & 0.170 \\ 
\bottomrule
\end{tabular}
%\end{adjustbox}
\end{table}
%\end{center}

%-----------------------------------------------------------------------
\section{Application of PL/PW Laws}
\label{M31}

\subsection{Data: M31 as a benchmark galaxy}
\label{M31_data}

\par In order to test the usage of T2Cs as a potential first ladder of the extragalactic distance scale, we consider the Andromeda galaxy, M31 (also known as NGC\,224) as a benchmark galaxy. M31 is our closest neighboring spiral galaxy, its disk featuring 
slightly sub-solar metallicities for the oldest stars to super-solar metallicities for relatively young stars \citep[e.g.,][]{gregersen2015_PanchromaticHubbleAndromeda}, while stars in its halo have [Fe/H] ranging from -1.0\,dex to -2.0\, dex \citep[e.g.,][]{wojno2023_ElementalAbundancesM31}. Its large distance makes it an ideal target to test the extragalactic distance scale, and its spiral nature ensures a good number of T2Cs in its halo and an even larger number of CCs in its disk. However, it is a difficult target for small ground-based telescopes, and a number of studies took advantage of the high resolution of the space-based HST.

\subsubsection{M31 data: previous studies}

%\par\bl{TOO MANY CITATIONS FOR DIRECT vs OTHERS?}
\par Several projects have carried out ground-based surveys of M31, for instance, the DIRECT (for "direct distances") project \citep{kaluzny1998_DIRECTDistancesNearby, kaluzny1999_DIRECTDistancesNearby, stanek1998_DIRECTDistancesNearby, stanek1999_DIRECTDistancesNearby, mochejska1999_DIRECTDistancesNearby, bonanos2003_DIRECTDistancesNearby}, which targeted M31 and M33 using small telescopes. They were interested in their Cepheids and eclipsing binaries and observed in various optical bands. The WeCAPP microlensing survey \citep{fliri2006_WendelsteinCalarAlto} monitored the bulge of M31 in the $R$ and $I$ bands and delivered catalogs of variable stars as a by-product. \citet{vilardell2006_EclipsingBinariesSuitable, vilardell2007_ComprehensiveStudyCepheid} did the same in the $B$ and $V$ bands for the north-eastern quadrant of M31. Finally, the PAndromeda project which stands for Pan-STARRS1 (PS1) survey \citep{kaiser2002_PanSTARRSLargeSynoptic} of Andromeda, investigated various types of variable stars, including Cepheids \citep[][]{kodric2013_PropertiesM31II, kodric2018_CepheidsM31PAndromeda} in the $gri$ bands of PS1.

\par On the other hand, the Panchromatic Hubble Andromeda Treasury (PHAT) space-based survey is an ongoing $HST$ Multi-Cycle Treasury program to image $\sim$1/3 of M31's disk \citep{dalcanton2012_PanchromaticHubbleAndromeda}. The data from this program provides various samples of M31 Cepheids: \cite{riess2012_CepheidPeriodLuminosityRelations} sample; \cite{kodric2015_M31NearinfraredPeriodLuminosity, kodric2018_M31PAndromedaCepheid}, and the \cite{wagner-kaiser2015_PanchromaticHubbleAndromeda} samples. The images are collected with the Wide Field Camera 3 (WFC3) and the Advanced Camera for Surveys (ACS) and they cover from the ultraviolet (UV) to the near-infrared (NIR) wavelength regime. In addition to CCs, \cite{kodric2015_M31NearinfraredPeriodLuminosity} provides the only T2Cs data in the HST NIR filters ($F110W$ and $F160W$), which consists of 36 T2Cs located in M31.

\subsubsection{K18 data}
\label{K18_data}

\par Out of all these samples, \citetalias{kodric2018_CepheidsM31PAndromeda} presents the largest and most homogeneous sample of Cepheids, based on the complete Pan-STARRS1 survey \citep{chambers2016_PanSTARRS1Surveys}. With a large field of view covering M31's disk in a single exposure, they secured up to 56 epochs in the $g$ band, up to 420 epochs in the $r$ band, and up to 262 epochs in the $i$ band (and each epoch consists of at least three frames for the Cepheid light curves).  Since the number of observations is highest in the $r$ band for all the pulsating stars, \citetalias{kodric2018_CepheidsM31PAndromeda} used the $r$ band light curves to determine the pulsation periods of these stars. \citetalias{kodric2018_CepheidsM31PAndromeda} also extinction-corrected their $gri$ bands mean magnitudes using the \citet{montalto2009_PropertiesM31Dust} reddening map which only covers the disk of M31 (see Fig. 17 and 18 in \citetalias{kodric2018_CepheidsM31PAndromeda}). They also included the foreground extinction ($E(B-V)_{fg}$ = 0.062 mag) towards M31 determined by \citet{schlegel1998_MapsDustInfrared} with a correction factor of 0.86 from \cite{schlafly2011_MeasuringReddeningSloan} in this extinction correction calculation (see  Eqn.~2 in \citetalias{kodric2018_CepheidsM31PAndromeda}). Furthermore, \citetalias{kodric2018_CepheidsM31PAndromeda} derived a $W_{ri}$ Wesenheit index with $R_{ri}$ = 3.86 based on the reddening law $R_{V}$ = 3.1 from \citet{fitzpatrick1999_CorrectingEffectsInterstellar}.\\

\par
\begin{figure}
\centering
\includegraphics[width=0.7\columnwidth, keepaspectratio]{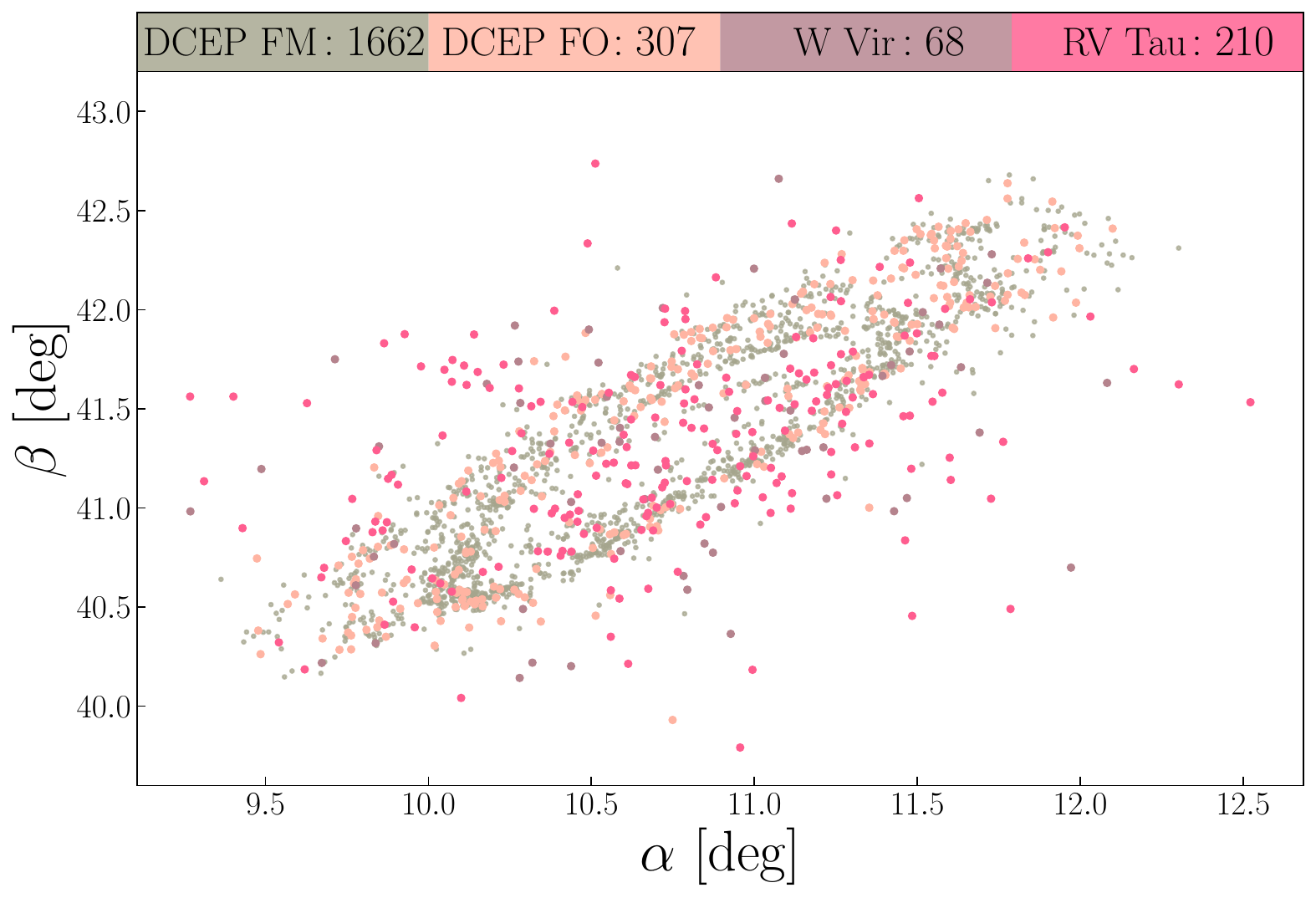}
\caption{2D distribution of M31 CCs, ACEPs, and T2Cs in the original
M31 data of \citetalias{kodric2018_CepheidsM31PAndromeda}. 
}
\label{Fig: m31_ra_dec}
\end{figure}

 By combining the Fourier parameters of the Cepheids' light curves with a color cut and with other selection criteria developed in \citet{kodric2013_PropertiesM31II}, \citetalias{kodric2018_CepheidsM31PAndromeda} classified their sample into the various types of Cepheid variables. They ended up with 1662 DCEP FM, 307 DCEP FO, 278 T2Cs, and 439 unclassified Cepheids. We divide these 278 T2Cs into their classes according to their empirical period range discussed in \ref{LMC_data}. While we do not find any BL Her, there are 68 W Vir and 210 RV Tau stars in the \citetalias{kodric2018_CepheidsM31PAndromeda} sample as displayed in Fig.~\ref{Fig: m31_ra_dec}. In addition to this, it is evident from Fig.~\ref{Fig: m31_ra_dec} that the T2Cs are majorly located in the halo of M31 while the CCs trace its disk.\\

\par In sect. \ref{LMC_method}, we derived PL/PW relations for these Cepheid variables in the SMSS $gri$ bands photometric system. These SMSS filters differ from the PS1 filters, especially in terms of efficiency curves. \cite{onken2019_SkyMapperSouthernSurvey} provide the transformation equations between SMSS and PS1, they contains a $(g-i)$ color term. They mention that these transformation equations are reasonably safe only for the main sequence and the sub-giant stars with spectral type F or G and luminosity class V or IV.
\par In our case, since pulsating stars do not fall into the categories listed above, bringing the PAndromeda photometry to the SkyMapper photometric system may induce an additional source of uncertainty. This is because: 1) The color and the metallicity are generally correlated, and different colors exhibit varying sensitivities to metallicity \citep{bono1999_TheoreticalModelsClassical}; 2) More importantly, for pulsating stars, the color term varies with the pulsation phase \citep{kanbur2010_Periodcolour_amplitude-colour_relations}.

\begin{figure}
\centering
\includegraphics[width=0.7\columnwidth, keepaspectratio]{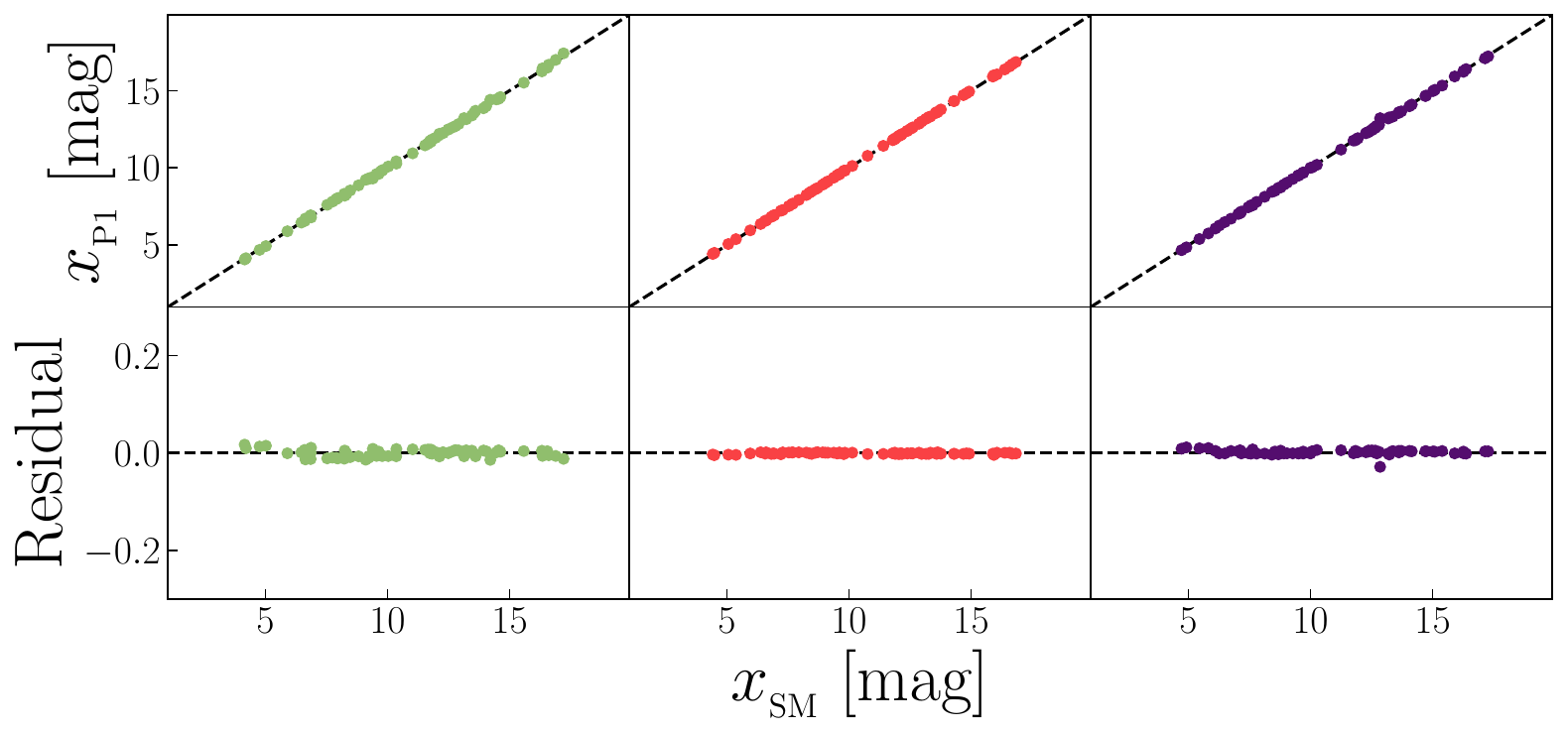}
\caption{Comparison between the SMSS and the PS1 photometric systems. The comparison is carried out using synthetic photometry of HST standard stars provided by SMSS.} 
\label{Fig: SM_PS_photo_compare}
\end{figure}

\par To illustrate the zero point difference between these two systems, we compare in Fig.~\ref{Fig: SM_PS_photo_compare} the SMSS and PS1 synthetic photometry of HST standard stars obtained from the second table provided on the SMSS website\footnote{\url{https://skymapper.anu.edu.au/filter-transformations/}}. For the $gri$ bands, the zero-point between these two systems does not differ significantly and the median absolute difference ($\Delta\tilde{\mathrm{ZP}}$) between them in the $gri$ bands are only $\sim$ 0.066 mag, $\sim$ 0.012 mag, and $\sim$ 0.031 mag respectively. Similarly, $\Delta\tilde{\mathrm{ZP}}$  for the subsequent $W_{ri}$ index is only $\sim$ 0.022 mag.\\

\par It is apparent that the $g$ band observations, in comparison to the $ri$ bands, are five to ten times scarcer in this \citetalias{kodric2018_CepheidsM31PAndromeda} data (see also Figs. \ref{Fig: lc_Panstarrs_DCEP_F_gri} and \ref{Fig: lc_Panstarrs_DCEP_1O_gri}). To aggravate this, amongst the total of 2686 stars in this M31 sample, 387 stars do not have mean $g$ band magnitudes and 789 stars carry a nonzero bit flag which indicates issues with the $g$ band data. As a consequence, we will estimate distances to M31 only in the $ri$ bands and the corresponding $W_{ri}$ index for various classes of Cepheid variables available in this data set. Moreover, in the further analysis to follow, we discard the 439 stars that remained unclassified by \citetalias{kodric2018_CepheidsM31PAndromeda}. Hence, we only focus on DCEP FM, DCEP FO, W Vir and RV Tau subtypes.

\subsubsection{Filtering the data}
\label{M31 Filtering the data}

\par Before we apply our $ri$ bands PL relations and $W_{ri}$ PW relation, we filter some potential contaminant stars contained in the \citetalias{kodric2018_CepheidsM31PAndromeda} data. We first reject PAndromeda stars that have fractional errors larger than 1\% on their pulsation period and mean $ri$ magnitudes. \citetalias{kodric2018_CepheidsM31PAndromeda} assigns Flag bit 128 to the 8 stars whose light curves do not align with a typical Cepheid-like light curve through visual inspection. In order to make this benchmark galaxy sample as clean as possible, we disregard these stars in our further analysis.
\par
\begin{figure}
\centering
\includegraphics[width=0.6\columnwidth, keepaspectratio]{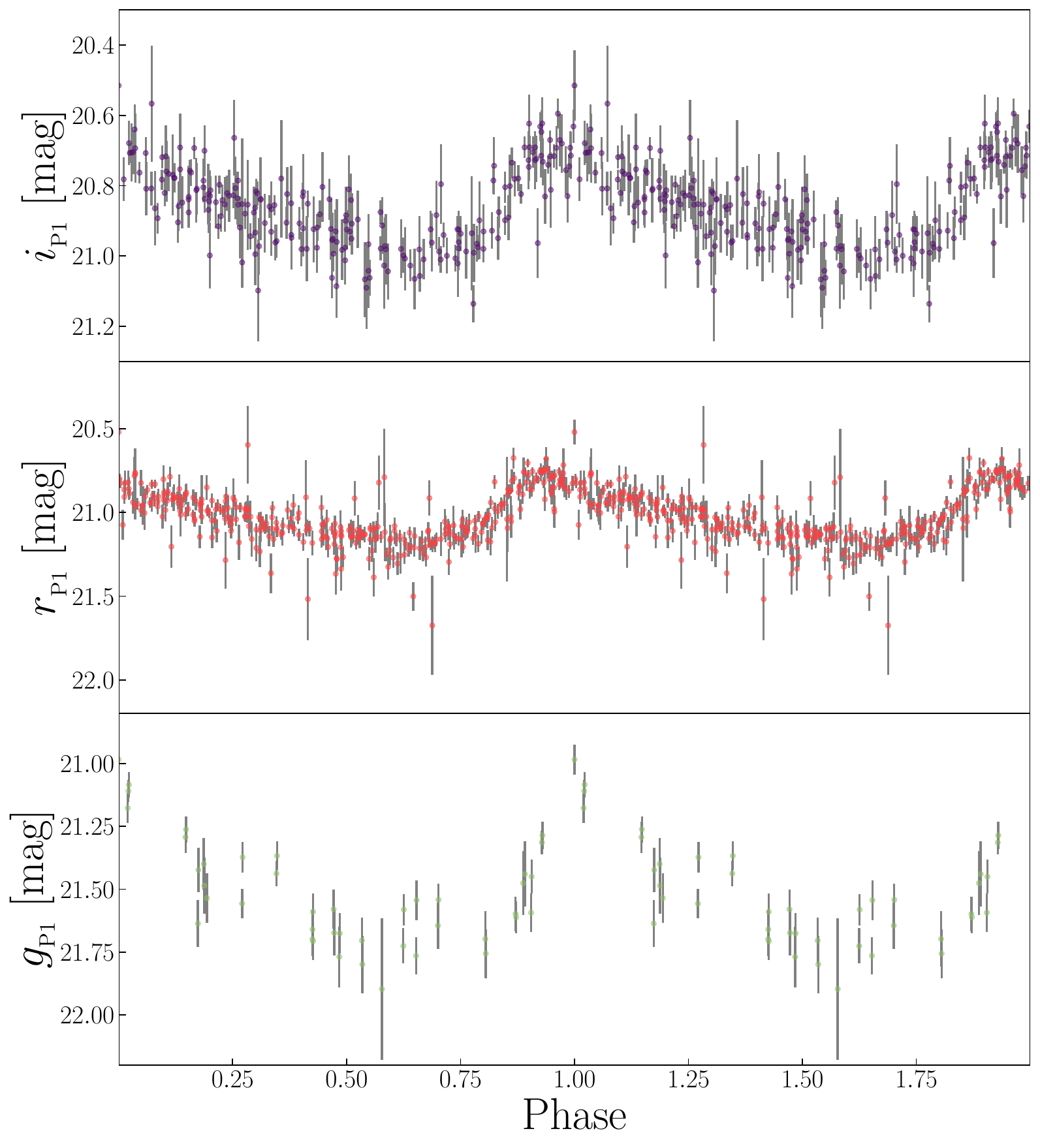}
\caption{PanSTARRS $gri$ bands light curves (bottom to top) of the DCEP FM (PAndromeda ID: 556411)}
\label{Fig: lc_Panstarrs_DCEP_F_gri}
\end{figure}

\begin{figure}
\centering
\includegraphics[width=0.6\columnwidth, keepaspectratio]{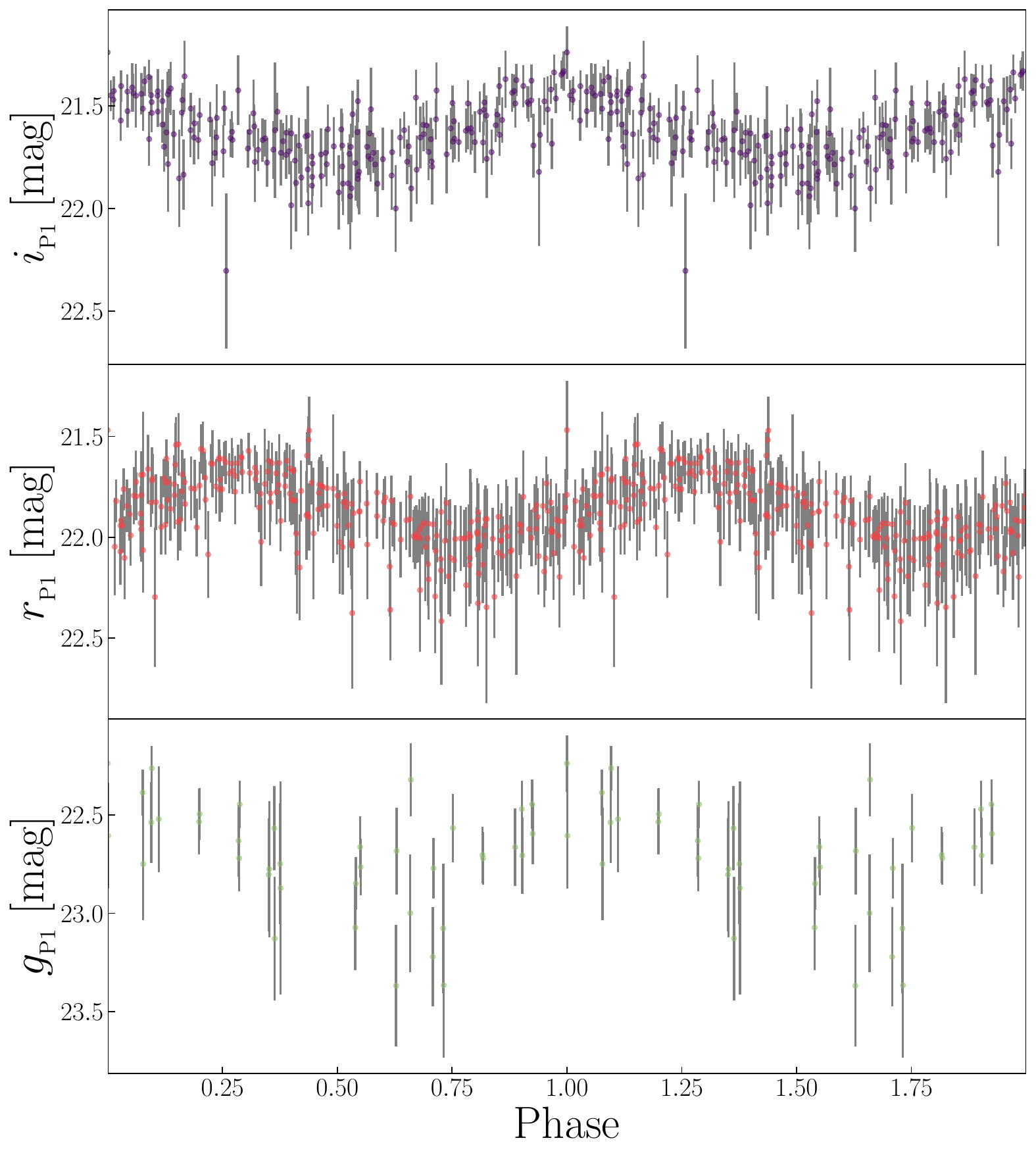}
\caption{PanSTARRS $gri$ bands light curves (bottom to top) of the DCEP FO (PAndromeda ID: 2925713)}
\label{Fig: lc_Panstarrs_DCEP_1O_gri}
\end{figure}

 Moreover, we notice that photometric error bars for individual measurements in the light curves are quite large in the \citetalias{kodric2018_CepheidsM31PAndromeda} data. We show the $gri$ bands light curves for representative stars of classes DCEP FM (PAndromeda ID: 556411) and DCEP FO (PAndromeda ID: 2925713), both having a pulsation period of $\sim$3.5\,d, in Fig. \ref{Fig: lc_Panstarrs_DCEP_F_gri} and Fig. \ref{Fig: lc_Panstarrs_DCEP_1O_gri}, respectively. Since there is an overlap between the period ranges of DCEP FM and DCEP FO stars at around $\sim$ 2 d to $\sim$ 6 d, larger uncertainties on individual light curves increase the risk of misclassification in the \citetalias{kodric2018_CepheidsM31PAndromeda} data.\\

\par  We screen these potentially misclassified stars by applying a sophisticated filtering scheme based on the Bhattacharyya distance. The Bhattacharyya distance \citep[][hereafter $D_{B}$]{1943_BulletinCalcuttaMathematical,bhattacharyya1946_MeasureDivergenceTwo} is an extension of the Mahalanobis distance which gauges the true separation between two distributions of any shape \citep{kailath1967_DivergenceBhattacharyyaDistance, aherne1998_BhattacharyyaMetricAbsolute} and quantifies the degree of proximity between these distinct statistical populations. More details on the Bhattacharyya distance and its comparison with Mahalanobis distance are outlined in Sect.~3.1 Lala et al. and their Appendix~A, respectively.\\

\par To estimate the $D_{B}$ for all the stars within this M31 sample, we use the following procedure:
\par 1. We first apply our derived $ri$ bands PL relations of DCEP FM, DCEP FO, and W Vir + RV Tau (shown in Table \ref{tab: CC PLPW relations} and \ref{tab: T2C PLPW relations}) on the entire set of M31 sample. Independent of their classes, this is executed on all the stars in a Monte Carlo sampling setup (same as mentioned in sect. \ref{LMC_result_subresult}) to estimate their model-predicted absolute magnitude/Wesenheit ($L_{\lambda}^{model}$) and its uncertainty. As discussed earlier, the M31 sample is deprived of BL Her stars, we then use the combined relations of W Vir + RV Tau shown in Table \ref{tab:T2C class PLPW relation} for this analysis.

\par 2. Furthermore, we aim to estimate the observation-based absolute magnitudes/Wesenheits ($L_{\lambda}^{obs}$) of stars in the \citetalias{kodric2018_CepheidsM31PAndromeda} sample. To achieve this, we estimate the corresponding pre-processed mean distance moduli of M31 by applying the $ri$ bands PL relations of DCEP FM, DCEP FO, and W Vir + RV Tau stars, but this time only to the respective classes of stars in the \citetalias{kodric2018_CepheidsM31PAndromeda} sample. It is now possible to estimate the $L_{\lambda}^{obs}$ of stars and the uncertainties on it by using the extinction corrected $ri$ apparent magnitudes from the \citetalias{kodric2018_CepheidsM31PAndromeda} data and this pre-processed estimated distance modulus of M31 $wrt$ various classes of Cepheid variables.

\par 3. With these $L_{\lambda}^{obs}$ and $L_{\lambda}^{model}$ along with their uncertainties in hand, we now have two distinct statistical populations in both the $ri$ bands. In step 1, since we are using three different PL relations in each $ri$ band, we gain the corresponding three $L_{\lambda}^{model}$ per star. This gives us the opportunity to calculate the associated three $D_{B}$ per star for each of the three pairs of $L_{\lambda}^{obs}$ and $L_{\lambda}^{model}$ by using the  Eqn.~1 from Lala et al.\\

\begin{figure}
\centering
\includegraphics[width=0.7\columnwidth, keepaspectratio]{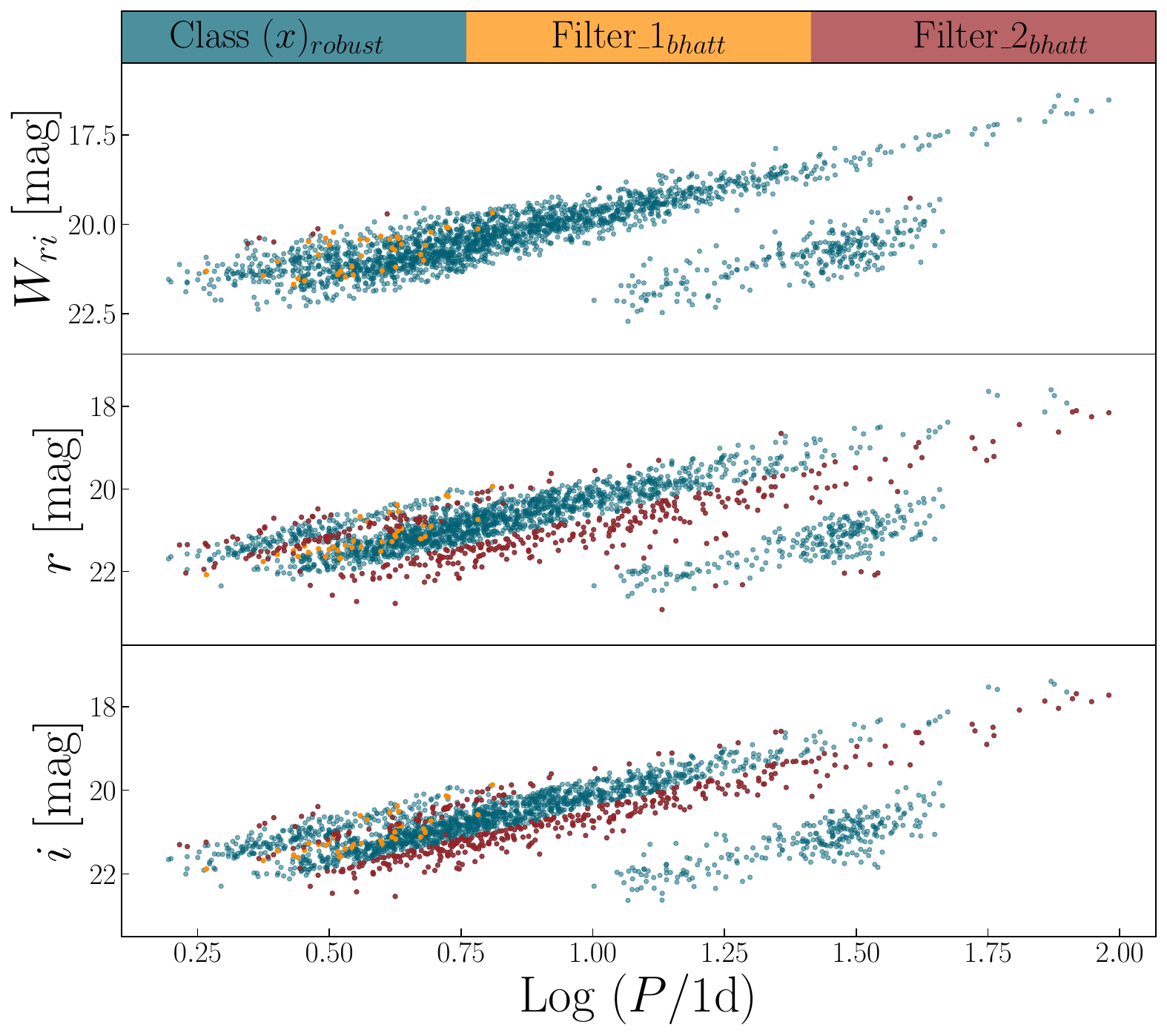}
\caption{Filtering the \citetalias{kodric2018_CepheidsM31PAndromeda} data to retain only the most robustly classified stars (in blue). The other two categories (Filter\_1$_{bhatt}$ in yellow and Filter\_2$_{bhatt}$ in red) are stars that obtained lower scores in our Bhattacharyya distance classification scheme. They were ignored when estimating the distance to M31.} 
\label{Fig: bhatt clean}
\end{figure}

\par We take a complete leverage of these estimated $D_{B}$ and derive the corresponding posterior probabilities $p_{match}$ according to  Eqn.~2 from Lala et al. In our case, $p_{match}$ denotes the probability for a given star to be classified in one of the five subclasses we consider. We then filter out those stars with $p_{match}$ $\geq$ 0.5 in both $ri$ bands for another class than the class assigned by \citetalias{kodric2018_CepheidsM31PAndromeda}. They are marked as Filter\_1$_{bhatt}$ in Fig.~\ref{Fig: bhatt clean}. We do not suggest here that these stars have been misclassified by \citetalias{kodric2018_CepheidsM31PAndromeda}, we simply ensure that we retain only those with the most robust classification. A Cepheid classified as DCEP FM by \citetalias{kodric2018_CepheidsM31PAndromeda} that would obtain probabilities $p_{match}$=0.3 as DCEP FM, $p_{match}$=0.5 as DCEP FO and $p_{match}$=0.2 as BL Her would not be reclassified as DCEP FO but simply ignored. Moreover, any star with a low probability ($p_{match}$ $\leq$ 0.05 in the $ri$ and $W_{ri}$ bands) in its own class as assigned by \citetalias{kodric2018_CepheidsM31PAndromeda}, for instance, an RV TAU star with $p_{match}$=0.03 in the RV Tau category, would similarly be ignored. Such stars are marked as Filter\_2$_{bhatt}$ in Fig.~\ref{Fig: bhatt clean}. We use only the remaining Class $(x)_{robust}$ stars shown in Fig.~\ref{Fig: bhatt clean} and enumerated in Table~\ref{tab:m31 filtered data} to estimate the M31 distances in the $ri$ bands and $W_{ri}$ index, the outcome of our analysis is described in the next section.\\

\begin{table}[!htbp]
 %\begin{adjustbox}{width=0.992\columnwidth, keepaspectratio}
 \centering
\caption{Number of M31 Cepheids in various classes after filtering the \citetalias{kodric2018_CepheidsM31PAndromeda} data (see Sections~\ref{K18_data} and \ref{M31 Filtering the data}).}
\label{tab:m31 filtered data}
\begin{tabular}{cccc}
\toprule\toprule
\textbf{Class}   & \textbf{$r$}  & \textbf{$i$}  & \textbf{$W_{ri}$} \\
\midrule
DCEP FM & 1364 & 1257 & 1634     \\
\midrule
DCEP FO & 175  & 236  & 265      \\
\midrule
W Vir   & 56   & 67   & 67       \\
\midrule
RV Tau  & 198  & 203  & 204   \\
\bottomrule
\end{tabular}
% \end{adjustbox}
 \end{table}

%----------------------------------------------------------------------
\subsection{Results: Estimations of M31 distance}
\label{M31_result}

\par In order to derive the multiple estimates of the M31 distance based on the filtered stars (the corresponding numbers are shown in Table \ref{tab:m31 filtered data}), we adopt the same Monte Carlo approach as described in Sect. \ref{LMC_result_validation}. Along the individual population distances estimated (for stars of class DCEP FM, DCEP FO, W Vir, and RV Tau) using the associated PL/PW relations, we estimate the median distances jointly for W Vir stars and RV Tau stars using our combined W Vir + RV Tau PL/PW relations and the global T2C PL/PW relations (shown in Table \ref{tab: T2C PLPW relations}). Similarly, we also estimate the median distances of W Vir stars using the combined PL/PW relations of BL Her + W Vir. All these CCs and T2Cs median distances ($\tilde{\mu}_{M31,x}$, where $x$ signifies various classes and their combinations) are listed in the third column of Table \ref{tab: m31 distances result} along with the corresponding uncertainty on this median distance ($\tilde{\mu}_{err}$) in column 4.

\par \citetalias{kodric2018_CepheidsM31PAndromeda} uses the reddening map from \cite{montalto2009_PropertiesM31Dust} in order to deredden the $gri$ band magnitudes (c.f. Eqn.~2 in \citetalias{kodric2018_CepheidsM31PAndromeda}). However, since \cite{montalto2009_PropertiesM31Dust} do not provide color excess uncertainties in their map, \citetalias{kodric2018_CepheidsM31PAndromeda} could not propagate these values into the dereddened magnitude errors. Following \cite{dalcanton2015_PanchromaticHubbleAndromeda}, we assume a 20\% uncertainty on the \cite{montalto2009_PropertiesM31Dust} reddening values. Consequently, we included the uncertainty of extinction ($\sigma_{A,x,y}$) for stars in class $x$ in each band $y$ in the statistical uncertainty, $\mu_{err}^{stat}$, following Eqn.~\ref{statistical uncertainty on distance modulus}. The values of $\mu_{err}^{stat}$ on M31 distances are listed in column 5 of Table \ref{tab: m31 distances result}.

\par Moreover, the median absolute difference in the zero-points ($\Delta\tilde{\mathrm{ZP}_{y}}$) of SMSS and PS1 photometry in Sect.~\ref{K18_data} are added in quadrature in Eqn.~\ref{systematic uncertainty on distance modulus}. They contribute to the systematic uncertainty ($\mu_{err}^{sys}$) concerning M31 distances in band $y$. These uncertainties are shown in column 6 of Table \ref{tab: m31 distances result}. The complete error budget on the determination of total formal uncertainty ($\mu_{err}^{tot}$) of M31 distances is outlined in Table \ref{tab: M31 distance error budget}. This uncertainty is computed using Eqn.~\ref{total unceratinty} and its values are listed in column 7 of Table~\ref{tab: m31 distances result}.
\\

\begin{figure}
\centering
\includegraphics[width=0.7\columnwidth, keepaspectratio]{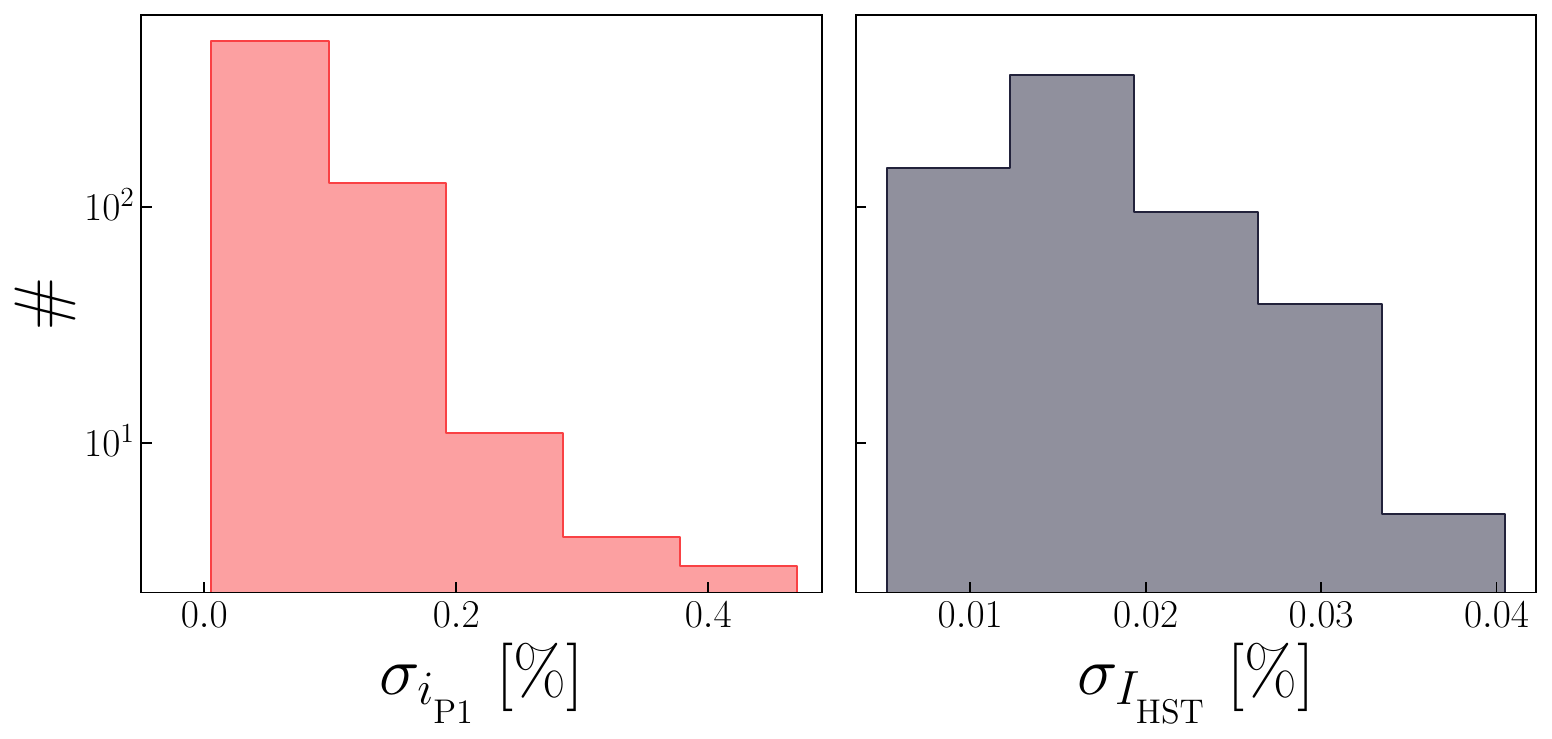}
\caption{Comparison of the photometric percentage errors in the PS1 $i$ band and the HST $F814W$ band for the stars in common in the \citetalias{kodric2018_CepheidsM31PAndromeda} PS1 sample and the \citet{kodric2018_M31PAndromedaCepheid} HST sample, illustrating the superior quality of HST photometry.}
\label{Fig: Photometric percentage error PS1 vs HST}
\end{figure}

\par Fig. \ref{Fig: Photometric percentage error PS1 vs HST} shows the photometric percentage errors in the PS1 $i$ band and the HST $F814W$ band for the stars that are common in the \citetalias{kodric2018_CepheidsM31PAndromeda} PS1 sample and in the \cite{kodric2018_M31PAndromedaCepheid} HST sample. It is evident from the figure that the photometric percentage error in the PS1 $i$ band is $\sim$ 10 times larger than the one in the $I$ band of the HST, as expected. Hence, in the following, we will only focus on the accuracy of our median distance estimates of M31 $wrt$ recent robust estimates of the M31 distance.

\par We mentioned earlier that the extinction correction carried out by \citetalias{kodric2018_CepheidsM31PAndromeda} relies on the \citet{montalto2009_PropertiesM31Dust} reddening map which only covers the disk of M31, leading to potential issues regarding the treatment of extinction for T2Cs in the halo of M31. We suspect that this is the reason why the distances based on T2Cs PL relations in the $ri$ reach lower accuracies ($\sim$90-95\%) than those based on the reddening free $W_{ri}$ PW relations ($>$98\% in general). Indeed, their accuracy is similar in the case of the CCs.\\

%--------------------------------------------------------------------------------------------------------------------

\begin{table}[!htbp]
\centering
\caption{Summary of the total error budget for various median distance determinations of M31.}
\label{tab: M31 distance error budget}
\begin{tabular}{lll}
\toprule\toprule
\multicolumn{3}{c}{$\mu_{err}^{tot}$ on the M31 distance modulus $\tilde{\mu}_{M31,x}$} \\
\midrule
\textbf{Uncertainty type} & \textbf{Contributing term} & \textbf{Description} \\
\midrule
\begin{tabular}[c]{@{}l@{}}Uncertainty on median \\ distance modulus \\ ($\tilde{\mu}_{err}$)\end{tabular} 
& $\sigma_{\mu,x}$ 
& \begin{tabular}[c]{@{}l@{}}Standard deviation of all the stellar \\ distance moduli ($\sigma_{\mu}$) in class $x$\end{tabular} \\[2mm]

\midrule
\multirow{3}{*}{\begin{tabular}[c]{@{}l@{}}Statistical uncertainty \\ ($\mu_{err}^{stat}$ per star $i$ \\ of class $x$ in band $y$)\end{tabular}} 
& $\sigma_{m,x,y}$ 
& Uncertainty on apparent magnitude $m$ \\[2mm]
& $\sigma_{A,x,y}$ 
& Uncertainty on interstellar extinction $A$ \\[2mm]
& $\beta_{sd,x,y}$ 
& \begin{tabular}[c]{@{}l@{}}Uncertainty on slope $\beta$ (coefficient of Log$P$)\\ of PL/PW relations\end{tabular} \\[3.5mm]

\midrule
\multirow{4}{*}{\begin{tabular}[c]{@{}l@{}}Systematic uncertainty \\ ($\mu_{err}^{sys}$)\end{tabular}} 
& $\sigma_{\mu,cal}$ 
& \begin{tabular}[c]{@{}l@{}}Uncertainty on 1\% precise LMC calibration \\ distance modulus\end{tabular} \\[3.5mm]
& $\Delta\tilde{\mathrm{ZP}_{y}}$ 
& \begin{tabular}[c]{@{}l@{}}Zero-point absolute median difference \\ between SMSS \& PS1\end{tabular} \\[3.5mm]
& $\alpha_{sd,x,y}$ 
& Uncertainty on intercept $\alpha$ of PL/PW relations \\[2mm]
& $\beta_{sd,x,y}$ 
& \begin{tabular}[c]{@{}l@{}}Uncertainty on slope $\beta$ (coefficient of Log$P_{0}$;\\ where Log$P_{0}$ = 1) of PL/PW relations\end{tabular} \\

\bottomrule
\end{tabular}
\end{table}
%--------------------------------------------------------------------------------------------------------------------

\subsubsection{Comparing M31 distances with recommended value}

\begin{figure}
\centering
\includegraphics[width=0.78\columnwidth, keepaspectratio]{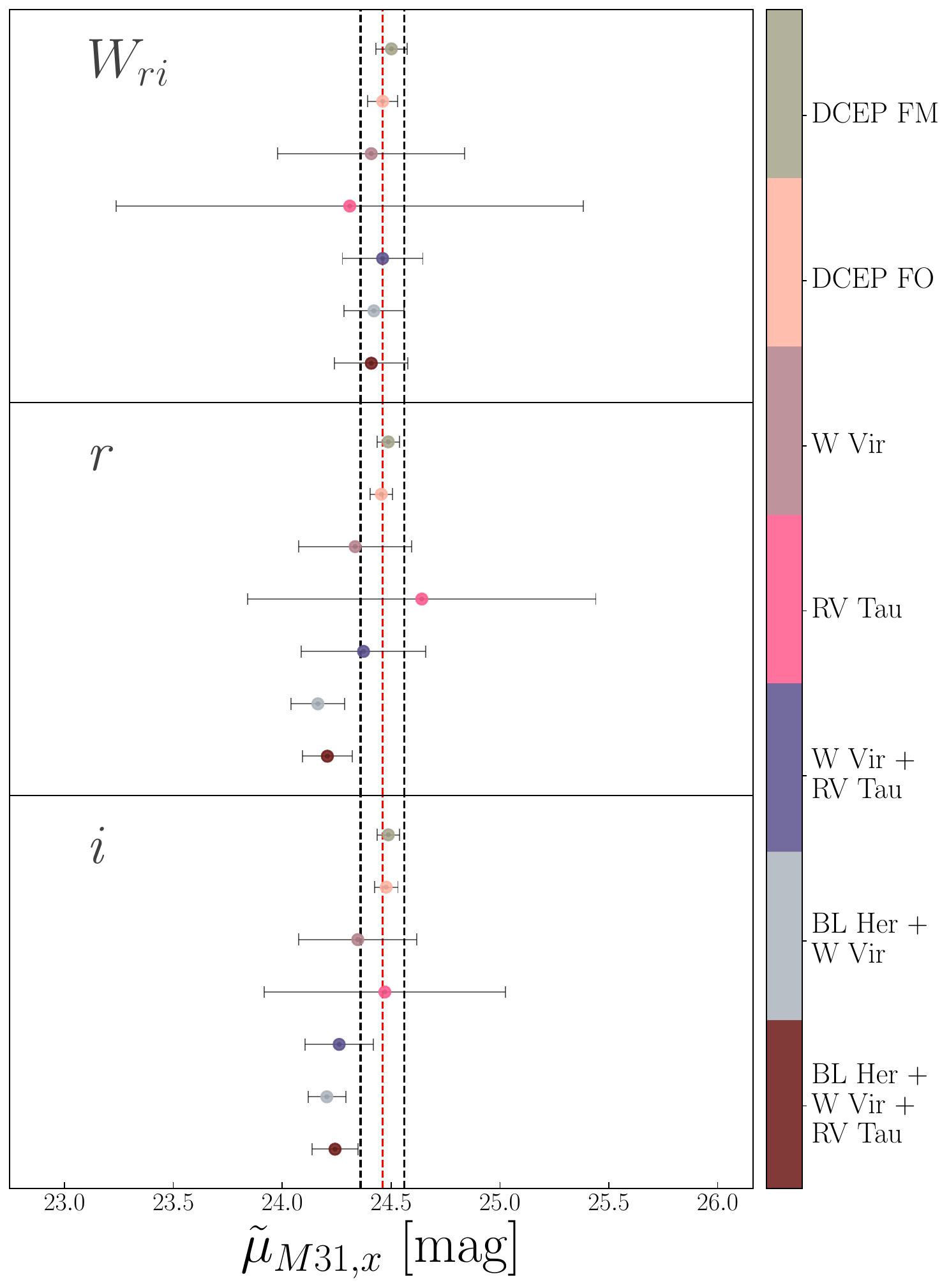}
\caption{Median value of the M31 distance modulus (colored dots) and the {\it total} uncertainty on this median value $\tilde{\mu}_{err}^{tot}$ (horizontal thin gray lines) for different classes of pulsating stars (in different colors) and different PL/PW relations (in different panels). The vertical thin red line marks the M31 recommended distance modulus by \cite{degrijs2014_ClusteringLocalGroup} and the vertical black lines the total uncertainty on this distance. The distance modulus of M31 derived by \citet{conn2012_BayesianApproachLocating} lies exactly on top of this recommended value.} 
\label{Fig: m31_result_de_grijs_boxplot}
\end{figure}

\par We first compare all these resultant median distances of M31 listed in Table \ref{tab: m31 distances result} with the recommended distance modulus of M31 ($\mu=24.46 \pm 0.10$ mag\footnote{In this paper, we consider $\mu$ as a true distance modulus.}) by \cite{degrijs2014_ClusteringLocalGroup}. Fig. \ref{Fig: m31_result_de_grijs_boxplot} shows all these estimated median distances of M31 ($\tilde{\mu}_{M31,x}$) in the form of a box plot in the $W_{ri}$ index and $ri$ bands. 

\par \cite{degrijs2014_ClusteringLocalGroup} provide a robust set of benchmark distance estimates for a selected number of different types of galaxies within the local volume. They are the outcome of the exhaustive meta-analysis of distance measurements available at that time for these galaxies. In particular, for M31, they exploited many of its distance estimates based on reliable stellar population tracers, for instance, CCs, RR Lyrae and TRGBs. While adopting a common LMC calibrating distance modulus of $\mu=18.5$\,mag for all these past literature estimates, they provide a benchmark value of $\mu = 24.46 \pm 0.10$ mag for the distance of M31 through a careful statistical weighting of the distance estimates they considered. Since our PL/ PW relations calibration distance of the LMC ($\mu=18.477 \pm 0.026$ mag) is very close to the one they adopted, we compare directly our median distances of M31 with this study.\\

\par In terms of accuracy $wrt$ this study, we see excellent alignment of all our distance estimates for both CCs and T2Cs in the $W_{ri}$ index, clearly visible in Fig.~\ref{Fig: m31_result_de_grijs_boxplot}. Note that we do not include the metallicity correction ($\gamma$, coefficient of [Fe/H] in the PLZ relations of CCs) in these median estimates of M31 distance for DCEP FM and DCEP FO stars (more discussion on a possible metallicity dependence of PL/PW relations for these stars is covered in Sect. \ref{Metallicity dependence}). Even without this correction, the accuracy of our distance estimates relating to the one from \cite{degrijs2014_ClusteringLocalGroup} is more than 98\% for DCEP FM and more
than 99\% in the case of DCEP FO stars (see Table \ref{tab: m31 distances result}). The accuracy however is relatively less for all the median distance estimates related to T2Cs, especially in $ri$ bands which is noticed in both Fig. \ref{Fig: m31_result_de_grijs_boxplot} and Table \ref{tab: m31 distances result} (except for the RV Tau estimate in $i$ band). This slightly low accuracy of these distances may be the consequence of the color excess map of \cite{montalto2009_PropertiesM31Dust} used by \citetalias{kodric2018_CepheidsM31PAndromeda} for extinction correction which only covers the disk of M31, while the T2Cs reside in its halo (see Fig.~\ref{Fig: m31_ra_dec}). On top of this, the treatment of these extinction corrections by \citetalias{kodric2018_CepheidsM31PAndromeda} according to  Eqn.~2 in their study assumes that all the Cepheid variables are only partially masked by the dust in M31, as \cite{montalto2009_PropertiesM31Dust} provide the line of sight color excess map which completely goes through the M31. As far as RV Tau stars are concerned, the presence of a circumbinary disk around a significant fraction of them might interfere with the reddening correction (see Sect.~\ref{Intrinsic properties of RV Tau stars}).\\

\par The precision of this recommended distance modulus of M31 is $\sim$4.6\%. The precision of our derived median distances in the $ri$ bands, tabulated in Table \ref{tab: m31 distances result} using DCEP FM, DCEP FO, and W Vir stars is either better or comparable with this benchmark study. Even though RV Tau stars are intrinsically brighter stars, leading to a higher S/N when compared to the other T2Cs, their estimated distance precision is the poorest of all. The main contributors to the large total uncertainty on the median distance estimates for this class are: 1) the relatively large uncertainty on the derived slope and intercept of their PL/PW relations; 2) the large dispersion of these PL/PW relations (see Table \ref{tab:T2C class PLPW relation}) resulting from the relatively large variation in their luminosities (also evident from Fig.\ref{Fig: RVTAU_lmc_PL_pp}). While the precision of the distances of W Vir stars obtained by applying the combined BL Her + W Vir PL/PW relations is better than 10\% in all cases, the median distances obtained for W Vir stars and RV Tau stars from the combined W Vir + RV Tau PL/PW relations and the global T2C PL/PW relations are less precise, which may also be linked with the intrinsic properties of RV Tau stars (see sect. \ref{Intrinsic properties of RV Tau stars} for a more comprehensive discussion). However, one feature that is common in the precision of all these distance estimates is that the $W_{ri}$ index portrays relatively less precise distances. This is due to the large dispersion in the $W_{ri}$ PW relations as compared to $ri$ bands PL relations.

\subsubsection{Comparing M31 distances with other robust estimates}
\label{Comparing M31 distances with other robust estimates}

\par In this section, we compare our distance estimates to recent robust estimates obtained by \citet{li2021_Sub2DistanceM31} using HST photometry of CCs, by \citet{conn2012_BayesianApproachLocating} using the TRGB method, and by \citet{savino2022_HubbleSpaceTelescope} using HST photometry of RR Lyrae.
\par \cite{li2021_Sub2DistanceM31} provide a 1.49\% precise distance estimate based on HST photometric data in one NIR ($F160W$) filter and two optical ($F555W$ and $F814W$) filters for DCEP\_FM stars located in M31. They use the LMC-based 1.28\% absolute calibration of PL/PW relations by \cite{riess2019_LargeMagellanicCloud} in the same filters in order to derive a distance modulus to M31 of $\mu=24.407 \pm 0.032$ mag, which is the most precise determination of its distance to date.
\par In the case of the T2Cs distances estimated with the $W_{ri}$ index, the agreement with \citet{li2021_Sub2DistanceM31} is exquisite, since our accuracy is $>$ 99\% for the distances obtained from individual W Vir, BL Her + W Vir and the global T2Cs PW relations, and remains satisfactory ($>$ 97\%) in the case of the W Vir + RV Tau PW relation. The accuracy of our M31 distance estimates based on CCs in the $ri$ bands and the $W_{ri}$ index $wrt$ to \cite{li2021_Sub2DistanceM31} is slightly less ($\sim$ 95\% - 97\%). We believe that the $\sim$ 3\% - 4\% inaccuracy accounts for the metallicity effect since our distances are not corrected while those of \citet{li2021_Sub2DistanceM31} include a corrective term of $\gamma = - 0.17 \pm 0.06$\,mag dex$^{-1}$ \citep{riess2019_LargeMagellanicCloud}.\\

\par Since the TRGB method requires the collection of a large number of RGB stars to derive robust distances \citep{madore1995_TipRedGiant}, the few HST observations of the halo of M31 do not provide a large enough sample. Hence, \cite{conn2012_BayesianApproachLocating} turned to the ground-based Pan-Andromeda Archaeological Survey (PAndAS) survey from the Canada-France-Hawaii Telescope (CFHT), and determined a distance modulus of $\mu=24.46 \pm 0.05$ mag for M31. Since the TRGB distance of M31 derived by \cite{conn2012_BayesianApproachLocating} falls exactly at the previously discussed recommended distance of M31, our accuracy $wrt$ this estimate is exactly the same.\\

\par Recently, taking advantage of HST photometry of RR Lyrae stars in M31, \cite{savino2022_HubbleSpaceTelescope} applied Period-Wesenheit-Metallicity (PWZ) relations calibrated on Gaia eDR3 parallaxes and obtained a distance modulus of $\mu=24.45 \pm 0.06$ mag for M31, precise to a level of $\sim$3\%. The accuracy of our estimated M31 distances when compared with the RR Lyrae-based one by \cite{savino2022_HubbleSpaceTelescope} is $>$ 98\% in the majority of the cases.\\

\par To conclude, the remarkable agreement (particularly for T2Cs) of our distances to M31 with the robust estimates from high-quality data for both young (CCs) and old (RR Lyrae and TRGB) population tracers provides for the first time the strong evidence that T2Cs can be used as standard candles to probe the extragalactic distance scale.\\

\subsubsection{M31 distances from T2Cs}

\par In addition to all these robust estimates of M31 distance with a variety of distance indicators from 
young or old populations, we compare our M31 distance estimates from T2Cs with the ones from \cite{ngeow2022_ZwickyTransientFacility}. They applied the global T2Cs PL/PW relations in the $gri$ bands that they derived from a small sample (37) of stars located in Galactic GCs to the same sample of M31 T2Cs available in \citetalias{kodric2018_CepheidsM31PAndromeda}. They report M31 distances of $\mu_{r} = 24.180 \pm 0.021$ mag in the $r$ band, $\mu_{i} = 24.249 \pm 0.020$ mag in the $i$ band, and $\mu_{ri} = 24.423 \pm 0.026$ mag in the $W_{ri}$ index. The M31 distances that we obtained from the global T2Cs PL/PW relations calibrated on our caLMC data, portray a solid consistency with the values of \cite{ngeow2022_ZwickyTransientFacility}.

\begin{figure}
\centering
\includegraphics[width=0.78\columnwidth, keepaspectratio]{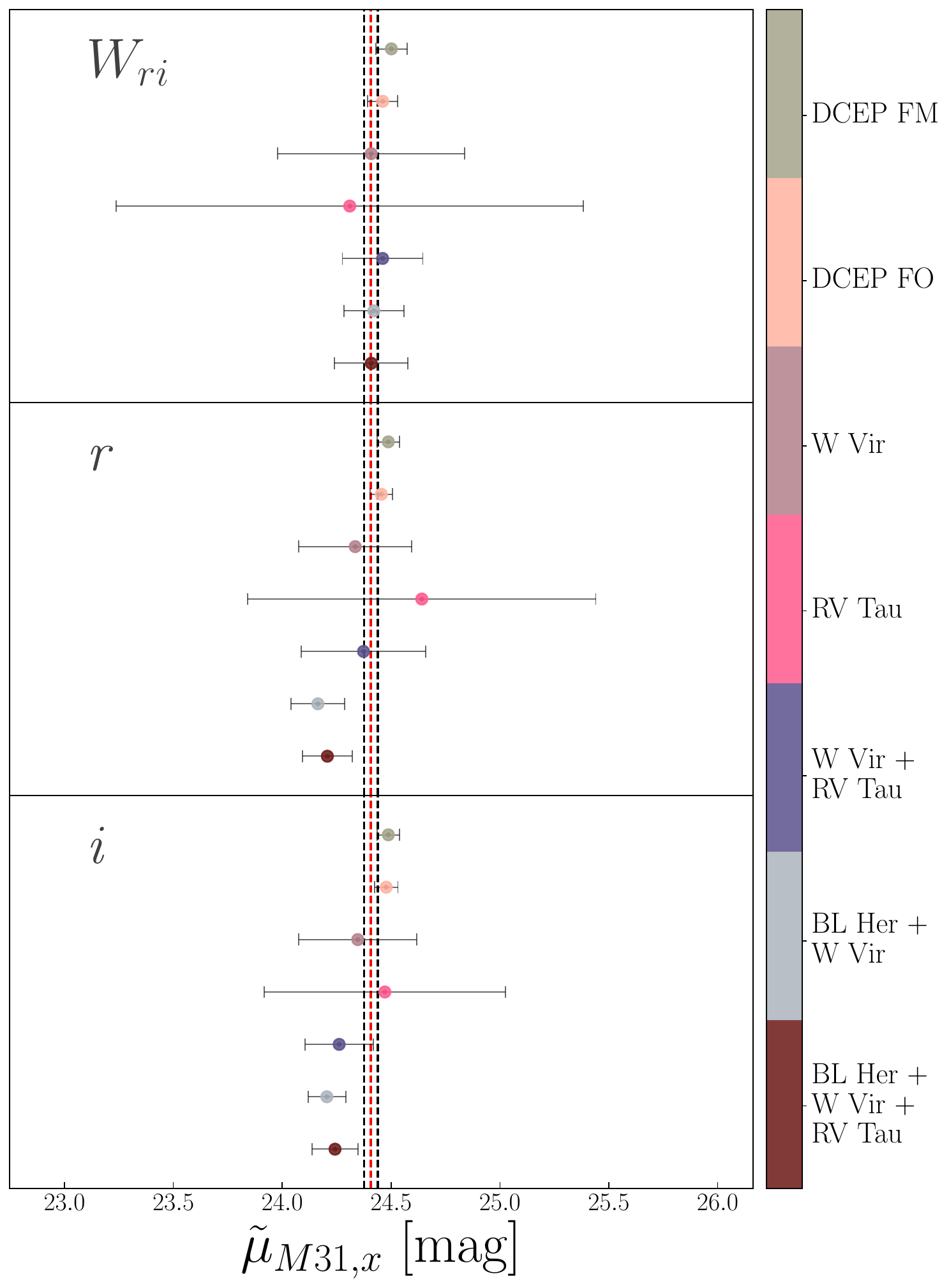}
\caption{Median value of the M31 distance modulus (colored dots) and the {\it total} uncertainty on this median value $\tilde{\mu}_{err}^{tot}$ (horizontal thin gray lines) for different classes of pulsating stars (in different colors) and different PL/PW relations (in different panels). The vertical thin red line marks the 1.49\% precise M31 distance modulus by \cite{li2021_Sub2DistanceM31} based on HST photometry of CCs, and the vertical black lines the total uncertainty on this distance.} 
\label{Fig: m31_result_li_boxplot}
\end{figure}

\begin{figure}
\centering
\includegraphics[width=0.78\columnwidth, keepaspectratio]{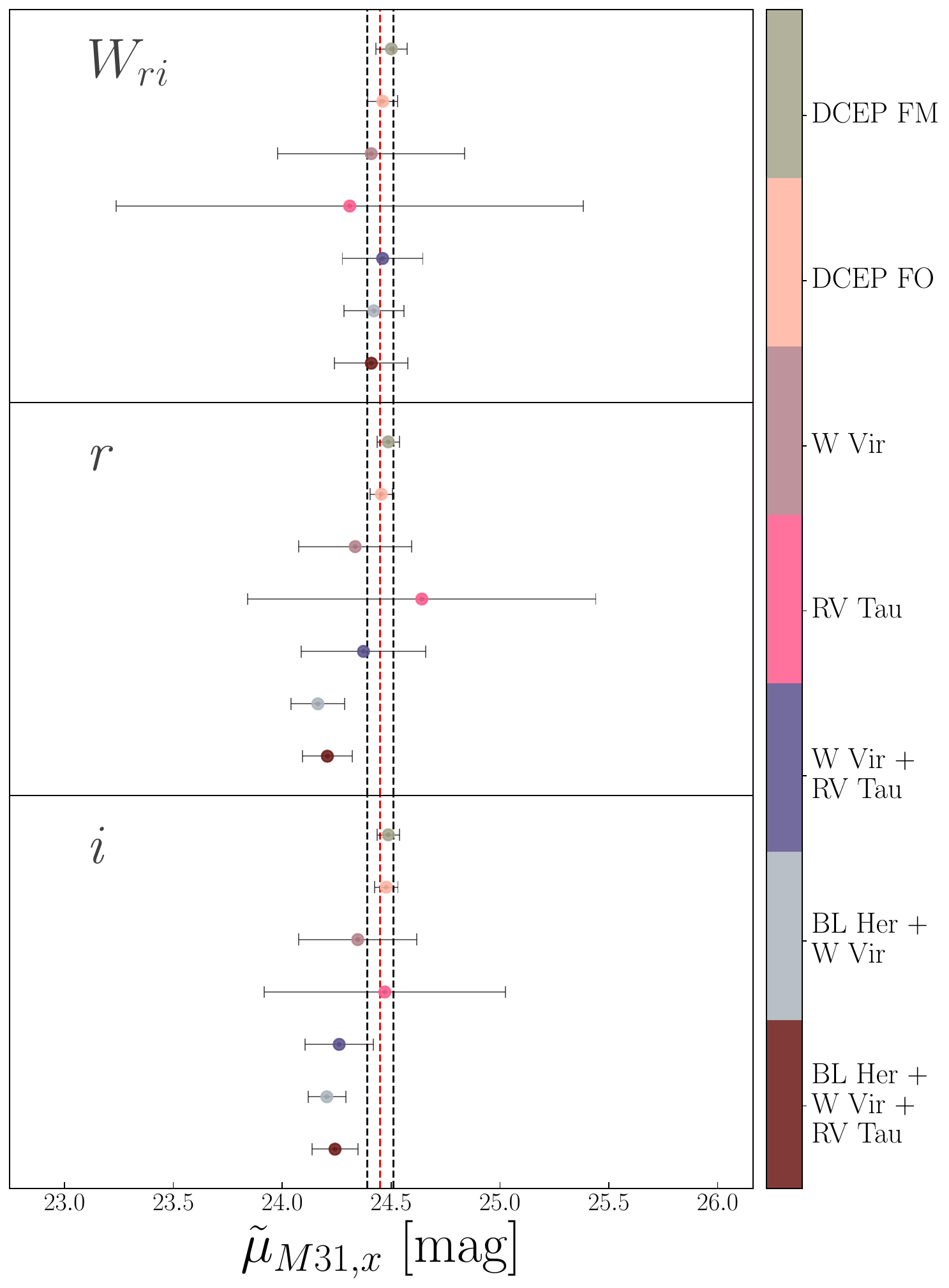}
\caption{Median value of the M31 distance modulus (colored dots) and the {\it total} uncertainty on this median value $\tilde{\mu}_{err}^{tot}$ (horizontal thin gray lines) for different classes of pulsating stars (in different colors) and different PL/PW relations (in different panels). The vertical thin red line marks the M31 distance modulus by \cite{savino2022_HubbleSpaceTelescope} based on HST photometry of RR Lyrae, and the vertical black lines the total uncertainty on this distance.} 
\label{Fig: m31_result_savino_boxplot}
\end{figure}

\begin{sidewaystable*}[!htbp]
\centering
\caption{Estimates of the distance to M31 for different Cepheid classes (or groups of classes). The table provides the median distance estimates ($\tilde{\mu}_{M31,x}$), the uncertainty on median distance moduli $\tilde{\mu}_{err}$, the statistical uncertainty $\mu_{err}^{stat}$, the systematic uncertainty $\mu_{err}^{sys}$, and the total uncertainty $\mu_{err}^{tot}$. Column 8 lists the precision of the distance provided by various PL/PW relations, and columns 9--11 give the accuracy with respect to: the recommended distance to M31 \citep{degrijs2014_ClusteringLocalGroup}$^{1}$, the TRGB distance \citep{conn2012_BayesianApproachLocating}$^{3}$, the most precise distance from HST photometry of CCs \citep{li2021_Sub2DistanceM31}$^{2}$, and the distance from HST photometry of RR Lyrae \citep{savino2022_HubbleSpaceTelescope}$^{4}$.}
\label{tab: m31 distances result}
\setlength{\tabcolsep}{4pt} % reduce column separation
\renewcommand{\arraystretch}{1.3} % increase row spacing
\begin{tabular}{ccccccccccc}
\toprule\toprule
\textbf{Class ($x$)} & \textbf{Filter} & \textbf{$\tilde{\mu}_{M31,x}$ [mag]} & \textbf{$\tilde{\mu}_{err}$ [mag]} & \textbf{$\mu_{err}^{stat}$ [mag]} & \textbf{$\mu_{err}^{sys}$ [mag]} & \textbf{$\mu_{err}^{tot}$ [mag]} & \textbf{Precision [\%]} & \textbf{Accuracy [\%] $^{1,3}$} & \textbf{Accuracy [\%] $^{2}$} & \textbf{Accuracy [\%] $^{4}$} \\
\midrule
\multirow{3}{*}{DCEP FM} & $r$      & 24.487 & 0.008 & 0.001 & 0.051 & 0.052 & 2.38  & 98.74 & 96.24 & 98.27 \\
                          & $i$      & 24.487 & 0.006 & 0.001 & 0.052 & 0.052 & 2.41  & 98.73 & 96.22 & 98.25 \\
                          & $W_{ri}$ & 24.501 & 0.010 & 0.002 & 0.071 & 0.071 & 3.29  & 98.08 & 95.56 & 97.60 \\
\midrule
\multirow{3}{*}{DCEP FO} & $r$      & 24.455 & 0.010 & 0.003 & 0.050 & 0.052 & 2.38  & 99.79 & 97.75 & 99.75 \\
                          & $i$      & 24.478 & 0.012 & 0.002 & 0.053 & 0.054 & 2.49  & 99.19 & 96.70 & 98.72 \\
                          & $W_{ri}$ & 24.462 & 0.021 & 0.006 & 0.065 & 0.069 & 3.18  & 99.92 & 97.44 & 99.45 \\
\midrule
\multirow{3}{*}{W Vir}   & $r$      & 24.335 & 0.033 & 0.034 & 0.255 & 0.259 & 11.94 & 94.42 & 96.76 & 94.86 \\
                          & $i$      & 24.347 & 0.047 & 0.032 & 0.265 & 0.271 & 12.47 & 94.94 & 97.28 & 95.38 \\
                          & $W_{ri}$ & 24.409 & 0.060 & 0.054 & 0.422 & 0.430 & 19.81 & 97.67 & 99.91 & 98.13 \\
\midrule
\multirow{3}{*}{RV Tau}  & $r$      & 24.641 & 0.033 & 0.047 & 0.797 & 0.799 & 36.80 & 91.30 & 88.61 & 90.79 \\
                          & $i$      & 24.471 & 0.029 & 0.032 & 0.553 & 0.554 & 25.53 & 99.47 & 96.99 & 99.00 \\
                          & $W_{ri}$ & 24.310 & 0.034 & 0.062 & 1.070 & 1.073 & 49.40 & 93.33 & 95.64 & 93.77 \\
\midrule
\multirow{3}{*}{\begin{tabular}[c]{@{}c@{}}W Vir +\\ RV Tau\end{tabular}} & $r$      & 24.374 & 0.022 & 0.018 & 0.285 & 0.287 & 13.20 & 96.10 & 98.48 & 96.55 \\
                          & $i$      & 24.262 & 0.024 & 0.009 & 0.154 & 0.156 & 7.21  & 91.28 & 93.53 & 91.70 \\
                          & $W_{ri}$ & 24.461 & 0.030 & 0.012 & 0.182 & 0.185 & 8.51  & 99.94 & 97.47 & 99.48 \\
\midrule
\multirow{3}{*}{\begin{tabular}[c]{@{}c@{}}BL Her +\\ W Vir\end{tabular}} & $r$      & 24.207 & 0.032 & 0.014 & 0.109 & 0.115 & 5.29  & 89.02 & 91.22 & 89.44 \\
                          & $i$      & 24.243 & 0.046 & 0.010 & 0.094 & 0.105 & 4.85  & 90.47 & 92.71 & 90.90 \\
                          & $W_{ri}$ & 24.409 & 0.060 & 0.025 & 0.156 & 0.169 & 7.78  & 97.69 & 99.89 & 98.15 \\
\midrule
\multirow{3}{*}{\begin{tabular}[c]{@{}c@{}}BL Her +\\ W Vir +\\ RV Tau\end{tabular}} & $r$      & 24.164 & 0.024 & 0.007 & 0.122 & 0.124 & 5.71  & 87.26 & 89.41 & 87.66 \\
                          & $i$      & 24.205 & 0.025 & 0.004 & 0.083 & 0.087 & 3.99  & 88.92 & 91.12 & 89.34 \\
                          & $W_{ri}$ & 24.421 & 0.030 & 0.010 & 0.134 & 0.137 & 6.33  & 98.23 & 99.34 & 98.69 \\
\bottomrule
\end{tabular}
\end{sidewaystable*}

%--------------------------------------------------------------------------------------------------------------------

\section{Discussion}
\label{discussion}

\subsection{Properties of RV Tau stars}
\label{Intrinsic properties of RV Tau stars}

\par Table~\ref{tab: m31 distances result} indicates that when compared to other studies, the PL/PW relations for RV Tau stars have a degraded accuracy ($\sim$90\% instead of $>$98\% for the other classes) and a poor precision (20-40\% while they reach 5-10\% for the other classes). Several times in this paper we mentioned that some specific aspects of RV Tau stars might hinder their use as distance indicators through PL/PW relations. Here we briefly discuss a few of them.
\par First, the light curves of RV Tau stars exhibit alternate deep and shallow minima, which have to be taken into account when determining the fundamental period \citep[e.g.,][]{preston1963_SpectroscopicPhotoelectricSurvey}. Moreover, this periodicity is subject to cycle-to-cycle variations \citep[][and references therein]{plachy2018_ChaoticDynamicsPulsation}. Second, and more problematic is the fact that a fraction of them \citep[RVb,][]{pollard1996_RVTauriStars} experience long-term variability of their mean luminosity over periods from a few hundred to a few thousand days, leading to a larger dispersion of PL/PW relations. For distant systems, only LSST \citep{ivezic2019_LSSTScienceDrivers} will allow us to tackle the issue of properly identifying those stars.
\par Third, numerous theoretical and empirical investigations have demonstrated that RV Tau stars are mostly post-AGB stars \citep{ bono2020_EvolutionaryPulsationProperties}, a significant fraction of them exhibiting an infrared excess emanating from a shell or a disk. In the latter case, the disk is the product of circumbinary evolution \citep[e.g.,][]{vanwinckel2009_PostAGBStarsHot,manick2017_EstablishingBinarityAmongst} and affects the infrared luminosities. \citet{manick2018_EvolutionaryNatureRV} reported in addition that dusty RV Tau stars are generally more luminous than their non-dusty counterparts, they speculate that the former evolved from more massive progenitors. Finally, \citet{bodi2019_PhysicalPropertiesGalactic} found that the instability strip of RV Tau stars is broader than the one of CCs (it extends further toward the cooler range), leading to broader PL relations.

\par The presence of a circumbinary disk also affects the chemical composition of RV Tau stars \citep[][and references therein]{giridhar2005_AbundanceAnalysesField}, their atmosphere is underabundant in refractory elements as the result of a mechanism first brought forward by \citet{waters1992_ScenarioSelectiveDepletion}. Since a similar chemical signature has been observed in some W Vir stars \citep{maas2007_ChemicalCompositionsType,lemasle2015_TypeIICepheids}, circumbinary disks might also affect the photometric properties of these stars.

\subsection{Linearity of PL/PW relations}
\label{Linearity of PL relations}

\par As far as CCs are concerned, many studies investigated the controversial linearity of PL/PW relations of DCEP FM stars using various photometric bands and methods. For instance, \cite{sandage2004_NewPeriodluminosityPeriodcolor} found a sudden, statistically significant break in slope at $P$ $\sim$ 10\,d using $BVI$ photometry. Conversely, \cite{testa2007_InfraredPhotometryCepheids} did not find any break or non-linearity in the slope in the NIR $JHK$ bands. In order to assert the (non-)linearity of PL/PW relations in optical ($BVI$) bands, NIR ($JHK$) bands, and in two Wesenheit indices ($W_{BI}$ and $W_{VI}$), \cite{ngeow2008_TestingNonlinearityBVIcJHKs} applied multiple statistical tests and drew a strong conclusion that there is indeed a break in the slope at P=10\,d in the $BVIJH$ PL relations, while the PL relation in the $K$ band and the period-Wesenheit relations P$W_{BI}$ and P$W_{VI}$ are linear. Moreover, \cite{inno2013_DistanceMagellanicClouds} investigate NIR and optical-NIR PW relations using the $VIJHK$ bands and find a complete linearity in all the associated PW relations. \cite{groenewegen2018_CepheidPeriodluminositymetallicityRelation} also did not find a significant non-linearity in the $VK$ bands PL relations and their corresponding PW relation. The spaced-based HST bands have also been investigated. For example, \cite{riess2016_DeterminationLocalValue} report no break in the slope in the PW relation of DCEP FM stars (where the associated Wesenheit index is of the form shown in  Eqn.~\ref{EqW123}). Through an independent investigation, the studies by \cite{kodric2018_M31PAndromedaCepheid} and \cite{li2021_Sub2DistanceM31} also found no broken slope in the PL/PW relations using various HST filters. As far as the $gri$ bands are concerned, only the \citetalias{kodric2018_CepheidsM31PAndromeda} study touched upon this topic to our knowledge. They also report a break in slope at 10\,d, although with a lower significance.

\par There are not many studies that sheds light on the break in the slope of PL/PW relations for DCEP FO stars. \cite{inno2013_DistanceMagellanicClouds} investigates this and finds no non-linearity in their PL/PW relations in $VIJHK$ bands. On the contrary, \cite{ripepi2022_VMCSurveyXLVIII} finds a break for the first time at $P$ = 0.58 d in their $YJK$ bands PL/PW relations.\\

\par Visually inspecting Figs.~\ref{Fig: WVIR_RVTAU_lmc_PL_pp} and \ref{Fig: T2C_lmc_PL_pp_study_comp}, suggests that in the T2C PL/PW relations of the LMC, RV Tau stars have comparatively steeper slopes than the BL Her and the W Vir stars, rendering the PL/PW relations for all T2Cs together non-linear. This has already been reported by \citealp[][]{harris1985_CatalogueFieldType,soszynski2008_OpticalGravitationalLensing,groenewegen2017_PeriodluminosityPeriodradiusRelations}. While studying for the first time in the 21$^{st}$ century, the period-bolometric luminosity relations for T2Cs, \cite{groenewegen2017_PeriodluminosityPeriodradiusRelations} report however purely linear relationships between BL Her, W Vir and RV Tau stars by excluding dusty RV Tau stars possessing IR excess. 

\par A possible non-linearity of PL/PW relation remains therefore quite inconclusive, especially in the case of DCEP FO and T2C stars. We have started to investigate this topic using a more sophisticated (Bayesian) approach, the results will be published in a forthcoming paper (Pipwala et al., in prep).

\subsection{Metallicity dependence}
\label{Metallicity dependence}

\par As shown in Sect. \ref{M31_result}, even though Cepheid variables deliver robust means to estimate distances, their PL/PW relations may depend on their metallicity, as their chemical composition influences their temperature, their evolutionary properties  and, in turn, their brightness. In this section, we briefly discuss the possible metallicity dependence of PL/PW relations of CCs and T2Cs\footnote{Since the samples of ACs from which PL/PW relations are determined are very small (see Table \ref{Table: photometric filtering scheme}), these relations are not robust enough to examine their possible metallicity dependence.}.\\

\par Many theoretical studies point out a metallicity dependence not only on the intercept but also on the slope of the PL/PW relations of CCs in the optical bands, with a focus on DCEP FM stars (for instance, see \citealp{fiorentino2007_ClassicalCepheidPulsation,fiorentino2013_CepheidTheoreticalModels, ngeow2012_TheoreticalCepheidPeriodLuminosity, dicriscienzo2013_PredictedPropertiesGalactic,  gieren2018_EffectMetallicityCepheid}). Observational studies such as \cite{storm2011_CalibratingCepheidPeriodluminositya,storm2011_CalibratingCepheidPeriodluminosity} and \cite{groenewegen2018_CepheidPeriodluminositymetallicityRelation} confirm these predictions.

A metallicity dependence of the slope of the PL relations in the $gri$ bands is also strongly supported by Fig.~\ref{Fig: CC_gri_slope_comparison}: the $gri$ bands slopes estimated by \cite{narloch2023_PeriodLuminosityRelationsGalactic} using galactic DCEP FM stars and the one by \citetalias{kodric2018_CepheidsM31PAndromeda} using the same stars in M31 are in good agreement, and significantly lower than our LMC-based slope estimates. 
We attribute this to a metallicity effect, since the MW and M31 young disk stars display overall a solar-like mean metallicity, whereas LMC Cepheids are more metal-poor by $\sim$ 0.2-0.5\,dex. In the same Figure, a similar trend but with an even more significant difference in the $gri$ band slopes is visible between our study and \citetalias{kodric2018_CepheidsM31PAndromeda} in the case of DCEP FO stars.\\

\par Theoretical studies such as \cite{dicriscienzo2007_SyntheticPropertiesBright} and \cite{das2021_TheoreticalFrameworkBL} found that in contrast to CCs, T2Cs show no significant metallicity dependence for the optical PL/PW relations of individual BL Her stars. Observationally, from a sample of T2Cs located in different Galactic GCs, \cite{matsunaga2006_PeriodluminosityRelationType} found a negligible metallicity dependence in the $JHK$ bands. However, \cite{wielgorski2022_AbsoluteCalibrationNearinfrared} report a significant value of $\gamma$ $\sim$ -0.2 mag/dex in each of these bands using a sample of T2Cs in the field of the MW. Using the abundance measurements of Galactic GCs from the GOTHAM survey \citep[][and references therein]{vasquez2018_HomogeneousMetallicitiesRadial}, \cite{ngeow2022_ZwickyTransientFacility} found instead $\gamma$ close to 0 in all the $VIJHK$ bands and also in the $gri$ bands (see Fig. 5 in their paper). However, they report a significant $\gamma$ term in the $B$ band.\\

\par Our M31 distances based on P$W_{ri}$ relations for T2Cs are in solid concordance (Sect.~\ref{Comparing M31 distances with other robust estimates}) with the robust, metallicity-corrected estimates by \cite{li2021_Sub2DistanceM31} using DCEP FM. This hints at a marginal metallicity dependence for T2Cs in the $ri$ bands, supporting the previous findings of \cite{ngeow2022_ZwickyTransientFacility}.

\section{Conclusions}
\label{conclusion}

\par In this paper we investigate the reliability of T2Cs as a possible means to calibrate the first ladder of the extragalactic distance scale. Choosing M31 as a benchmark galaxy is a very natural choice for this experiment, as it contains large numbers of T2Cs and, especially, CCs, from which the most accurate distance could be derived \citep{li2021_Sub2DistanceM31}. The sample by \citet{kodric2018_M31PAndromedaCepheid} is the largest, most homogeneous, public sample of Cepheids (including thick disk and halo T2Cs). It was built using Pan-STARRS $gri$ photometry, leading us to calibrate our PL/PW relation using the only public data in the same photometric bands, the SkyMapper Southern Survey. 

\par These data are however poorly suited for our purpose, providing $<$10 data points in the light curves, hence a limited accuracy of the Cepheids' mean magnitudes. The large number of stars available and the use of a Bayesian robust regression method compensate for this drawback, and the LMC distances we compute to validate the method show, in general, an accuracy $>$ 99\% when compared to the LMC distance established by \citet{pietrzynski2019_DistanceLargeMagellanic} using double eclipsing binaries.

\par After filtering the M31 sample from potential contaminants thanks to a classification routine based on Bhattacharyya distances, we find a distance to M31 of
24.487$\pm$0.001 (statistical) $\pm$0.052 (systematic) mag using CCs and of 24.409$\pm$0.025 (statistical) $\pm$0.156 (systematic) mag using T2Cs. The excellent agreement between our results and those obtained from the meta-analysis of \citep{degrijs2014_ClusteringLocalGroup}, from HST photometry of CCs \citep{li2021_Sub2DistanceM31} and of RR Lyrae \citep{savino2022_HubbleSpaceTelescope}, and from the TRGB method \citep{conn2012_BayesianApproachLocating} demonstrates that T2Cs provide accurate distances and are therefore reliable tracers for calibrating the extragalactic distance scale. Hence, they allow us to investigate older systems without a young population, and reach farther than the fainter RR Lyrae.\\   

Since M31 is $\sim 70\degree$ inclined \citep{dalcanton2012_PanchromaticHubbleAndromeda}, the pulsating stars may be closer or farther than the average M31's disk plane. For the young population tracers such as DCEP FM and DCEP FO, this may result in only a difference of a few hundredths of a magnitude between the true brightness of the star and their observed brightness \citep{ripepi2012_VMCSurveyFirst, pietrzynski2019_DistanceLargeMagellanic,riess2019_LargeMagellanicCloud}, depending on their projected distance from the line of node of M31. However, since T2Cs have a more halo-like distribution since they are older (see Fig.~\ref{Fig: m31_ra_dec}), the resulting geometric correction on their observed brightness may be larger. Even though \cite{walterbos1988_OpticalStudyStars, seigar2008_RevisedCDMMass,tempel2010_DustcorrectedSurfacePhotometry,courteau2011_LuminosityProfileStructural}, and \cite{dalcanton2012_PanchromaticHubbleAndromeda,dalcanton2023_PanchromaticHubbleAndromeda} provide the 3D geometrical parameters for the M31 disk plane where young stars are concentrated, there is no robust geometrical model available as per our knowledge that also constrains the halo of M31. For the same reason, while calibrating our PL/PW relations using the caLMC data, we assume all stars in this data to be at the same LMC distance of $\mu$ = 18.477 $\pm$ 0.026 mag \citep{pietrzynski2019_DistanceLargeMagellanic}, rather than correcting for this geometric effect.\\

\par Another source of uncertainty that is yet to be explored is the stellar metallicity differences. Indeed, the metallicity effect on PL/PW relations of CCs has an impact on the extragalactic distance scale. \citet{breuval2023_DistanceM33HST} found a uniform $\gamma$ in all bands, with a value of $\sim$ -0.28 mag\,dex$^{-1}$ in agreement with recent studies. They achieved this result by fixing the slope to the LMC one and pushing the metallicity dependence into the $\gamma$ coefficient in the intercept. This is partially because the chemical composition of Galactic, and in particular external, Cepheids is still very sparse \citep[see][for instance]{romaniello2022_IronOxygenContent}. Moreover, we still lack firm constraints on the difference in metallicity between the MW and LMC cepheids, and their metallicity effect. And maybe the depth of LMC is of the same order of magnitude. As far as T2Cs are concerned, the census of their chemical composition is notably worse. In addition, we are dealing with smaller samples of stars (a few 100s instead of a few 1000s) in each galaxy, and they are located in the halo where geometrical corrections are likely more necessary. Notwithstanding, our study supports the absence of a metallicity effect for T2Cs PL/PW relations already proposed in earlier studies.

%\section*{Acknowledgements}

\bibliographystyle{aasjournal}
\bibliography{t2c_paper}

\end{document}